\documentclass[prd,aps,a4paper,twocolumn,eqsecnum,nofootinbib,floatfix]{revtex4}  

\newif\ifusesec
\usesectrue  
   
\usepackage{graphicx} 
\usepackage{amsmath,amsfonts,amssymb}

\newcommand{\beq}{\begin{equation}}
\newcommand{\eeq}{\end{equation}}
\newcommand{\bea}{\begin{eqnarray}}
\newcommand{\eea}{\end{eqnarray}}
\newcommand{\eq}{\eqref}
\newcommand{\g}{{\gamma}}
\newcommand{\e}{\widehat{\mathcal E}_{\rm eff}}

\newcommand{\pinf}{p_{\infty}}
\newcommand{\ie}{{\it i.e.}}

\begin{document}

\title{Sixth post-Newtonian nonlocal-in-time dynamics of binary systems}

\author{Donato Bini$^{1,2}$, Thibault Damour$^3$, Andrea Geralico$^1$}
  \affiliation{
$^1$Istituto per le Applicazioni del Calcolo ``M. Picone,'' CNR, I-00185 Rome, Italy\\
$^2$INFN, Sezione di Roma Tre, I-00146 Rome, Italy\\
$^3$Institut des Hautes \'Etudes Scientifiques, 91440 Bures-sur-Yvette, France
}

\date{\today}

\begin{abstract}
We complete our previous derivation, at the sixth post-Newtonian (6PN) accuracy, of the local-in-time dynamics of a gravitationally 
interacting two-body system by giving two gauge-invariant characterizations of its complementary nonlocal-in-time 
dynamics. On the one hand,  we compute the nonlocal part of the scattering angle for hyberboliclike motions; and, on the
other hand, we compute the nonlocal part of the averaged (Delaunay) Hamiltonian for ellipticlike motions. The former is computed as a 
large-angular-momentum expansion (given here to next-to-next-to-leading order), while the latter is given as a small-eccentricity
 expansion (given here to the tenth order). We note the appearance of $\zeta(3)$ in the nonlocal part of the scattering angle. The averaged Hamiltonian for ellipticlike motions then yields two more gauge-invariant
 observables: the energy and the periastron precession as  functions of orbital frequencies.
 We point out the existence of a hidden simplicity 
in the mass-ratio dependence of the gravitational-wave energy loss of a two-body system. 
\end{abstract}

\maketitle

\section{Introduction}

A new strategy for deriving to higher post-Newtonian (PN) accuracy the  conservative dynamics of gravitationally interacting two-body  
systems has been recently introduced \cite{Bini:2019nra}. This strategy combines, in a new way, various analytical approximation methods:
post-Newtonian (PN), post-Minkowskian (PM), multipolar-post-Minkowskian (MPM), effective-field-theory (EFT), gravitational self-force (SF),
effective one-body (EOB), and Delaunay averaging. In Ref. \cite{Bini:2020wpo}, we have shown how to use this new methodology to derive
the two-body dynamics at the fifth post-Newtonian (5PN), and fifth-and-a-half post-Newtonian (5.5PN) levels. The latter results
were then extended to the sixth post-Newtonian (6PN) level in Ref. \cite{Bini:2020nsb}. 

A basic aspect of our new method is to split the Hamiltonian describing the dynamics of binary systems into two separate parts:
a local-in-time Hamiltonian, $H_{\rm loc, f}$ (which starts at the Newtonian level), and a nonlocal-in-time one, $H_{\rm nonloc, f}$
(which starts at the fourth post-Newtonian, 4PN, level \cite{Damour:2014jta}). The total Hamiltonian,
\beq \label{decompHtotf}
H^{\rm tot} = H^{\rm loc, f}+ H^{\rm nonloc, f}\,,
\eeq
is independent of the choice of the flexibility factor $f(t)$.  The latter enters the nonlocal Hamiltonian via
a multiplicative renormalization of the time scale $\Delta t^f= f(t) \Delta t^h$ used as ultraviolet cutoff in the (external) nonlocal tail action, 
so that one has
\beq \label{decompHnonlocf}
H^{\rm  nonloc, f}(t) =  H^{\rm  nonloc, h}(t) + \Delta^{\rm f-h} H(t)\,,
\eeq
where $H^{\rm  nonloc, h}(t)$ is (uniquely\footnote{We work here at the second-post-Newtonian (2PN) fractional accuracy,
where harmonic coordinates are uniquely defined and lead to a finite higher-order action \cite{DD1981a,D1982}.}) defined by choosing the {\it harmonic-coordinate} cutoff $\Delta t^h=2 r_{12}^h/c$
(where $r_{12}^h$ denotes the two-body radial separation in harmonic coordinates), while
\beq \label{deltaHlnf}
\Delta^{\rm f-h} H(t)= + 2\frac{G H_{}}{c^{5}}{\mathcal F}^{\rm GW}(t)  \ln \left( f(t)\right)\,,
\eeq
is an additional contribution which involves the gravitational wave (GW) energy flux ${\mathcal F}^{\rm GW}(t)$,
and which vanishes when $f(t)=1$. An element of our new method is to choose a flexibility factor $f(t)$ 
such that decomposition  \eq{decompHtotf} of the total Hamiltonian $H^{\rm tot}$ into local and nonlocal parts implies that the two
corresponding parts of the total scattering angle, say, 
\beq \label{chitot}
\chi^{\rm tot}(E,J)= \chi^{\rm loc, f}(E,J)+ \chi^{\rm nonloc, f}(E,J)\,,
\eeq
{\it separately} satisfy the simple mass-ratio dependence proven in Ref. \cite{Damour:2019lcq} for $\chi^{\rm tot}$.
[Here, $\chi$ is considered as a function of the center-of-mass (c.m.) energy, $E$, and 
c.m. angular momentum, $J$, of the binary system.]

In our previous work \cite{Bini:2020nsb} we computed the {\it local-in-time} part of the Hamiltonian, $H^{\rm loc, f}$, at the 6PN accuracy. 
We  gave two  gauge-invariant characterizations of  $H^{\rm loc, f}$. First, we explicitly derived 
the 6PN-accurate contribution to the scattering angle, say  $\chi^{\rm loc, f}_{6 \rm PN}(E,J)$,  coming from $H^{\rm loc, f}$. 
Second, we computed the 6PN-accurate radial action,
\beq
I_R^{\rm loc, f}(E,J)= \frac1{2\pi} \oint dR \, P_R\,,
\eeq
along ellipticlike motions (with energy $E$ and angular momentum $J$) described by $H^{\rm loc, f}$.

The aim of the present work is to complete  the results of 
Ref. \cite{Bini:2020nsb} by deriving the explicit 6PN-accurate values of the complementary  contributions,
both to $\chi$ and to $I_R$,  coming from the  {\it nonlocal-in-time} dynamics, $H^{\rm nonloc, f}$.
More precisely, we shall compute here both  $\chi^{\rm nonloc, f}_{6 \rm PN}(E,J)$ and ${I_R}^{\rm nonloc, f}_{6 \rm PN}(E,J)$,
such that the quantities
\beq \label{chitot6pn}
\chi^{\rm tot}_{6 \rm PN}(E,J)= \chi^{\rm loc, f}_{6 \rm PN}(E,J)+ \chi^{\rm nonloc, f}_{6 \rm PN}(E,J)\,,
\eeq
and
\beq \label{Irtot6pn}
I_{ R \,6 \rm PN}^{\rm tot}(E,J)= I_{ R \,6 \rm PN}^{\rm loc, f}(E,J)+ I_{ R \,6 \rm PN}^{\rm nonloc, f}(E,J)\,,
\eeq
give the  scattering angle (for hyperboliclike motions), and the radial action (for ellipticlike motions) described by the
total Hamiltonian \eq{decompHtotf}, considered at the 6PN accuracy. Because of the nonlocal-in-time nature of 
$H^{\rm nonloc, f}$, it seems impossible to derive (for general motions) closed-form expressions for 
$\chi^{\rm nonloc, f}_{6 \rm PN}(E,J)$ and $I_{ R \,6 \rm PN}^{\rm nonloc, f}(E,J)$. We will compute them in the
form of expansions in a relevant small parameter. For hyperboliclike motions, the expansion parameter is the inverse
eccentricity $\frac1e$, or equivalently the inverse impact parameter $\frac1b$, or the inverse angular momentum $\frac1J$.
For ellipticlike motions, the expansion parameter is the (unperturbed) squared eccentricity $e^2_{\rm loc}(E,J)$, or, 
equivalently, the (unperturbed) radial action $I_R^{\rm loc}(E,J)$. We will also give the 6PN-accurate value of the
energy, and of the periastron advance, along circular orbits.

Let us stress that both quantities Eqs. \eq{chitot6pn}, and 
\eq{Irtot6pn} are  gauge-invariant characteristics of the (6PN-accurate) two-body dynamics. In addition, the left-hand sides of 
Eqs. \eq{chitot6pn}, and \eq{Irtot6pn} are completely independent of the choice of the flexibility factor $f$. It is only
the decomposition into the two parts ($\chi^{\rm loc, f}$ versus $\chi^{\rm nonloc, f}$, 
and  $ I_{ R \,6 \rm PN}^{\rm loc, f}$ versus $ I_{ R \,6 \rm PN}^{\rm nonloc, f}$) which depends on the choice of $f(t)$. Finally, we will
derive below the explicit form of the constraints that must be satisfied by $f(t)$, so that the specific separability condition
(between local and nonocal) that we assumed in our previous work \cite{Bini:2020nsb} is satisfied. The gauge-invariant
content of the corresponding Hamiltonian contribution $\Delta^{f-h} H$ will be explicitly displayed.

The possibility of characterizing (in a gauge-invariant manner) the conservative dynamics of  binary systems by means of
the functional relation between  the radial action, $I_R$, and the energy and angular momentum, $E$, $J$ (or, equivalently,
the functional relation $E(I_R, I_\phi)$, with $I_\phi=\frac1{2\pi} \oint P_\phi d\phi=J$)
is well known in classical mechanics (particularly since the work of Delaunay on the averaging of action-angle Hamiltonians),
and has been emphasized many years ago in the general-relativistic context \cite{Damour:1988mr}.
By contrast,  the possibility of fully characterizing (in a gauge-invariant manner) the conservative dynamics of  binary systems by means of
the functional relation between  the (c.m.) scattering angle $\chi$ and $E$ and $J$ has only been recently emphasized
\cite{Damour:2016gwp,Damour:2017zjx}.  Many different aspects of the physics of classical and quantum scattering (and of the
relation between the two) have been recently explored
\cite{Bini:2017wfr,Bini:2017pee,Bini:2017xzy,Vines:2017hyw,Bini:2018ywr,Bini:2018zxp,Bjerrum-Bohr:2018xdl,Collado:2018isu,Cheung:2018wkq,Caron-Huot:2018ape,Kosower:2018adc,Vines:2018gqi,Guevara:2018wpp,Bern:2019nnu,KoemansCollado:2019lnh,Antonelli:2019ytb,Bautista:2019tdr,KoemansCollado:2019ggb,Cristofoli:2019neg,Cristofoli:2019ewu,Plefka:2019hmz,Maybee:2019jus,Guevara:2019fsj,Bern:2019crd,Bjerrum-Bohr:2019nws,Damgaard:2019lfh,Siemonsen:2019dsu,Kalin:2019rwq,Kalin:2019inp,Bjerrum-Bohr:2019kec,Blumlein:2019bqq,Damour:2019lcq,DiVecchia:2019myk,DiVecchia:2019kta,Bini:2020flp,Abreu:2020lyk,Chung:2020rrz,Cheung:2020gyp,Cristofoli:2020uzm,Bern:2020buy,Parra-Martinez:2020dzs,Kalin:2020mvi,Kalin:2020fhe}.

Let us summarize the current state of the art in the theoretical knowledge of the conservative dynamics of gravitationally interacting two-body  
systems. The PN-expanded dynamics is fully known at the 4PN level (corresponding to $1/c^8$ fractional corrections to the Newtonian description) \cite{Damour:2014jta,Jaranowski:2015lha,Bernard:2015njp,Damour:2016abl,Marchand:2017pir,Foffa:2019rdf,Foffa:2019yfl,Blumlein:2020pog}. At the 5PN level, our new method~\cite{Bini:2019nra} has allowed us to derive, in a gauge-invariant way, the
full dynamics modulo {\it two undetermined} numerical parameters, denoted $\bar d_5^{\nu^2}$ and $a_6^{\nu^2}$.
These coefficients  parametrize terms of the (sketchy) form 
\beq \label{H5pnundet}
\Delta H_{\rm 5PN}^{\rm  loc} \sim \bar d_5^{\nu^2}\frac{G^5 m_1^3 m_2^3}{c^{10} R^5} p_r^2 + a_6^{\nu^2} \frac{G^6 m_1^3 m_2^3(m_1+m_2)}{c^{10} R^6}\,,
\eeq
in the (c.m. frame) local 5PN Hamiltonian.
Here  $m_1$ and $m_2$ denote the two masses, $R= | {\bf x}_1 - {\bf x}_2|$ their radial distance, while $p_r= P_R/\mu$ 
denotes the radial momentum
$ P_R={\bf n}_{12} \cdot {\bf P}_1=- {\bf n}_{12} \cdot {\bf P}_2$, rescaled by the  reduced mass of the system $\mu\equiv m_1 m_2/(m_1+m_2)$. [Note that $p_r$ has the dimension of a velocity, and, actually, is equal, in lowest approximation, to the relative radial
velocity $dR/dt$.]
Recent progress in the (EFT-based) computer-aided evaluation of the PN-expanded interaction potential 
 of binary systems \cite{Foffa:2019hrb,Blumlein:2019zku,Blumlein:2020pog,Blumlein:2020znm}  
gives hope that the two missing coefficients
$\bar d_5^{\nu^2}$ and $a_6^{\nu^2}$ might be soon derived. This would lead to a complete knowledge of the 5PN dynamics.

The 5.5PN Hamiltonian is entirely nonlocal, and it is fully known \cite{Bini:2020wpo}. 
At the 6PN level, our method has allowed us to to derive \cite{Bini:2020nsb}, in a gauge-invariant way, 
the full 6PN dynamics modulo {\it four undetermined} numerical parameters, denoted  $q_{45}^{\nu^2}$,  $\bar d_6^{\nu^2}$,
$ a_7^{\nu^2}$, and $a_7^{\nu^3}$. These coefficients  parametrize terms of the (sketchy) form 
\bea \label{H6pnundet}
\Delta H_{\rm 6PN}^{\rm  loc} &\sim& q_{45}^{\nu^2}\frac{G^5 m_1^3 m_2^3}{c^{12} R^5} p_r^4 + \bar d_6^{\nu^2} \frac{G^6 m_1^3 m_2^3(m_1+m_2)}{c^{12} R^6}p_r^2 \nonumber\\
&+&  a_7^{\nu^2} \frac{G^7 m_1^3 m_2^3(m_1+m_2)^2}{c^{12} R^7}\nonumber\\
&+&  a_7^{\nu^3} \frac{G^7 m_1^4 m_2^4}{c^{12} R^7}\,,
\eea
in the (c.m. frame) local 6PN Hamiltonian.

Besides this knowledge of the PN-expanded dynamics (i.e., its expansion in powers of $\frac1c$), 
one  has also recently acquired the knowledge of the first three terms in
the (conservative) PM-expanded dynamics, i.e., its expansion in powers of the gravitational coupling constant $G$ (keeping the 
velocity dependence exact). Hamiltonian formulations of the first post-Minkowskian (1PM, i.e. $O(G)$) dynamics 
have been derived in various gauges \cite{Ledvinka:2008tk,Damour:2016gwp}. The second post-Minkowskian (2PM, i.e. $O(G^2)$) dynamics,
whose equations of motion had been known for many years  \cite{Westpfahl:1979gu,Bel:1981be,Westpfahl:1985}, was expressed only
recently in Hamiltonian form \cite{Damour:2017zjx,Cheung:2018wkq}. The third post-Minkowskian (3PM, i.e. $O(G^3)$) dynamics 
has been derived in Refs. \cite{Bern:2019nnu,Bern:2019crd} (see also Refs. \cite{Antonelli:2019ytb,Damour:2019lcq} for its simpler EOB formulation). Confirmations of the 3PM dynamics of Refs. \cite{Bern:2019nnu,Bern:2019crd} have been obtained
in Refs. \cite{Bini:2019nra} (5PN level),  \cite{Blumlein:2020znm,Cheung:2020gyp,Bini:2020wpo} (6PN level)
and \cite{Kalin:2020fhe} (3PM level).

Eqs. \eq{H5pnundet}, \eq{H6pnundet} clearly display the fact that the parts of the 5PN and 6PN dynamics left undetermined 
by our new method belong to the fifth, sixth and seventh post-Minkowskian (5PM, 6PM, 7PM) approximations. This shows, in particular,
that our current work leads to a complete knowledge of the fourth post-Minkowskian (4PM; $O(G^4)$) dynamics up to the 6PN level included.
However, in order to explicate this knowledge (in a gauge-invariant way)  from our current results \cite{Bini:2020wpo,Bini:2020nsb},
one needs to explicitly derive the (f-route) {\it nonlocal} contribution, $\chi^{\rm nonloc, f}_{6 \rm PN}(E,J)$, to the
total scattering angle, $\chi^{\rm tot}_{6 \rm PN}(E,J)$, Eq. \eq{chitot}, so as to complete the explicit expression for the (f-route) local
contribution $\chi^{\rm loc, f}_{6 \rm PN}(E,J)$ given in Ref. \cite{Bini:2020nsb}.

Our basic tool for deriving the nonlocal contribution to the scattering angle will be the general, simple formula, derived in Ref. \cite{Bini:2017wfr},
that computes the additional contribution $\delta \chi(E,J)$ to $\chi(E,J)= \chi_0(E,J) + \delta \chi(E,J)$ induced by an additional contribution $\delta H$ to the
Hamiltonian ($H(q,p)= H_0(q,p) + \delta H(q,p)$), namely
\beq \label{deltachigen}
\delta \chi(E,J) = \frac{\partial}{\partial J}  W_{\rm hyp}(E,J)+ O\left[(\delta H)^2 \right]\,,
\eeq
where
\beq \label{Wgen}
 W_{\rm hyp}(E,J) \equiv  \int_{- \infty}^{+ \infty} dt \,\delta H\,,
\eeq
is integrated along the unperturbed hyperboliclike motion (with energy $E$ and angular momentum $J$) defined by the unperturbed Hamiltonian $H_0$. Note the important point that Refs. \cite{Damour:2014jta,Damour:2015isa,Bini:2017wfr}  
have shown that the relation \eq{deltachigen}, which is easily derived for usual {\it local} Hamiltonians, holds also in the present case of a {\it nonlocal} Hamiltonian.

Similarly, it is easy to relate the elliptic-motion analog of \eq{Wgen},  say
\beq \label{Well}
 W_{\rm ell}(E,J) \equiv  \oint dt \,\delta H\,,
\eeq
where, now, the integral is taken over one radial period of an ellipticlike motion, to the (first-order) perturbation $\delta I_R(E,J)$ of the
radial action, 
\beq
I_R(E,J)= I_R^0(E,J)+ \delta I_R(E,J)\,, 
\eeq
corresponding to a general perturbation $H= H_0(q,p) + \delta H(q,p)$ of the Hamiltonian. 
Indeed, the fundamental property of Delaunay averaging (for ellipticlike motions) is that the perturbation $\delta {\bar H}(I_R, I_\phi)$
of the angle-averaged Delaunay Hamiltonian, 
\beq
{\bar H}(I_R, I_\phi) = \frac1{\oint dt} \oint dt H={\bar H}_0(I_R, I_\phi)+ \delta {\bar H}(I_R, I_\phi)\,,
\eeq
is simply given by averaging the perturbation of the Hamiltonian\footnote{This fundamental result of classical mechanics
played an important role in the development of Quantum Mechanics, where it got transmuted into the well-known 
Hellman-Feynman theorem.} , so that
\bea \label{deltabarH}
\delta {\bar H}(I_R, I_\phi) &=&  \frac1{\oint dt} \oint dt \delta H(q,p) = \frac{\Omega_R}{2\pi}\oint dt \delta H(q,p)\nonumber\\
&=&  \frac{\Omega_R}{2\pi}  \left[W_{\rm ell}(E,J)\right]_{E\mapsto {\bar H}_0(I_R, I_\phi)}\,.
\eea
Here, $\Omega_R= \frac{2\pi}{T_R} = \partial {\bar H}(I_R, I_\phi)/ \partial I_R$ denotes the radial angular frequency 
($T_R= \oint dt$ denoting the radial period). Note that in the last equation \eq{deltabarH} one can use the leading-order replacement 
$E\mapsto {\bar H}_0(I_R, I_\phi)$ to express $\delta {\bar H}$ as a function of $I_R$, and $I_\phi$, instead of the natural
variables  $E$, $J$ entering the integrated action $W_{\rm ell}(E,J)$, \eq{Well}.
Writing that $I_R(E,J)$ is the inverse function of ${\bar H}(I_R, I_\phi)$, and using $\Omega_R= \partial {\bar H}(I_R, I_\phi)/ \partial I_R$, 
also leads to the result that the perturbation $\delta I_R(E,J)$ of the
radial action, $I_R(E,J)= I_R^0(E,J)+  \delta I_R(E,J)$ is simply given by 
\beq \label{deltaIRvsW}
\delta I_R(E,J) = -\frac1{2\pi}  W_{\rm ell}(E,J) + O\left[(\delta H)^2\right]\,,
\eeq
where $W_{\rm ell}(E,J)$ is again the integrated elliptic-motion action defined in Eq. \eq{Well}.
Note in passing that by combining the result \eq{deltaIRvsW} with the standard general result for the periastron advance $\Phi$
(see, e.g., \cite{Damour:1988mr})
\beq
\frac{\Phi(E,J)}{2\pi}= - \frac{\partial I_R(E,J)}{\partial J}\,,
\eeq
one finds that the perturbation $\delta \Phi(E,J)$ of the  periastron advance $\Phi(E,J)=\Phi_0(E,J) + \delta \Phi(E,J)$ is given by
\beq
\delta \Phi(E,J) = + \frac{\partial  W_{\rm ell}(E,J)}{\partial J}\,.
\eeq
 
In the present paper we shall apply the general results of Eqs. \eq{deltachigen}, \eq{Wgen}, \eq{Well}, \eq{deltabarH}, to the perturbed 
dynamics $H=H_0+ \delta H$ with
\bea
 H_0 &=&H^{\rm loc, f}\,,\nonumber\\
 \delta H &=& H^{\rm nonloc, f}= H^{\rm nonloc, h} + \Delta^{f-h} H\,.
\eea
As we have derived in Refs. \cite{Bini:2020wpo,Bini:2020nsb} the contributions of $ H_0 =H^{\rm loc, f}$ both to
the scattering angle, $\chi^{\rm loc, f}_{6 \rm PN}(E,J)$ (see Section VIII in \cite{Bini:2020nsb}), and to
the Delaunay averaged Hamiltonian $H^{\rm loc, f}_{6 \rm PN}(I_R, I_\phi)$, or equivalently $I^{\rm loc, f}_{\rm R \,6  PN}(E,J)$ (see Tables
X and XI  in \cite{Bini:2020wpo} and section IX in \cite{Bini:2020nsb}), we only need now to compute the complementary
contributions 
\beq
\delta \chi(E,J) = \chi^{\rm nonloc, f}_{6 \rm PN}(E,J)=\frac{\partial}{\partial J}  W^{\rm nonloc, f}_{\rm hyp}(E,J)\,,
\eeq
and
\bea
\delta {\bar H }(I_R, I_\phi) &=& {\bar H}^{\rm nonloc, f}_{6 \rm PN}(I_R, I_\phi) \nonumber\\
&=& \frac{\Omega_R}{2\pi}  \left[W^{\rm nonloc, f}_{\rm ell}(E,J)\right]_{E\mapsto {\bar H}_0(I_R, I_\phi)}\,.
\eea
From the latter result, we shall then be able to deduce the nonlocal contribution to the periastron advance
\beq
\delta^{\rm nonloc, f} \Phi(E,J) = + \frac{\partial  W^{\rm nonloc, f}_{\rm ell}(E,J)}{\partial J}\,.
\eeq
 Our first task will then be to compute the f-route, nonlocal perturbed action along hyperbolic motions, i.e.
\beq
W^{\rm nonloc, f}_{\rm hyp}(E,J)=  \int_{- \infty}^{+ \infty} dt \, H^{\rm nonloc, f}(t)\,.
\eeq
In view of the linear decomposition \eq{decompHnonlocf} of the  f-route nonlocal Hamiltonian, $H^{\rm nonloc, f}$, we have a corresponding
linear decomposition of  $W^{\rm nonloc, f}_{\rm hyp}(E,J)$, namely
\beq \label{decompWnonlocf}
W^{\rm nonloc, f}_{\rm hyp}(E,J)= W^{\rm nonloc, h}_{\rm hyp}(E,J)+\Delta^{\rm f-h}_{\rm hyp} W(E,J)\,,
\eeq
where
\beq
W^{\rm nonloc, h}_{\rm hyp}(E,J)=  \int_{- \infty}^{+ \infty} dt \, H^{\rm nonloc, h}(t)\,,
\eeq
and
\beq
\Delta^{\rm f-h}_{\rm hyp} W(E,J)= \int_{- \infty}^{+ \infty} dt \, \Delta^{\rm f-h} H(t)\,,
\eeq
both integrals being evaluated along an hyperbolic motion of $H_0 =H^{\rm loc, f}_{\rm 6 PN}$ with energy $E$ and
angular momentum $J$. Actually, as nonlocal effects start at the 4PN level, it is enough to use as $H_0$ in this calculation
the 2PN-accurate Hamiltonian (whose Delaunay form was given in \cite{Damour:1988mr}; see Appendix \ref{PN}).

While $\Delta^{\rm f-h}_{\rm hyp} W(E,J)$ can be (and will be) computed in closed form, it does not seem possible to compute
$W^{\rm nonloc, h}_{\rm hyp}(E,J)$ in closed form. But, it will be enough for our purposes
to compute the first three terms  in the large-$J$ (or large eccentricity) expansion of
the function, $W^{\rm nonloc, h}_{\rm hyp}(E,J)$, namely 
\bea \label{Wnonlochxp}
W^{\rm nonloc, h}_{\rm hyp}(E,J)&=& W_4(E)\frac{(Gm_1m_2)^4}{J^3}\nonumber\\
&+& W_5(E) \frac{(Gm_1m_2)^5}{J^4} \nonumber\\
&+& W_6(E)\frac{(Gm_1m_2)^6}{J^5}\nonumber\\
&+& O\left(\frac{G^7}{J^6}\right).
\eea
As displayed here, this expansion in powers of $\frac1J$ is also a PM expansion in powers of $G$.
In view of Eq. \eq{deltachigen}, the corresponding expansion for the (h-route) nonlocal contribution to the scattering angle reads
\bea \label{chinonlochxp0}
\chi^{\rm nonloc, h}(E,J) &=& - 3 W_4(E)\frac{(Gm_1m_2)^4}{J^4}\nonumber\\
&-& 4 W_5(E) \frac{(Gm_1m_2)^5}{J^5} \nonumber\\
&-& 5W_6(E)\frac{(Gm_1m_2)^6}{J^6}\nonumber\\
&+& O\left(\frac{G^7}{J^7}\right).
\eea
While we will be able to analytically compute closed-form expressions for the first two expansion coefficients $W_4(E)$ and $W_5(E)$, 
we will only be able to write down integral expressions for  the third expansion coefficient $W_6(E)$. We did not succeed in analytically
computing the latter integral expressions, but we could estimate then numerically.

Our next task will be to use the mass-ratio dependence of the coefficients $W_4(E)$, $W_5(E)$ and $W_6(E)$ to constrain
the choice of the flexibility factor $f(t)$. Indeed, as recalled above, the choice of $f(t)$ is constrained, within our method, by
requiring that the two parts, $\chi^{\rm  loc, f} $ and $\chi^{\rm  nonloc, f}=\chi^{\rm  nonloc, h}+  \chi^{\rm f- h}$ of the total scattering angle $\chi^{\rm tot}$, Eq. \eq{chitot},
{\it separately} satisfy the simple mass-ratio dependence proven in Ref. \cite{Damour:2019lcq} for $\chi^{\rm tot}$.

Finally, we will complete our 6PN-accurate description of the dynamics of ellipticlike motions by computing the elliptic analog
of Eq. \eq{decompWnonlocf}, namely
\beq \label{decompWnonlocfell}
W^{\rm nonloc, f}_{\rm ell}(E,J)= W^{\rm nonloc, h}_{\rm ell}(E,J)+\Delta^{\rm f-h}_{\rm ell} W(E,J)\,,
\eeq
with
\beq
W^{\rm nonloc, h}_{\rm ell}(E,J)=  \oint dt \, H^{\rm nonloc, h}(t)\,,
\eeq
and
\beq
\Delta^{\rm f-h}_{\rm ell} W(E,J)= \oint dt \, \Delta^{\rm f-h} H(t)\,,
\eeq
both integrals being now evaluated along {\it one radial period} of an elliptic motion, with given energy $E$ and
angular momentum $J$  of $H_0 =H^{\rm loc, f}_{\rm 6 PN}$. As before, it is enough to use  $H_0 \approx H_{\rm 2 PN}$
in this calculation. 

\subsection*{Notation}

We use a mostly plus signature. We define the symmetric mass ratio $\nu$ as the ratio of 
the reduced mass $\mu\equiv m_1 m_2/(m_1+m_2)$ to the total mass $M=m_1+m_2$:
\beq \label{defnu}
\nu\equiv \frac{\mu}{M}=\frac{m_1m_2}{(m_1+m_2)^2}\,.
\eeq
We use several different measures of the total energy $E_{\rm tot} = Mc^2 + \cdots$ of the binary system 
(considered in the c.m. frame). Of particular importance is the  EOB effective energy, ${\mathcal E}_{\rm eff}$,
which is  defined by
\beq
{\mathcal E}_{\rm eff}= \frac{E_{\rm tot}^2- m_1^2 c^4 - m_2^2 c^4}{2 (m_1+m_2) c^2}\,.
\eeq
Equivalently, we have
\bea \label{eobmap}
E_{\rm tot} &=& M c^2\sqrt{1 + 2 \nu \left( \frac{{\mathcal E}_{\rm eff}}{\mu c^2}-1\right)}\nonumber\\
&\equiv& M c^2 \sqrt{1 + 2 \nu(\e-1)}\,,
\eea
where
\beq
\e \equiv \frac{{\mathcal E}_{\rm eff}  }{\mu \,c^2}\,.
\eeq
We also use the dimensionless specific binding energy 
\beq \label{defbarE}
\bar E \equiv \frac{E_{\rm tot}-Mc^2}{\mu c^2}\,.
\eeq
The total c.m. angular momentum $J$ will often be measured by its dimensionless rescaled version
\beq \label{defj2}
 j\equiv \frac{c J}{G m_1 m_2}=\frac{c J}{G M \mu}\,.
 \eeq
[The definitions used in the present work for $\bar E $ and $j$ differ by respective factors $\frac1{c^2}$ and $c$ from those
 used in our last work \cite{Bini:2020nsb}.]
The latter equation shows that one can formally consider that $j = O\left(\frac{c}{G}\right)$,
so that a term or order $\frac1{j^n}$ is of order $\frac{G^n}{c^n}$.

In the following, we shall often use the shorthand notations
\beq
\g \equiv \e\,,
\eeq
\beq
\pinf \equiv \sqrt{\g^2-1}, \; {\rm so \; that}\; \gamma =\sqrt{1+\pinf^2}\,,
\eeq
and
\beq
h(\g, \nu) \equiv \sqrt{1+2\nu(\gamma-1)}\,.
\eeq
We shall often find convenient to work with 
dimensionless rescaled orbital parameters, such as $r_{12} \equiv c^2 r_{12}^{\rm phys}/(GM)$, or 
$a \equiv c^2 a^{\rm phys}/( GM)$. The context should make it clear whether we use physical or rescaled quantities.

Most of our final results will be expressed in terms of dimensionless quantities, such as $\bar E$, $j$, $\pinf$, and
$a \equiv c^2 a^{\rm phys}/( GM)$. In other words, we essentially use units where $c$ and $G$ (and sometimes also
$GM$) are set to unity. However, in some formulas we indicate the powers of $G$ (or $GM$) that they originally contain.
Concerning the powers of $c$, and the corresponding {\it absolute} PN order, we will not explicitly keep track of them.
However, we will keep track of the {\it fractional} PN order of various contributions to PN-expanded quantities
by using $\eta \sim \frac1c$ (to be set to one at the end) as a bookkeeping device for PN 
orders beyond the leading-order term in a quantity. E.g.,
we will write $Q=Q^{\rm LO} (1+ \eta^2 q_2 + \eta^4 q_4)$ for a quantity $Q$ which is expanded to
fractional 2PN accuracy beyond its leading order PN contribution. To help the reader keeping track of the absolute PN order
of the quantities we shall compute, let us note that: (i) nonlocal effects in the dynamics start at the absolute 4PN order,
and (ii) one can use the formal scalings  $\frac1j=O\left( \frac{G}{c}\right)$, $\bar E= O(\frac1{c^2})= \g-1$, 
  and $\pinf=O(\frac1c)$ to recover the powers of $G$ and $c$.

\section{Brief reminder about the nonlocal part of the action}

Let us consider in more detail the structure of the nonlocal part of the action, $S_{\rm nonloc, f}$.
As discussed in  Ref. \cite{Bini:2020wpo}, at the 6PN accuracy the nonlocal  action can be linearly decomposed into its 4+5+6PN piece, 
and its   5.5PN piece,
\beq \label{Snonloc00}
S_{\rm nonloc, f}^{ \leq 6 \rm PN}= S_{\rm nonloc, f}^{ 4+5+6 \rm PN}+  S_{\rm nonloc}^{ 5.5 \rm PN}\,,
\eeq
where each piece is a time-nonlocal functional of the two worldlines (considered in the center-of-mass frame)
\beq 
\label{Snonloc_0}
S_{\rm nonloc, f}^{ 4+5+6\rm PN}[x_1(s_1), x_2(s_2)]= -\int dt \, H_{\rm nonloc, f}^{4+5+6\rm PN}(t)\,, 
\eeq
and 
\beq \label{S5.5pn}
S_{\rm nonloc}^{\rm 5.5PN}[x_1(s_1), x_2(s_2)]=-\int dt H_{\rm nonloc}^{\rm 5.5 PN}(t)\,.
\eeq
The two nonlocal Hamiltonians $H_{\rm nonloc, f}^{4+5+6\rm PN}(t)$ and $H_{\rm nonloc}^{\rm 5.5 PN}(t)$ are given by 
integrals over a shifted time $t'\equiv t+\tau$. The  $\tau$ integral entering $H_{\rm nonloc, f}^{4+5+6\rm PN}(t)$ is
 logarithmically divergent when $\tau \to 0$, and is defined by introducing a specific (Hadamard Partie finie, Pf)
 time-scale $\Delta t_f=2 r_{12}^f(t)/c$. By contrast, the  $\tau$ integral entering $H_{\rm nonloc}^{\rm 5.5 PN}(t)$
 is  convergent when $\tau \to 0$, and therefore involves no regularization scale.
 
More precisely, the  4+5+6PN piece reads
\begin{eqnarray} 
\label{Hnonloc0}
H_{\rm nonloc, f}^{4+5+6 \rm PN}(t) &=& \frac{G {\cal M}}{c^3} {\rm Pf}_{2 r_{12}^f(t)/c} \int  \frac{ dt'}{|t-t'|} {\cal F}_{ \rm 2PN}^{\rm split}(t,t')\,.\nonumber\\
\end{eqnarray} 
Here, ${\cal M}$ denotes the total ADM conserved mass-energy of the binary system;
\beq
r_{12}^f(t)= f(t) r_{12}^h(t)\,,
\eeq
is a flexed version of the radial distance between the two bodies ($ r_{12}^h(t)$ denoting the harmonic-coordinate distance
and $f(t)$ being a function of the instantaneous state of the system), while ${\cal F}_{ \rm 2PN}^{\rm split}(t,t')$ is
the time-split version of the fractionally 2PN-accurate gravitational-wave energy flux (absorbed and) emitted  by the (conservative) system.

On the other hand, the 5.5 PN  Hamiltonian is given by the following nonlocal (second-order tail) expression
\begin{eqnarray} \label{H5.5}
H_{\rm nonloc}^{\rm 5.5 PN}(t)&=&\frac{B}{2}\left(\frac{G {\cal M}}{c^3} \right)^2 \int_{-\infty}^\infty \frac{d\tau}{\tau}[{\mathcal G}^{\rm split}(t,t+\tau)\nonumber\\
&&
-{\mathcal G}^{\rm split}(t,t-\tau)]\,,
\end{eqnarray}
with $B=-\frac{107}{105}$.
Similarly to the first-order tail effect entering $H_{\rm nonloc, f}^{4+5+6 \rm PN}(t)$, this action
 involves a time-split bilinear  function of the multipole moments that is closely linked to the gravitational-wave flux,
 namely
 \beq
{\mathcal G}^{\rm split}(t,t')=\frac{G}{5c^5} I_{ij}^{(3)}(t) I_{ij}^{(4)}(t') + \ldots\,.
\eeq
 At the present 6PN accuracy, it is enough to use the leading-order version of the time-split function ${\mathcal G}^{\rm split}(t,t')$, 
 obtained by keeping only  the  quadrupolar contribution (neglecting higher multipole terms), and by evaluating  $I_{ij}(t)$ at the Newtonian level.

Up to the 7PN-accuracy included, each piece of the nonlocal action can be treated as a first-order perturbation of the (local) 3PN
dynamics, and their contributions to the scattering angle can be treated separately, and then linearly added together.

The  4+5+6PN nonlocal Hamiltonian can be further decomposed into its purely harmonic, unflexed contribution 
$ H_{\rm  nonloc, h}^{4+5+6 \rm PN}$ (defined by using $\Delta t_h=2 r_{12}^h(t)/c$ as Pf scale), and a contribution 
$\Delta^{\rm f-h} H(t)$ proportional to $\ln f(t)$:
\beq \label{Hnonlocf}
H_{\rm  nonloc, f}^{4+5+6 \rm PN}(t) =  H_{\rm  nonloc, h}^{4+5+6 \rm PN} + \Delta^{\rm f-h} H(t)\,.
\eeq
Replacing ${\cal M} = \frac{E_{\rm tot}}{c^2}=\frac{H}{c^2}$ where $H$ is the (2PN-accurate, as needed for the present computation) Hamiltonian, and 
 introducing an intermediate length scale $s$, we have
\begin{eqnarray} 
\label{Hnonloch}
 H_{\rm nonloc, h}^{4+5+6 \rm PN}(t)&=&
-\frac{G H_{}}{c^{5}}{\rm Pf}_{2s/c}\int \frac{d\tau}{|\tau|}{\mathcal F}^{\rm split}_{\rm 2PN}(t,t+\tau)\nonumber\\
&+&2\frac{G H_{}}{c^{5}}{\mathcal F}^{\rm split}_{\rm 2PN}(t,t)  \ln \left( \frac{r_{12}^h(t)}{s}\right)
\,,
\end{eqnarray} 
and
\beq \label{DHfh}
\Delta^{\rm f-h} H(t)= + 2\frac{G H_{}}{c^{5}}{\mathcal F}^{\rm split}_{\rm 2PN}(t,t)  \ln \left( f(t)\right)\,.
\eeq

\subsection{Scattering angle}

As already mentioned the ``f-route"  local Hamiltonian $ H_{\rm loc, f}$ is defined so that
\beq
H_{\rm tot}= H_{\rm loc, f}+ H_{\rm nonloc, f}\,,
\eeq
where $H_{\rm nonloc, f}$ is defined by Eqs. \eq{Hnonlocf}, \eq{Hnonloch}, \eq{DHfh}.
Refs. \cite{Bini:2020wpo,Bini:2020nsb} have determined $H_{\rm loc, f}$ at the 6PN accuracy.
In order to complete the derivation of the f-route  6PN dynamics we need to compute 
the h-route nonlocal part of the scattering angle, say $\chi^{\rm nonloc, h}$, at the 6PN accuracy, \ie,
at order $\frac1{c^{\leq 12}}$, and at the 6PM accuracy, \ie, at order $G^{\leq 6}$. Indeed, it is
the $\nu$-dependence of $\chi^{\rm nonloc, h}$ which constrains the additional, f-dependent
contribution $\chi^{\rm f-h}$ needed to render $\chi^{\rm nonloc, f}= \chi^{\rm nonloc, h}+ \chi^{\rm f-h}$
compatible with the particular  $\nu$-dependence of $\chi^{\rm tot}$ pointed out in Ref. \cite{Damour:2019lcq}.

The leading-order (LO) contribution to $\chi^{\rm nonloc, f}$ is at the 4PN and 4PM levels (i.e.,  of order $\frac{G^4}{c^8}$). In view of the PN and PM scalings of $\pinf$ and $\frac1j$
recalled above, this means that the LO contribution to $\chi^{\rm nonloc, f}$ starts by a contribution of order  $\frac{\pinf^4}{j^4}$.
Beyond this order, we can (see Eq. \eq{chinonlochxp0}, which concerned $\chi^{\rm nonloc, h}$) 
write an expansion for $\chi^{\rm nonloc, f}$ of the type
\bea 
\label{chinonlocxp}
\chi^{\rm nonloc, f}(\pinf,j;\nu) &=& \nu \frac{\pinf^4}{j^4} \left( A_0(\pinf;\nu) + \frac{A_1(\pinf;\nu)}{\pinf j}\right. \nonumber\\
&&\left.+ \frac{A_2(\pinf;\nu)}{(\pinf j)^2} + \cdots \right),
\eea
where $A_0(\pinf;\nu)$, $A_1(\pinf;\nu)$, $A_2(\pinf;\nu)$, etc, are further PN-expanded in powers of $\pinf$.
Namely:
\bea \label{Axp}
A_0(\pinf;\nu)&=&A_0^{\rm N}+ \eta^2 A_0^{\rm 1PN}+ \eta^4 A_0^{\rm 2PN}+\cdots,\nonumber\\
A_1(\pinf;\nu)&=&A_1^{\rm N}+ \eta^2 A_1^{\rm 1PN}+ \eta^3 A_1^{\rm 1.5PN}+ \eta^4 A_1^{\rm 2PN}+\cdots, \nonumber\\
A_2(\pinf;\nu)&=&A_2^{\rm N}+ \eta^2 A_2^{\rm 1PN}+ \eta^3 A_2^{\rm 1.5PN}+\eta^4 A_2^{\rm 2PN}+\cdots.\nonumber\\
\eea
Here the label N (standing for Newtonian) denotes a term of order $\pinf^0$, modulo a $\ln \pinf$ correction,
while the label 1PN (respectively, 1.5PN or 2PN) denotes a term of order $\pinf^2$ (respectively, $\pinf^3$ or $\pinf^4$), 
modulo $\ln \pinf$ corrections. [As explained in the Introduction, $\eta (=1)$ is used as a bookeeping parameter
for counting the {\it fractional} PN orders.]
As we shall see the 1.5PN fractional corrections come from the 5.5PN nonlocal action,
and only contribute at orders $\frac1{j^{\geq 5}}$ (\ie, $G^{\geq 5}$).
We recall that the powers of $\frac1j$ count the powers of $G$, \ie, the PM order. It should also be noted that
the product $\pinf j$ in the denominators entering Eq. \eq{chinonlocxp} scales like $c^0$, \ie, is of Newtonian order.
Actually, at the Newtonian level, the quantity 
\beq
\label{ecc_newt}
e_{\rm N}\equiv \sqrt{1+ \pinf^2 j^2}\,,
\eeq
measures the eccentricity of the hyperbolic trajectory of a scattering motion. The PM-expansion in powers of 
$\frac1j \sim G$ used in Eq. \eq{chinonlocxp} is also a large-eccentricity expansion.

As already explained in the Introduction, the combined PN and PM expansion of $\chi^{\rm nonloc, f}$, 
Eq. \eq{chinonlocxp}, will be obtained 
by computing the various contributions to the integrated nonlocal action, 
\beq
W^{\rm nonloc, f}_{\rm hyp}(\pinf,j;\nu)= \int_{-\infty}^{\infty} dt H_{\rm nonloc, f}(t)\,,
\eeq 
and then by differentiating it with respect to $j$. We can rewrite Eq. \eq{deltachigen} (setting $c=1$) as
\beq \label{chivsW}
\chi^{\rm nonloc, f}(\pinf,j;\nu)= \frac1{G M^2\nu} \frac{\partial W^{\rm nonloc, f}_{\rm hyp}(\pinf,j;\nu)}{\partial j}\,.
\eeq
In the following, we shall use the shorthand notation\footnote{Beware of distinguishing the use of the notation 
$\langle \cdots \rangle_\infty$ for an hyperbolic-motion {\it integral} from the use of $\langle \cdots \rangle$ 
for denoting an elliptic motion {\it average}. }
\beq
\langle H_{\rm nonloc, X}\rangle_\infty \equiv  \int_{-\infty}^{\infty} dt\, H_{\rm nonloc, X}(t)\,,
\eeq
for the various time-integrated contributions to the nonlocal Hamiltonian (where $X$ is a label for these contributions).

The total nonlocal potential $W^{\rm nonloc, f}_{\rm hyp}(\pinf,j;\nu)= \langle \sum_X H_{\rm nonloc, X}\rangle_\infty$ is then
decomposed as
\beq \label{Wnonloctot}
W^{\rm nonloc, f}_{\rm hyp}= W^{\rm tail,h} + W^{\rm tail,f-h}+ W^{\rm 5.5PN},
\eeq
where 
\bea \label{decompWnonloc}
W^{\rm tail,h}&\equiv& \langle H_{\rm nonloc,h}^{4+5+6 \rm PN} \rangle_\infty ,\nonumber\\
W^{\rm tail,f-h}&\equiv& \langle \Delta^{\rm f-h} H(t) \rangle_\infty ,\nonumber\\
W^{\rm 5.5PN}&\equiv& \langle H_{\rm nonloc}^{\rm 5.5PN} \rangle_\infty.
\eea
For brevity, we used the label \lq\lq tail"  to denote the (4+5+6PN) first-order tail contribution of Eq. \eq{Hnonloc0},
which is proportional to $\frac{G {\cal M}}{c^3}$. The second-order tail contribution of Eq. \eq{H5.5} (which is
proportional to $\left(\frac{G {\cal M}}{c^3} \right)^2$) is simply denoted by the label 5.5PN because it will be
evaluated at this accuracy.
 
 In the following sections, we shall successively compute $W^{\rm tail,h}$, $W^{\rm tail,f-h}$ and
 $W^{\rm 5.5PN}$. Of particular importance will be to control the $\nu$-dependence of these quantities.
 As the split fluxes ${\mathcal F}^{\rm split}(t,t')$ and ${\mathcal G}^{\rm split}(t,t')$ contain an overall
 factor $\nu^2$ (coming from $I_{ij} = \mu x^{\langle i} x^{j \rangle}+ \cdots$, etc.), each contribution to $W^{\rm nonloc, f}$
 will contain an overall factor $\nu^2$. [This applies also to $W^{\rm tail,f-h}$ whose role is to compensate some terms in $W^{\rm tail,h}$.]
 We then see from Eq. \eq{chivsW} that $\chi^{\rm nonloc, f}$ contains
 an overall factor $\nu^1$, which has been factored out in Eq. \eq{chinonlocxp}. The $\nu$-dependence of the
 coefficients $A_n(\pinf;\nu)$ entering the large-eccentricity expansion \eq{Axp} will then be generated by the
 $\nu$-dependence of the solution of the hyperbolic motion $x^i(t)$ inserted in the computation of the 
 ($\nu$-dependent) multipole moments $I_{ij}(t)$, etc.

\section{Computation of $W^{\rm tail,h}\equiv \langle H_{\rm nonloc,h}^{4+5+6 \rm PN} \rangle_\infty$}

Let us start with the computation of  the time integral of $H_{\rm nonloc,h}^{4+5+6 \rm PN}(t)$ along a
2PN-accurate hyperboliclike motion in harmonic coordinates.
The time-split version of the fractionally 2PN-accurate gravitational-wave energy flux ${\cal F}_{ \rm 2PN}^{\rm split}(t,t')$ emitted by the system can be written as 
\begin{eqnarray}
{\cal F}_{ \rm 2PN}^{\rm split}(t,t')&=&\frac{G}{c^5} \left[F_{ I_2}^{\rm split}(t,t')+\eta^2F_{I_3, J_2}^{\rm split}(t,t')\right. \nonumber\\
&& \left.+\eta^4F_{ I_4, J_3}^{\rm split}(t,t')\right]\,,
\end{eqnarray}
where
\begin{eqnarray}
\label{flux2PNdef}
F_{  I_2}^{\rm split}(t,t')&=& \frac15 I_{ab}^{\rm (3)}(t) I_{ab}^{\rm (3)}(t') \,,\nonumber\\
F_{ I_3, J_2}^{\rm split}(t,t')&=& \frac1{189 } I_{abc}^{\rm (4)}(t) I_{abc}^{\rm (4)}(t') +\frac{16}{45 } J_{ab}^{\rm (3)}(t) J_{ab}^{\rm (3)}(t')\,,\nonumber\\
F_{I_4, J_3}^{\rm split}(t,t')&=& \frac{1}{9072}I_{abcd}^{\rm (5)}(t) I_{abcd}^{\rm (5)}(t') \nonumber\\
&&+\frac{1}{84}J_{abc}^{\rm (4)}(t) J_{abc}^{\rm (4)}(t')\,.
 \end{eqnarray}
Here $\eta\equiv1/c$ and the superscript in parenthesis  denotes  repeated time-derivatives. The multipole moments $I_L$, $J_L$ denote
here the values of the canonical moments $M_L$, $S_L$  entering the PN-matched~\cite{Blanchet:1987wq,Blanchet:1989ki,Damour:1990ji,Blanchet:1998in,Poujade:2001ie} multipolar-post-Minkowskian (MPM) formalism~\cite{Blanchet:1985sp}, when they are
reexpressed as explicit functionals of the instantaneous state of the binary system. These multipole moments parametrize
 (in a minimal, gauge-fixed way) the exterior gravitational field
(and therefore the relevant coupling between the system and a long-wavelength external radiation field).

\subsection{The 2PN-accurate n-polar moments}

At the 2PN accuracy, we need the 2PN-accurate value of the quadrupole moment expressed in terms of the material source~\cite{Blanchet:1995fr,Blanchet:1995fg}. The other moments (the electric octupole moment $I_{ijk}$, electric hexadecapole moment, $I_{ijkl}$, 
the magnetic quadrupole moment, $J_{ij}$, and the magnetic octupole moment, $J_{ijk}$) need only to be known at the 1PN fractional
accuracy \cite{Blanchet:1989ki,Damour:1990ji,Damour:1994pk}. They have the following explicit expressions (in the c.m. harmonic
coordinate frame) \cite{Arun:2007rg}
\begin{eqnarray} \label{Moments}
I_{ij}&=&C_1 x_{\langle ij \rangle}+C_2 v_{\langle ij \rangle}+C_3 x_{\langle i}v_{j \rangle}\,,\nonumber\\
I_{ijk}&=&B_1 x_{\langle ijk \rangle}+B_2x_{\langle ij}v_{k\rangle}+B_3x_{\langle i}v_{jk\rangle}\,,\nonumber\\
I_{ijkl}&=&\nu M (1-3\nu)x_{\langle ijkl \rangle}\,,\nonumber\\
J_{ij}&=& D_1 L_{\langle i}x_{j\rangle}+D_2L_{\langle i}v_{j\rangle}\,,\nonumber\\
J_{ijk}&=&\nu M (1-3\nu)L_{\langle i}x_{jk\rangle}\,,
\end{eqnarray}
where the various coefficients (as well as the notation) have been summarized in Table \ref{table_flux_relations}. [See also Refs. \cite{Bini:2020wpo,Bini:2020nsb}.]
%
%
%
\begin{table*}
\caption{\label{table_flux_relations} Coefficients entering the multipolar moments \eq{Moments} used in the 2PN flux. Here, $x^i$ and $v^i \equiv \frac{ d x^i}{dt}$ denote the harmonic-coordinate  relative  center-of-mass
 position and  velocity of a two-body system, whereas $L_i \equiv\epsilon_{ijk}x^jv^k$. We assume $m_1\leq m_2$.
}
\begin{ruledtabular}
\begin{tabular}{ll}
$C_1$ &$ 1+ \eta^2 \left[\frac{29}{42}  (1-3\nu) v^2 -\frac17 (5-8\nu) \frac{G M}{r}\right]$\\
&$+
 \eta^4  \left[\frac{G M}{r} v^2 \left(\frac{2021}{756}-\frac{5947}{756}\nu-\frac{4833}{756}\nu^2\right)
+\frac{G^2 M^2}{r^2} \left(\frac{355}{252}-\frac{953}{126}\nu+\frac{337}{252}\nu^2\right)\right. $\\
&$ \left.
+v^2 \left(\frac{253}{504}-\frac{1835}{504}\nu+\frac{3545}{504}\nu^2\right)
+\frac{G M}{r} \dot r^2 \left(-\frac{131}{756}+\frac{907}{756}\nu-\frac{1273}{756}\nu^2\right)\right]$\\
$C_2$ &  $
 \eta^2 r^2\left\{\frac{11}{21} (1-3\nu)  
+ \eta^2  \left[\frac{G M}{r} \left(\frac{106}{27}-\frac{335}{189}\nu-\frac{985}{189}\nu^2\right)+v^2\left(\frac{41}{126}-\frac{337}{126}\nu+\frac{733}{126}\nu^2\right)
+\dot r^2 \left(\frac{5}{63}-\frac{25}{63} \nu+\frac{25}{63}\nu^2\right)\right]\right\}$\\
$C_3$ &$ 2 \eta^2 r\dot r  
\left\{ -\frac{2}{7}+\frac{6}{7}\nu+ \eta^2 
\left[v^2 \left(-\frac{13}{63}+\frac{101}{63}\nu-\frac{209}{63}\nu^2\right)
+\frac{G M}{r} \left(-\frac{155}{108}+\frac{4057}{756}\nu+\frac{209}{108}\nu^2\right)\right]\right\}$\\
$B_1$ & $\sqrt{1-4\nu}\left\{-1+ \eta^2  \left[\frac{GM}{r} \left(\frac56 -\frac{13}{6}\nu\right)+v^2 \left(-\frac56 +\frac{19}{6}\nu\right)\right]\right\}$\\
$B_2$ &$ \sqrt{1-4\nu}(1-2\nu)\,  \eta^2  r \dot r $\\
$B_3$ &$- \sqrt{1-4\nu}(1-2\nu)\,  \eta^2  r^2 $\\
$D_1$ &$ \sqrt{1-4\nu}\left\{-1+ \eta^2  \left[\frac{GM}{r} \left(-\frac{27}{14}-\frac{15}{7}\nu\right)+v^2\left(-\frac{13}{28}+\frac{17}{7}\nu\right)\right]\right\}$\\
$D_2$ &$ \sqrt{1-4\nu}  r \dot r \left(-\frac{5}{28}-\frac{5}{14}\nu\right) \eta^2 $\\
\end{tabular}
\end{ruledtabular}
\end{table*}

\subsection{The harmonic-coordinate quasi-Keplerian parametrization of the hyperbolic  motion}
 
We need also to use  the 2PN-accurate dynamics of a binary system in harmonic coordinates \cite{DD1981a,D1982},
and the corresponding  quasi-Keplerian parametrization \cite{Damour:1990jh} of the hyperbolic motion \cite{Cho:2018upo} 
(which we checked against the 2PN equations of motion given in Ref. \cite{Iyer:1995rn}):
\begin{eqnarray} \label{hypQK2PN}
r&=& \bar a_r (e_r \cosh v-1)\,,\nonumber\\
\ell &=& \bar n (t-t_P)=e_t \sinh v-v + f_t V+g_t \sin V\,,\nonumber\\
\bar \phi &=&\frac{\phi-\phi_P}{K}=V+f_\phi \sin 2V+g_\phi \sin 3V\,.
\end{eqnarray}
Here, we use adimensionalized variables (and $c=1$), notably $r=r^{\rm phys}/(GM)$, $t= r^{\rm phys}/(GM)$,
while $V=V(v)$ is given by
\beq
\label{Vdef}
V=2\, {\rm arctan}\left[\Omega_{e_\phi}\tanh \frac{v}{2}  \right]\,,
\eeq
with the notation
\beq
\Omega_{e_\phi}=\sqrt{\frac{e_\phi+1}{e_\phi-1}}\,.
\eeq
The 2PN-accurate expressions of the orbital parameters $\bar n$, $\bar a_r$, $K$, $e_t,e_r,e_\phi$, $f_t,g_t,f_\phi, g_\phi$
 are given in Appendix \ref{PN} as functions of the specific binding energy $\bar E \equiv (E_{\rm tot}-Mc^2)/(\mu c^2)$, Eq. \eqref{defbarE}, and of the dimensionless  angular momentum $j=c J/(GM\mu)$, Eq. \eq{defj2}, of the system, and in harmonic coordinates (modified harmonic coordinates, according to the notation of Ref. \cite{Arun:2007rg}). 
Note that, as discussed in Ref. \cite{Cho:2018upo}, the analytic continuation from the ellipticlike to the hyperboliclike case 
(namely from $\bar E<0$ to $\bar E>0$)
cannot be performed in as simple a way at 2PN than at 1PN \cite{dd}. 
As a consequence, the  orbital parameters
entering the hyperbolic-motion representation \eq{hypQK2PN} (notably $\bar n$, $e_t$, $f_t$ and $g_t$) are not directly related
to the analytic continuation in $\bar E$ of the orbital parameters, denoted in a similar way (namely $n$, $e_t$, $f_t$ and $g_t$), 
entering the elliptic-motion quasi-Keplerian representation.

\subsection{$W^{\rm tail,h}$ along hyperbolic orbits: computational details}

Consider the h-route nonlocal Hamiltonian in units $G=1=c$
\begin{eqnarray} 
\label{Hnonloch2}
 H_{\rm nonloc, h}^{4+5+6 \rm PN}(t)&=& - H_{\rm tot}\, {\rm Pf}_{2s/c}\, \int_{-\infty}^\infty
 \frac{dt'}{|t-t'|}{\mathcal F}^{\rm split}_{\rm 2PN}(t,t')
\nonumber\\
&+&2 H_{\rm tot}{\mathcal F}^{\rm split}_{\rm 2PN}(t,t)  \ln \left( \frac{r_{12}^h(t)}{s}\right)
\,,
\end{eqnarray}
where ${\mathcal F}^{\rm split}_{\rm 2PN}(t,t')$ was defined above.
We need to compute the  integral of  $ H_{\rm nonloc, h}^{4+5+6 \rm PN}(t)$ along a 2PN-accurate hyperbolic motion:
\begin{eqnarray}
\label{Wtail12}
&&W^{\rm tail,h}(E,j)=\int_{-\infty}^\infty\,   H_{\rm nonloc,h}^{4+5+6 \rm PN}(t) dt\,.
\end{eqnarray} 
Following the decomposition, displayed in Eq. \eq{Hnonloch2}, of $H_{\rm nonloc,h}^{4+5+6 \rm PN}(t)$ in two terms,
we correspondingly decompose $W^{\rm tail,h}(E,j)$ in two integrals, namely
\beq \label{W1W2}
W^{\rm tail,h}(E,j)= W_1^{\rm tail,h}(E,j) +W_2^{\rm tail,h}(E,j)\,,
\eeq
where
\bea \label{W1}
W_1^{\rm tail,h}(E,j)&\equiv&- H_{\rm tot} \int_{-\infty}^\infty dt\, {\rm Pf}_{2s/c}\nonumber\\ 
&&\times
\int_{-\infty}^\infty
 \frac{dt'}{|t-t'|}{\mathcal F}^{\rm split}_{\rm 2PN}(t,t')\,,
\eea
while
\beq \label{W2}
W_2^{\rm tail,h}(E,j) \equiv 2 H_{\rm tot} \int_{-\infty}^\infty dt {\mathcal F}^{\rm split}_{\rm 2PN}(t,t)  \ln \left( \frac{r_{12}^h(t)}{s}\right)\,.
\eeq
Let us consider first the term $W_1^{\rm tail, h}$.
A crucial role is played by the  measure
\beq
\label{measure}
d{\mathcal M}_{(t,t')} \equiv\frac{dt\, dt'}{|t-t'|}\,.
\eeq
In order to compute the double integral  Pf$\int d{\mathcal M}_{(t,t')} {\mathcal F}^{\rm split}_{\rm 2PN}(t,t')$,
it is useful to replace the integral over $t$ and $t'$ by an integral over the variables
\beq
T \equiv\tanh \frac{v}{2}\; ; \;  T' \equiv \tanh \frac{v'}{2}\,,
\eeq
where $v$ is the hyperbolic eccentric anomaly entering the quasi-Keplerian parametrization of the 2PN hyperbolic motion
given above. This change of variables maps the original  integration domain $(t,t') \in {\mathbb R}\times {\mathbb R}$ onto the 
compact domain  $(T,T') \in [-1,1]\times [-1,1]$.
It also transforms the singular line $t=t'$ into $T=T'$, together with  a transformation of the constant  cutoff $|t'-t| = 2 s/c$ implied
by the Pf operation into a corresponding $T$-dependent cutoff (see below).

We succeeded in computing, with 2PN accuracy, the first three terms in the large-eccentricity expansion of $W^{\rm tail}$, i.e.,
\bea
W_1^{\rm tail, h}&=& {W_1}^{\rm tail, h \, LO} + { W_1}^{\rm tail, h \, NLO} \nonumber\\
&&+   {W_1}^{\rm tail, h \, NNLO} + O(e_r^{-6})\,,
\eea
where we used the fact that the  leading order (LO) term ${W_1}^{\rm tail, h \,\rm LO}$ starts at order $ O(e_r^{-3})$ (see below).
At the LO, and the next-to-leading order (NLO) in $\frac1{e_r}$ (and $\frac1{\pinf j}$), both integrals in $T'$ (with Pf) and in $T$ 
can be analytically performed. At the next-to-next-to-leading order (NNLO) in  $\frac1{e_r}$, we could explicitly write down
the integrand to be integrated, but we could only analytically compute part of the integral, and we had to resort to
numerical integration to evaluate the rest. During the various computational steps we keep as fundamental eccentricity $e_r$,
but, at the end, we express the final result in terms of an expansion in powers of $\frac1{\pinf j}$ (as in Eq. \eq{Axp}).
Some details follow. 

The 2PN-exact relation $t$ vs $T$ is given by
\bea \label{tvsT}
\frac{t^{\rm phys}}{M}&\equiv& t= \frac{2}{\bar n}\left[ e_t \frac{T}{(1-T^2)}- {\rm arctanh}(T)\right.\nonumber\\
&+& f_t {\rm arctan}\left(\Omega_{e_\phi} T\right) 
\left.  
+g_t \frac{\Omega_{e_\phi}T}{1+\Omega^2_{e_\phi}T^2}
\right]\,,
\eea
with a corresponding expression for $t'$ vs $T'$. One then forms  $|t-t'|$, whose eccentricity expansion reads
\bea
|t-t'|&=&|T-T'|\frac{1+TT'}{(1-T^2)(1-T'^2)}\bar a_r^{3/2}e_r\nonumber\\
&&\times 
\left[2-(1+2\nu)\frac{\eta^2}{\bar a_r}+\frac{8\nu^2-8\nu-1}{4}\frac{\eta^4}{\bar a_r^2}\right]\nonumber\\
&&\times
\left[1+\frac{1}{e_r}{\mathcal P}_1+\frac{1}{e_r^2}{\mathcal P}_2  +O\left(\frac1{e_r^3}\right)\right],
\eea
with ${\mathcal P}_1$ and ${\mathcal P}_2$ of the form
\bea
{\mathcal P}_1&=&{\mathcal P}_{10}(T,T')+{\mathcal P}_{12}(T,T')\frac{\eta^2}{\bar a_r}+{\mathcal P}_{14}(T,T')\frac{\eta^4}{\bar a_r^2}, \nonumber\\
{\mathcal P}_2&=&{\mathcal P}_{24}(T,T')\frac{\eta^4}{\bar a_r^2}\,.
\eea
The coefficients ${\mathcal P}_{nm}(T,T')$ entering ${\mathcal P}_1$ and ${\mathcal P}_2$ read
\begin{widetext}
\begin{eqnarray}
{\mathcal P}_{10}(T,T')&=&-\frac{(1-T'^2)(1-T^2)}{(T T'+1)(T-T')} K(T,T')
\,,\nonumber\\
{\mathcal P}_{12}(T,T')&=&-\frac{1}{8}\frac{(1-T'^2)(1-T^2)(12\nu-32) }{(T T'+1)(T-T')} K(T,T') 
\,,\nonumber\\
{\mathcal P}_{14}(T,T')&=& -\frac{1}{8}\frac{(1-T'^2)(1-T^2)}{(T T'+1) (T-T')} (3\nu^2-29\nu) K(T,T')
+\frac{1}{8}\frac{(-15+\nu)\nu (T T'-1)(1-T'^2)(1-T^2)}{(1+T'^2)(1+T^2)(T T'+1)} 
\,,\nonumber\\
{\mathcal P}_{24}(T,T')&=& -\frac{3}{2}\frac{(1-T'^2)(1-T^2)(-5+2\nu) }{ (T T'+1) (T-T')}\kappa(T,T')\nonumber\\
&&-\frac{1}{8(1+T'^2)^2 (1+T^2)^2} (16-2\nu^2 T^2+\nu^2 T^4-2\nu^2 T'^2+\nu^2 T'^4-172\nu T^2 T'^2\nonumber\\
&&-26\nu T^2 T'^4-26\nu T^4 T'^2-43\nu T^4 T'^4-43\nu T^4-4\nu^2 T T'-26\nu T'^2\nonumber\\
&&-43\nu T'^4-26\nu T^2+60\nu T T'+4\nu^2 T^3 T'+4\nu^2 T T'^3-4\nu^2 T^3 T'^3+\nu^2-43\nu+64 T^2 T'^2\nonumber\\
&&+32 T^2 T'^4+32 T^4 T'^2+16 T^4 T'^4+32 T^2+16 T^4+32 T'^2+16 T'^4+4\nu^2 T^2 T'^2-2\nu^2 T^2 T'^4\nonumber\\
&&-2\nu^2 T^4 T'^2+\nu^2 T^4 T'^4-60\nu T T'^3-60\nu  T^3 T'+60\nu T^3 T'^3) \,.
\end{eqnarray}
\end{widetext}
Here, we used the notation
\bea \label{defkkappa}
\kappa(T,T')&\equiv&{\rm arctan}(T)-{\rm arctan}(T')\,,\nonumber\\
K(T,T')&\equiv&{\rm arctanh}(T)-{\rm arctanh}(T')\,.
\eea
These relations imply  for the reexpression of the measure, Eq. \eqref{measure}, in the $T-T'$ 
plane\footnote{Note that the measure $d{\mathcal M}_{(T,T')}$ is a symmetric function of $T$ and $T'$.}, 
\beq
d{\mathcal M}_{(T,T')}=\frac{1}{|t(T)-t'(T')|}\frac{dt}{dT}\frac{dt'}{dT'}dTdT'\,,
\eeq
the following (schematic) expression 
\begin{eqnarray}
d{\mathcal M}_{(T,T')}&=& 2e_r\bar a_r^{3/2}\left[1-\frac{1+2\nu}{2\bar a_r}\eta^2-\frac{1+8\nu-8\nu^2}{8\bar a_r^2}\eta^4\right]\nonumber\\
&&\times
\frac{(1+T'^2) (1+T^2)dT dT'}{(1-T'^2)(1-T^2)(1+TT')|T-T'|} \nonumber\\
&&\times 
\left(1+\frac{{\mathcal M}_1}{e_r}+\frac{{\mathcal M}_2}{e_r^2}  +O\left(\frac1{e_r^3}\right) \right)\,,
\end{eqnarray}
where we have explicitly shown only the LO contribution in the large-eccentricity expansion. The  NLO and NNLO contributions
(described by the coefficients ${\mathcal M}_1(T,T';\nu,\eta)$ and ${\mathcal M}_2(T,T';\nu,\eta)$) have large expressions that we do
not explicitly display here. Let us simply note that  ${\mathcal M}_1(T,T';\nu,\eta)$ involves the function $K(T,T')$ linearly, while 
${\mathcal M}_2(T,T';\nu,\eta)$ involves $K(T,T')$,  $K^2(T,T')$ and $\kappa(T,T')$ (defined in Eq. \eq{defkkappa}). 

Similarly to the measure $d{\mathcal M}_{(T,T')}$, we  expand, in the following, many quantities  
in inverse powers of the eccentricity $e_r$. For instance, the first three terms of the large-eccentricity
expansion of the 2PN-accurate split-flux integrand ${\mathcal F}^{\rm split}_{\rm 2PN}(T,T')$ will be denoted as
\beq
{\mathcal F}^{\rm split}_{\rm 2PN}(T,T')= {\mathcal F}^{\rm LO}_{\rm 2PN}+ {\mathcal F}^{\rm NLO}_{\rm 2PN}+ {\mathcal F}^{\rm NNLO}_{\rm 2PN}+\cdots\,.
\eeq
In the following, we reserve the notation LO, NLO, NNLO to the first three terms in expansion in $e_r^{-1}$. Note that each term
in this expansion is itself PN-expanded in powers of $\eta=\frac1c$ up to the 2PN fractional accuracy, so that we have (for $n=0,1,2$)
\beq
{\mathcal F}^{\rm N^nLO}_{\rm 2PN}={\mathcal F}^{\rm N^nLO}_0 + \eta^2 {\mathcal F}^{\rm N^nLO}_2+ \eta^4 {\mathcal F}^{\rm N^nLO}_4 + O(\eta^6)\,.
\eeq
The LO term in the eccentricity expansion of  ${\mathcal F}^{\rm split}_{\rm 2PN}(T,T')$ is of order $e_r^{-4}$.
Therefore, the full structure of the double expansion in  $\eta=\frac1c$ and in $e_r^{-1}$ of the split-flux reads
\beq
\label{fluxdecomp}
{\mathcal F}^{\rm split}_{\rm 2PN}(T,T')= \sum_{k=0}^{2} \sum_{m=4}^6 \eta^{2k}e_r^{-m}{\mathcal F}_{(2k,-m)}\,,
\eeq
where  $k=0,1,2$ counts the (fractional) PN order, while $m=4,5,6$ indicates the eccentricity order.
Let us note in passing that the 1PN terms ${\mathcal F}_{(2,-m)}$ are linear in $\nu$, while the 2PN ones  
${\mathcal F}_{(4,-m)}$ are quadratic in $\nu$. 

The structure of the expansion coefficients ${\mathcal F}_{(2k,-m)}\equiv {\mathcal F}_{(2k,-m)}(T,T')$ is described in Table \ref{flux_split_relations}.
The explicit expressions of the polynomials  $P_N^{(n,-m)}(T,T')$, appearing as coefficients in Table \ref{flux_split_relations}, are given 
in Table \ref{flux_split_relations2a} for the Newtonian-level case ($n=0$). All these polynomials are either symmetric in $T, T'$,
or antisymmetric when they appear multiplied by $\kappa(T,T')$. 
%
%
%
%
\begin{table*}
\caption{\label{flux_split_relations} Structure of the terms ${\mathcal F}_{(n,-m)}\equiv {\mathcal F}_{(\eta^n,e_r^{-m})}$ entering the
large-eccentricity expansion of the 2PN-accurate harmonic coordinate flux, Eq. \eqref{fluxdecomp}.
Here, $P_{N}^{(n,-m)}(T,T')$ denotes an $N$-degree polynomial in $T$ and $T'$ entering ${\mathcal F}_{(n,-m)}$.}
\begin{ruledtabular}
\begin{tabular}{ll}
${\mathcal F}_{(0,-4)} $&$\frac{32}{15}\nu^2\frac{(1-T^2)^2(1-T'^2)^2}{\bar a_r^5 (1+T^2)^5 (1+T'^2)^5}P_{12}^{(0,-4)}(T,T')
$\\
${\mathcal F}_{(0,-5)} $&$-\frac{64}{5}\nu^2 \frac{(1-T^2)^2(1-T'^2)^2}{\bar a_r^5 (1+T^2)^6 (1+T'^2)^6} 
[(TT')^2-1]P_{12}^{(0,-5)}(T,T')
$\\
${\mathcal F}_{(0,-6)} $&$ \frac{32}{5}\nu^2 \frac{(1-T^2)^2(1-T'^2)^2}{\bar a_r^5 (1+T^2)^7 (1+T'^2)^7}
P_{20}^{(0,-6)}(T,T')
$\\
\hline
${\mathcal F}_{(2,-4)} $&$ -\frac{2}{105}\nu^2 \frac{(1-T^2)^2(1-T'^2)^2}{\bar a_r^6 (1+T^2)^7 (1+T'^2)^7}P_{20}^{(2,-4)}(T,T') $\\
${\mathcal F}_{(2,-5)} $&$ \frac{16}{105}\nu^2 \frac{(1-T^2)^2(1-T'^2)^2}{\bar a_r^6 (1+T^2)^8  (1+T'^2)^8}P_{24}^{(2,-5)}(T,T') $\\ 
${\mathcal F}_{(2,-6)} $&$ -\frac{4}{105}\nu^2 \frac{(1-T^2)^2(1-T'^2)^2}{\bar a_r^6 (1+T^2)^9 (1+T'^2)^9}[P_{27}^{(2,-6)}(T,T') \kappa(T,T')+P_{28}^{(2,-6)}(T,T')] $\\    
\hline 
${\mathcal F}_{(4,-4)} $&$ -\frac{4}{315}\nu^2 \frac{(1-T^2)^2(1-T'^2)^2}{\bar a_r^7 (1+T^2)^9 (1+T'^2)^9 } P_{26}^{(4,-4)}(T,T')  $\\ 
${\mathcal F}_{(4,-5)} $&$ \frac{8}{945}\nu^2 \frac{(1-T^2)^2(1-T'^2)^2}{\bar a_r^7 (1+T'^2)^{10} (1+T^2)^{10}} P_{32}^{(4,-5)}(T,T')$\\
${\mathcal F}_{(4,-6)} $&$ \frac{2}{2835}\nu^2 \frac{(1-T^2)^2(1-T'^2)^2}{\bar a_r^7 (1+T^2)^{11}(1+T'^2)^{11}} [P_{35}^{(4,-6)}(T,T')\kappa(T,T')+P_{36}^{(4,-6)}(T,T')]  $\\
\end{tabular}
\end{ruledtabular}
\end{table*}
%
%
%
%
\begin{table*}
\caption{\label{flux_split_relations2a} Explicit expressions of the various polynomials 
$P_{N}^{(n,-m)}(T,T')$ parametrizing the structure displayed in Table \ref{flux_split_relations} in  the Newtonian limit $n=0$. We have checked the symmetry property of these polynomials when exchanging $T$ and $T'$.}
\begin{ruledtabular}
\begin{tabular}{l|c|l}
$P_{12}^{(0,-4)}(T,T')$& sym &$(-3-15 T'^4+3T'^6+15T'^2)T^6+(37T'^5+37T'-52T'^3)T^5+(-75T'^2-15T'^6+15+75T'^4)T^4$\\
&&$+(-52T'-52T'^5+76T'^3)T^3+(75T'^2-75T'^4-15+15T'^6)T^2+(37T'^5+37T'-52T'^3)T$\\
&&$+3-15T'^2+15T'^4-3T'^6 $\\
$P_{12}^{(0,-5)}(T,T')$ &sym & $ (-10-22T'^4+3T'^6+25T'^2)T^6+(-90T'^3+51T'^5+69T')T^5+(-122T'^2-22T'^6+25+131T'^4)T^4$\\
&&$+(-90T'-90T'^5+120T'^3)T^3+(131T'^2-122T'^4-22+25T'^6)T^2+(69T'^5+51T'-90T'^3)T$\\
&&$+3-22T'^2+25T'^4-10T'^6$\\
$P_{20}^{(0,-6)}(T,T') $& sym &$ (20T'^{10}-2+26T'^2-182T'^4+290T'^6-200T'^8)T^{10}$\\
&&$+(-128T'^3+21T'-1024T'^7+904T'^5+443T'^9)T^9$\\
&&$+(-910T'^6+262T'^2+1420T'^8-200T'^{10}+26-134T'^4)T^8$\\
&&$+(-1024T'^9-128T'+1120T'^5-1700T'^3+228T'^7)T^7$\\
&&$+(-134T'^2-910T'^8+2388T'^4-1740T'^6-182+290T'^{10})T^6$\\
&&$+(904T'^9+1120T'^7-1728T'^5+1120T'^3+904T')T^5$\\
&&$+(-182T'^{10}+2388T'^6-134T'^8-1740T'^4-910T'^2+290)T^4$\\
&&$+(-1024T'+1120T'^5-128T'^9+228T'^3-1700T'^7)T^3$\\
&&$+(-200-134T'^6+262T'^8+1420T'^2-910T'^4+26T'^{10})T^2$\\
&&$+(904T'^5+21T'^9-1024T'^3-128T'^7+443T')T$\\
&&$+20+290T'^4-182T'^6-200T'^2+26T'^8-2T'^{10}$\\
\end{tabular}
\end{ruledtabular}
\end{table*}

Multiplying $d{\mathcal M}_{(T,T')}$ and ${\mathcal F}^{\rm 2PN}(T,T')$ yields an integrand that we denote as  
\beq
d{\mathcal M}_{(T,T')} {\mathcal F}^{\rm split}_{\rm 2PN}(T,T')={\mathcal G}(T,T') \frac{dT dT'}{|T-T'|}\,,
\eeq
where the (2PN-accurate) function ${\mathcal G}(T,T')$ is expanded only up to the (fractional) second order in $e_r^{-1}$,
say
\bea \label{eexpG}
{\mathcal G}(T,T')&=& {\mathcal G}^{\rm LO}(T,T')+ {\mathcal G}^{\rm NLO}(T,T')\nonumber\\
&&
+{\mathcal G}^{\rm NNLO}(T,T')+ \cdots\,.
\eea
As before, the notation LO, NLO, NNLO refers to the expansion in powers of $e_r^{-1}$ (starting at order $e_r^{-3}$
and extending up to order  $e_r^{-5}$). 

The function ${\mathcal G}(T,T')$ (which should not be confused with the time-split function entering Eq. \eq{H5.5})
is symmetric in $T$ and $T'$. In addition, we recall that each term in Eq. \eq{eexpG} is itself PN-expanded with fractional 2PN
accuracy, so that $ {\mathcal G}^{\rm N^nLO}=  {\mathcal G}^{\rm N^nLO}_0 + \eta^2{\mathcal G}^{\rm N^nLO}_2 +
\eta^4 {\mathcal G}^{\rm N^nLO}_4 +O(\eta^6)$.

The original integral was singular at $t=t'$, i.e., along the bisecting line of the $t-t'$ plane.
This singular line   becomes the bisecting line in the plane $T-T'$, but endowed with a $T-$dependent slit, which is identified from the relation $dT =f(T) dt$. Here, the Jacobian $f(T)=dT/dt$ admits also an expansion in powers of $e_r^{-1}$,
say $f(T)=f^{\rm LO}(T) + f^{\rm NLO}(T) +  f^{\rm NNLO}(T)+  O(e_r^{-4})$,
where, for example, the LO term in the  expansion in $e_r^{-1}$ reads
\beq
f^{\rm LO}(T)=\frac{\bar n} {2e_r }\frac{(1-T^2)^2} {1+T^2}\,.
\eeq
Note that, because of our use of the quasi-Keplerian representation, this Newtonian-looking $O(e_r^{-1})$ expression  is  PN exact.
The fractional PN corrections only enter the higher-order corrections in $e_r^{-1}$. 

As explained in Ref. \cite{Bini:2020wpo}, when considering the partie-finie integral 
giving the first part $W_1^{\rm tail,h}$ of $W^{\rm tail,h}$, Eq. \eq{W1W2},
the width of the slit around the bisecting line $T=T'$ defined by the partie-finie procedure is
$dT =f(T) \epsilon$ where $\epsilon$ is initially considered as being infinitesimal, before replacing it by the finite value $2s/c$ 
at the end. This leads to the following partie-finie integral
\begin{eqnarray}
{\mathcal I}(T)&=& {\rm Pf}_{2sf(T)/c}\int_{-1}^1 dT' \frac{{\mathcal G}(T,T')}{|-T'+T|}\nonumber\\
&=& \int_{-1}^1 dT' \frac{{\mathcal G}(T,T')-{\mathcal G}(T,T)}{|-T'+T|}\nonumber\\
&&+  {\mathcal G}(T,T)\left[\int_{-1}^{T-\epsilon f(T)} \frac{dT'}{T-T'}\right.\nonumber\\
&& \left. + \int_{T+\epsilon f(T)}^1 \frac{dT'}{T'-T}\right]_{\epsilon \mapsto 2s/c}\,.
\end{eqnarray}
Defining
\beq
q(T,T')={\mathcal G}(T,T')-{\mathcal G}(T,T)\,,
\eeq
the above expression can be rewritten as
\begin{eqnarray}
\label{calIdef}
{\mathcal I}(T)
&=& \int_{-1}^1 dT' \frac{q(T,T')}{|-T'+T|}\nonumber\\
&&+  {\mathcal G}(T,T)  \ln\left( \frac{1-T^2}{\epsilon^2 f^2(T)} \right) \,,
\end{eqnarray}
where $\epsilon=2s/c$.
Note that $q(T,T')$ is not (contrary to ${\mathcal G}(T,T')$) a symmetric function of $T$ and $T'$. As we are going to integrate 
${\mathcal I}(T)$ over a  $(T,T')$-symmetric domain, one could replace $q(T,T')$ by its  symmetric part.

Further integration in $T$ gives 
\begin{eqnarray}
{\mathcal J}&\equiv & \int_{-1}^1 {\mathcal I}(T) dT \nonumber\\
&=& \int_{-1}^1 dT \int_{-1}^1 dT' \frac{q(T,T')}{|-T'+T|}\nonumber\\
&& -2 \ln \left(\frac{2s}{c}\right)  \int_{-1}^1 dT  {\mathcal G}(T,T)\nonumber\\
&& +\int_{-1}^1 dT   {\mathcal G}(T,T)  \ln\left( \frac{1-T^2}{f^2(T)}\right)\,.
\end{eqnarray}

So far, we only discussed the first term $W_1^{\rm tail, h}$ in $W^{\rm tail}$. The second term is easier to evaluate because it
is given by a simple integral, namely (using as above $G=1=c$)
\beq \label{W2gen}
W_2^{\rm tail, h}=2 H_{} \int_{- \infty}^{+ \infty}dt {\mathcal F}^{\rm split}_{\rm 2PN}(t,t)  \ln \left( \frac{r_{12}^h(t)}{s}\right)\,.
\eeq
Again, we use the quasi-Keplerian representation of the hyperbolic motion, and therefore replace the integration with respect to $t$
by an integration with respect to $T$, using the explicit functional link $t= F(T)$ given in Eq. \eq{tvsT}.

Because of the presence of the logarithm of $r_{12}^h(t)$, the integral $W_2^{\rm tail, h}$, Eq. \eq{W2gen}, cannot be computed
as an exact function of the energy and angular momentum. However, like $W_1^{\rm tail, h}$, one can
compute the first three terms in its expansion in powers of $e_r^{-1}$, i.e.,
\bea
W_2^{\rm tail, h}&=& {W_2}^{\rm tail, h \, LO} + { W_2}^{\rm tail, h \, NLO} \nonumber\\
&&+   {W_2}^{\rm tail, h \, NNLO} + O(e_r^{-6})\,.
\eea
The LO term ${W_2}^{\rm tail, h \,\rm LO}$ starts at order $ O(e_r^{-3})$.

In  the next subsection we illustrate the results of the computation of $W^{\rm tail,h}$.

\subsection{Value of $W^{\rm tail, h}$ up to the NNLO in $e_r^{-1}$ and at the fractional 2PN accuracy}

Let us recap the methodology used in the present Section. We  have been considering 
 the time integral of the first-order-tail harmonic, nonlocal Hamiltonian,
 \begin{eqnarray}
W^{\rm tail, h} &=& \int_{- \infty}^{+ \infty}dt H_{\rm nonloc,h}^{4+5+6 \rm PN}(t)\nonumber\\
&=& W_1^{\rm tail, h} +W_2^{\rm tail, h}\,,
\end{eqnarray}
where the decomposition in the two contributions, $W_1^{\rm tail, h}$ and $W_2^{\rm tail, h}$ 
was defined in Eqs.  \eq{W1W2}, \eq{W1},\eq{W2}. 

Using the quasi-Keplerian representation of the hyperbolic motion, we could compute the first three terms in the large-$e_r$ expansion of $W^{\rm tail, h} $, say
\bea
W^{\rm tail, h}&=& {W}^{\rm tail, h \, LO} + { W}^{\rm tail, h \, NLO}  \nonumber\\
&&+   {W}^{\rm tail, h \, NNLO}+ O(e_r^{-6})\,,
\eea
where each contribution, ${W}^{\rm tail, h \, N^nLO}$, is itself computed as a fractionally 2PN-accurate expression:
\bea
{W}^{\rm tail, h \, N^nLO}&=& {W}_{\eta^0}^{\rm tail, h \, N^nLO} + \eta^2 {W}_{\eta^2}^{\rm tail, h \, N^nLO}\nonumber\\
&&+{W}_{\eta^4}^{\rm tail, h \, N^nLO}+ O(\eta^6)\,.
\eea

The 2PN-accurate values of the two contributions to $W^{\rm tail, h}$  at the LO in $e_r^{-1}$ are easily computed. They read,
respectively,
\begin{widetext}
\begin{eqnarray} 
\label{W1LO0}
W_1^{\rm tail,h\,LO}&=&\frac{2}{15} \frac{\pi M\nu^2}{e_r^3\bar a_r^{7/2}}H_{\rm real}\left\{ 
100 + 37 \ln \left(\frac{s}{4e_r\bar a_r^{3/2}}\right) 
+\left[\frac{685}{4}-\frac{1017}{14}\nu+\left(\frac{3429}{56}-\frac{37}{2}\nu\right)\ln \left(\frac{s}{4e_r\bar a_r^{3/2}}\right)\right] \frac{\eta^2}{\bar a_r}\right.\nonumber\\
&&\left.
+\left[
\frac{3656939}{8064}-\frac{18181}{72}\nu+\frac{235453}{4032}\nu^2
+\left(\frac{114101}{672}-\frac{7055}{112}\nu+\frac{111}{8}\nu^2\right)\ln \left(\frac{s}{4e_r\bar a_r^{3/2}}\right)
\right] \frac{\eta^4}{\bar a_r^2}
\right\}\,,
\end{eqnarray}
and
\begin{eqnarray}
\label{W2LO0}
W_2^{\rm tail,h\, LO}
&=& \frac{2}{15} \frac{\pi M\nu^2}{e_r^3\bar a_r^{7/2}}H_{\rm real}\left\{
-\frac{85}{4}-37\ln \left(\frac{s}{2e_r\bar a_r}\right)
+\left[
-\frac{9679}{224}+\frac{981}{56}\nu+\left(-\frac{3429}{56}+\frac{37}{2}\nu\right)\ln \left(\frac{s}{2e_r\bar a_r}\right)
\right] \frac{\eta^2}{\bar a_r}\right.\nonumber\\
&&\left.
+\left[
-\frac{1830565}{16128}+\frac{54899}{1152}\nu-\frac{29969}{4032}\nu^2
+\left(-\frac{114101}{672}+\frac{7055}{112}\nu-\frac{111}{8}\nu^2\right)\ln \left(\frac{s}{2e_r\bar a_r}\right)
\right] \frac{\eta^4}{\bar a_r^2}
\right\}\,.
\end{eqnarray}
\end{widetext}
 It is easily seen that the intermediate scale $s$ cancells between the two contributions. The same will hold at the
NLO and NNLO levels.

The extension of these results to the NLO level is significantly more involved (especially for the 1PN and 2PN corrections). 
In fact, though the first integration over $T'$ 
 in the  split-flux integrals is found to be relatively straightforward (in spite of the presence of the transcendental functions
 $K(T,T')$ and $\kappa(T,T')$, Eq. \eq{defkkappa}), the subsequent integration over $T$ is significantly more difficult
 because of the  presence of poly-logarithmic functions. However, we have 
 been able to analytically compute the 2PN-accurate NLO value of $W_1^{\rm tail,h}$. 
The 2PN-accurate NLO values of $W_1^{\rm tail, h}$ and $W_2^{\rm tail, h}$  read, respectively,
\begin{widetext}
\begin{eqnarray} 
\label{W1NLO0}
W_1^{\rm tail,h\,NLO}&=&\frac{2}{15} \frac{M\nu^2}{e_r^4\bar a_r^{7/2}}H_{\rm real}\left\{ 
\frac{2224}{9} + \frac{1568}{3} \ln \left(\frac{4s}{e_r\bar a_r^{3/2}}\right) 
+\left[-\frac{28072}{225}-\frac{38872}{63}\nu+\left(\frac{944}{105}-\frac{1136}{3}\nu\right)\ln \left(\frac{4s}{e_r\bar a_r^{3/2}}\right)\right] \frac{\eta^2}{\bar a_r}\right.\nonumber\\
&&\left.
+\left[
-\frac{67489874}{77175}-\frac{3115726}{3675}\nu+\frac{165086}{315}\nu^2
+\left(\frac{419036}{735}-\frac{3244}{7}\nu+\frac{764}{3}\nu^2\right)\ln \left(\frac{4s}{e_r\bar a_r^{3/2}}\right)
\right] \frac{\eta^4}{\bar a_r^2}
\right\}\,,
\end{eqnarray}
and
\begin{eqnarray}
W_2^{\rm tail,h\, NLO}
&=& \frac{2}{15} \frac{M\nu^2}{e_r^4\bar a_r^{7/2}}H_{\rm real}\left\{
\frac{2768}{9}-\frac{1568}{3}\ln \left(\frac{2s}{e_r\bar a_r}\right)
+\left[
-\frac{64904}{225}-\frac{5992}{45}\nu+\left(-\frac{944}{105}+\frac{1136}{3}\nu\right)\ln \left(\frac{2s}{e_r\bar a_r}\right)
\right] \frac{\eta^2}{\bar a_r}\right.\nonumber\\
&&\left.
+\left[
-\frac{2925494}{77175}-\frac{542014}{2205}\nu+\frac{145498}{735}\nu^2
+\left(-\frac{419036}{735}+\frac{3244}{7}\nu-\frac{764}{3}\nu^2\right)\ln \left(\frac{2s}{e_r\bar a_r}\right)
\right] \frac{\eta^4}{\bar a_r^2}
\right\}\,.
\end{eqnarray}
\end{widetext}
By contrast, at the  NNLO level, we encountered integrals of the type
\bea
&&\int_{-1}^1 dT\int_{-1}^1  \frac{dT'}{|T-T'|}\left[f_2(T,T') K^2(T,T')\right.\nonumber\\
&&\qquad \left.+ f_1(T,T') K(T,T')+ f_0(T,T')\right]\,,
\eea 
where $K(T,T')$ is defined in Eq. \eqref{defkkappa}, 
and where the $f_j(T,T')$ are rather complicated rational functions of $T$ and $T'$.
For instance, even at the Newtonian level (\ie, the lowest order in $\eta$) we had to deal with the
integrand $q^{\rm N}(T,T')= f_2^{\rm N}(T,T') K^2(T,T')
+ f_1^{\rm N}(T,T') K(T,T')+ f_0^{\rm N}(T,T')$ with  rational coefficients
$ f_2^{\rm N}(T,T')$,  $f_1^{\rm N}(T,T')$ and  $f_0^{\rm N}(T,T')$ given by
\begin{widetext}
\begin{eqnarray} \label{q_newt}
f_2^{\rm N}(T,T')&=& 
\frac{64}{15}\frac{(1-T'^2)^3 (1-T^2)^3}{(1+T^2)^4(1+T'^2)^4(T-T')^2(TT'+1)^3}
P_{10}^{f_2}(T,T')
\,,\nonumber\\
f_1^{\rm N}(T,T')&=& 
-\frac{128}{15}\frac{(1-T^2)^2(1-T'^2)^2(TT'-1)}{(1+T'^2)^5(1+T^2)^5(T-T')(TT'+1)}P_{12}^{f_1}(T,T')
\,,\nonumber\\
f_0^{\rm N}(T,T')&=&
-\frac{64}{15}\frac{(1-T^2)}{(1+T'^2)^6(1+T^2)^7(TT'+1)}P_{24}^{f_0}(T,T')
\,.
\end{eqnarray}
\end{widetext}
where the polynomials $P_{10}^{f_2}(T,T')$, $P_{12}^{f_1}(T,T')$ and $P_{24}^{f_0}(T,T')$ are displayed
in Table \ref{f012_funct}. We had to resort to numerical integration to evaluate some terms.

%
%
%
%
 \begin{table*}
 \caption{\label{f012_funct} Polynomial expressions entering the Newtonian level integral of Eq. \eqref{q_newt}. }
 \begin{ruledtabular}
 \begin{tabular}{ll}
 $P_{10}^{f_2}(T,T') $& $3-15T^2+15T^4-3T^6-15T'^2+15T'^4-3T'^6+37TT' +75T^4T'^4-15T^6T'^4-75T'^4T^2+15T'^6T^2$\\
&$-75T^4T'^2+75T'^2T^2-15T^4T'^6+3T^6T'^6+76T^3T'^3-52T^3T'-52T^5T'^3+15T^6T'^2+37T^5T'+37T^5T'^5$\\
&$-52T'^3T-52T^3T'^5+37T'^5T]$\\ 
$P_{12}^{f_1}(T,T')$ &$6-51T^2+60T^4-27T^6-51T'^2+60T'^4-27T'^6+116TT'+318T^4T'^4-51T^6T'^4-291T'^4T^2$\\
&$+60T'^6T^2-291T^4T'^2+318T'^2T^2-51T^4T'^6+6T^6T'^6+284T^3T'^3-218T^3T'-218T^5T'^3+60T^6T'^2$\\
&$+170T^5T'+116T^5T'^5-218T'^3T-218T^3T'^5+170T'^5T$\\
$P_{24}^{f_0}(T,T')$ &$15-411T^7T'^{13}+411T^5T'^{13}-207T^3T'^{13}+42TT'^{13}+207T^9T'^{13}-42T^{11}T'^{13}+117T^2$\\
&$+180T^4-579T^6+558T^8+630T'^2-303T'^4+1836T'^6+153T'^8-712TT'+12636T^4T'^4-9357T^6T'^4$\\
&$+651T'^4T^2-4836T'^6T^2+1716T^4T'^2-4050T'^2T^2-1920T^4T'^6-8220T^6T'^6-4365T^2T'^8$\\
&$+10356T'^8T^4-2973T'^8T^6-1086T'^4T^8+14280T'^6T^8+432T'^2T^8-3366T^8T'^8$\\
&$+60T'^{12}T^4+438T^8T'^{12}+27T^{12}T'^{12}-102T^{10}+3T^{12}+318T'^{10}+39T'^{12}-147T^2T'^{12}-243T'^{12}T^6$\\
&$-366T'^{12}T^{10}+630T^{10}T'^4+106T^{11}T'^3+4788T^9T'^3-68T^{11}T'+513T^9T'+272T'^9T-6157T^7T'^5$\\
&$-12292T^7T'^7-876T'^2T^{10}-144T'^6T^{10}+2556T'^{10}T^{10}-4650T'^{10}T^6-618T'^{10}T^2+3056T'^7T+326T^7T'^3$\\
&$-1963T^7T'-66T^{12}T'^2+477T^{12}T'^4-4911T^9T'^5-2846T^{11}T'^5-3181T^7T'^9-4386T^{10}T'^8+1992T^8T'^{10}$\\
&$+2895T^9T'^9+9804T^9T'^7+933T^{12}T'^8-3428T^{11}T'^9+3088T^{11}T'^7-996T^{12}T'^6-2626T^7T'^{11}-48T^9T'^{11}$\\
&$-378T^{12}T'^{10}+502T^{11}T'^{11}+278T'^{11}T+8212T^5T'^7+3876T^3T'^7+3276T^4T'^{10}+3373T^5T'^9$\\
&$-6651T^3T'^9+4018T^5T'^{11}-1548T^3T'^{11}-1032T^3T'^3+1083T^3T'-518T^5T'^3-282T^6T'^2+571T^5T'$\\
&$+10237T^5T'^5+3050T'^3T-8769T^3T'^5-3298T'^5T$\\
 \end{tabular}
 \end{ruledtabular}
 \end{table*}

Our final results for the 2PN-accurate NNLO values of $W_1^{\rm tail, h}$ and $W_2^{\rm tail, h}$  read, respectively,
\begin{widetext}
\begin{eqnarray} 
W_1^{\rm tail,h\,NNLO}&=&\frac{2}{15} \frac{\pi M\nu^2}{e_r^5\bar a_r^{7/2}}H_{\rm real}\left\{ 
\frac{5997}{8}+\frac{2529}{4}\ln(2)-\frac{15}{2}c_{00} + \frac{843}{2} \ln \left(\frac{s}{4e_r\bar a_r^{3/2}}\right)\right.\nonumber\\
&&
+\left[-\frac{400845}{448}-\frac{200997}{224}\ln(2)-\frac{15}{2}c_{20}+\left(-\frac{51711}{112}-\frac{5481}{8}\ln(2)-\frac{15}{2}c_{21}\right)\nu\right.\nonumber\\
&&\left.
+\left(-\frac{66999}{112}-\frac{1827}{4}\nu\right)\ln \left(\frac{s}{4e_r\bar a_r^{3/2}}\right)\right] \frac{\eta^2}{\bar a_r}\nonumber\\
&&
+\left[
-\frac{165424487}{96768}-\frac{442237}{896}\ln(2)-\frac{15}{2}c_{40}
+\left(\frac{22110289}{16128}+\frac{86205}{64}\ln(2)-\frac{15}{2}c_{41}\right)\nu\right.\nonumber\\
&&\left.\left.
+\left(-\frac{321757}{896}+\frac{13491}{32}\ln(2)-\frac{15}{2}c_{42}\right)\nu^2
+\left(-\frac{442237}{1344}+\frac{28735}{32}\nu+\frac{4497}{16}\nu^2\right)\ln \left(\frac{s}{4e_r\bar a_r^{3/2}}\right)
\right] \frac{\eta^4}{\bar a_r^2}
\right\}\,, \nonumber\\
\end{eqnarray}
and
\begin{eqnarray}
W_2^{\rm tail,h\, NNLO}
&=& \frac{2}{15} \frac{\pi M \nu^2}{e_r^5\bar a_r^{7/2}}H_{\rm real}\left\{
-\frac{3419}{8}-\frac{843}{2}\ln \left(\frac{s}{2e_r\bar a_r}\right)\right.\nonumber\\
&&
+\left[
\frac{103645}{448}+\frac{56559}{112}\nu+\left(\frac{66999}{112}+\frac{1827}{4}\nu\right)\ln \left(\frac{s}{2e_r\bar a_r}\right)
\right] \frac{\eta^2}{\bar a_r}\nonumber\\
&&\left.
+\left[
\frac{2467109}{13824}-\frac{3706175}{5376}\nu-\frac{1577635}{8064}\nu^2
+\left(\frac{442237}{1344}-\frac{28735}{32}\nu-\frac{4497}{16}\nu^2\right)\ln \left(\frac{s}{2e_r\bar a_r}\right)
\right] \frac{\eta^4}{\bar a_r^2}
\right\}\,.\nonumber\\
\end{eqnarray}
\end{widetext}
The coefficients $c_{00}$, $c_{20}$, $c_{21}$, $c_{40}$, $c_{41}$, and $c_{42}$ entering $W_1^{\rm tail,h\,NNLO}$ have been numerically computed. Our estimates of their values are listed in Table \ref{num_res}. From some numerical studies 
(increasing the working precision used in the computation), and by comparing with the exact values of $c_{00}$ and $c_{21}$ given below,
we estimate that the latter values have an absolute numerical error of order $1 \times 10^{-8}$. We accordingly  cite eight digits after
the decimal point.

%
%
%
%

\begin{table}
\caption{\label{num_res} Detailed results from numerical integration. See, however, below for the exact values of $c_{00}$ and $c_{21}$.}
\begin{ruledtabular}
\begin{tabular}{ll}
coefficient&numerical value\\
\hline
$c_{00}$&$-49.20484109 $ \\
$c_{20}$&$+115.95128578$ \\
$c_{21}$&$+161.90919858$ \\
$c_{40}$&$+22.31105671$ \\
$c_{41}$&$-116.85535736 $ \\
$c_{42}$&$-209.81006553$\\
\end{tabular}
\end{ruledtabular}
\end{table}

\subsection{Computation of $W^{\rm tail,h}(E,j)$ in the frequency domain}

So far, we have discussed the direct time-domain approach to the computation of the integrated tail action $W^{\rm tail,h \, NNLO}$.
It was shown in Ref. \cite{Bini:2017wfr} that $W^{\rm tail,h }$ has a simple expression in the frequency-domain.
Let us now briefly discuss the method we used to tackle, in parallel, the computation of $W^{\rm tail,h}(E,j)$ in the frequency domain.
The use of this method allowed us to go beyond the results obtained by the direct time-domain approach presented in the
previous subsection. In particular, we succeeded in analytically computing two of the NNLO integrals 
entering $W_1^{\rm tail,h \, NNLO}$ (namely $c_{00}$ and $c_{21}$) by working in the frequency domain, while we  could not compute them in the time domain. There remain four other integrals (one at the 1PN level, and three at the 2PN level)
in $W_1^{\rm tail,h \, NNLO}$ that we were still unable to compute analytically.

Let us start by presenting the analytical values we obtained for the two parameters $c_{00}$ and $c_{21}$, which could
only be numerically estimated in the time-domain, but which could be obtained in the frequency domain. Namely,
\bea
\label{solc00}
c_{00}&=&\frac{1039}{60}+\frac{843}{10}\ln(2) -\frac{2079}{20}\zeta(3)\nonumber\\
&=& -49.2048410955697697167634473834\ldots\,,
\eea
and
\bea
\label{solc21}
c_{21} &=& -\frac{1827}{20}\ln(2)+\frac{21867}{280}+\frac{612}{5}\zeta(3)\nonumber\\
&=&161.909198574011907946235225245\ldots\,.
\eea
Note the remarkable presence of $\zeta(3)$ in these expressions.
These analytical results are in agreement (within $\pm 1\times 10^{-8}$) with the numerical ones listed in Table \ref{num_res}. 
 
Let us now sketch our frequency-domain approach, relegating most details to  Appendices \ref{A} and \ref{B}.
We recall that the tail potential $W^{\rm tail,h}(E,j)$, Eq. \eqref{Wtail12}, computed along hyperbolic motion, can be split into two terms, Eq. \eqref{W1W2}, namely $ W_1^{\rm tail,h}(E,j)$, Eq. \eq{W1}, and  $ W_2^{\rm tail,h}(E,j)$, Eq. \eq{W2}.
Both $ W_1^{\rm tail,h}(E,j)$ and $ W_2^{\rm tail,h}(E,j)$ can  be evaluated in the frequency domain, after Fourier-transforming  the various multipolar moments. This frequency-domain approach turns out to be more convenient in the case of $W_1^{\rm tail,h}(E,j)$ because 
the logarithmic term in  $W_2^{\rm tail,h}(E,j)$ complicates matters.

The first step is to Fourier transform\footnote{In the following, we use $GM=1$, i.e., we work with $GM$-rescaled
time and frequency variables.} the multipolar moments. For example,
\beq
\label{I_ab_fourier}
I_{ab}(t)=\int \frac{d\omega}{2\pi}e^{-i \omega t}\hat I_{ab}(\omega)\,,
\eeq
where
\beq
\hat I_{ab}(\omega)=\int \frac{dt}{dv}e^{i\omega t(v)}I_{ab}(t)|_{t=t(v)} \, dv \,,
\eeq
with the associated PN expansion 
\bea
\hat I_{ab}(\omega)&=&\hat I_{ab}^{\rm N}(\omega)+\eta^2\hat I_{ab}^{\rm 1PN}(\omega)\nonumber\\
&&+\eta^4\hat I_{ab}^{\rm 2PN}(\omega)
+O(\eta^6)\,.
\eea
The PN expansion\footnote{We take $e_r$ and $\bar a_r$, defined in Table II, as fundamental variables.} of the exponential term $e^{i\omega t(v)}$ gives
\beq
e^{i\omega t(v)}=e^{q\sinh v -p v}(1+b_2\eta^2+b_4 \eta^4)\,,
\eeq
where
\beq \label{defupq}
 u\equiv \omega e_r \bar a_r^{3/2} \; ; \; q \equiv i \, u\; ; \;  p \equiv \frac{q}{e_r }= i \frac{u}{e_r }\,,
\eeq 
 and 
\bea
b_2&=&-\frac{i \omega\sqrt{\bar a_r}}{2} \left[(2\nu+1)e_r \sinh v +(\nu  -9)v
\right] \,,\nonumber\\
b_4&=&\frac{i\omega}{8 (e_r^2-1)\sqrt{\bar a_r}} \nonumber\\
&\times& 
\left[b_{40}+b_{41}{\rm arctan}\left(\sqrt{\frac{e_r-1}{e_r+1}}\tanh \frac{v}{2}\right)\right]\,,\nonumber\\
\eea
with
\bea
b_{40}&=&(e_r^2-1)\left[-4i\frac{\sqrt{\bar a_r}}{ \omega }b_2^2+v(\nu^2+11\nu-15)\right.\nonumber\\
&-&\left.
\nu(\nu-15)e_r\frac{\sinh v}{e_r\cosh v-1}\right]
+12v(4-7\nu)\nonumber\\
&+&
8e_r \left[\left(\nu^2-\nu-\frac18  \right)e_r^2-\nu^2+\frac92 \nu-\frac{15}{8}\right]\sinh v
\,,\nonumber\\
b_{41}&=& -48 \sqrt{e_r^2-1}\left(\nu-\frac{5}{2}\right) \,.
\eea
Moreover,
\beq
\frac{dt}{dv}=(e_r\cosh v-1)\bar a_r^{3/2} +c_2\eta^2 +c_4\eta^4\,,
\eeq
with 
\bea
c_2&=&-\frac12 \sqrt{\bar a_r} [(2\nu+1)e_r\cosh v +\nu-9] 
\,,\nonumber\\
c_4&=&\frac{1}{8\sqrt{\bar a_r}}\bigg\{
\nu^2+11\nu-15-e_r\cosh v(1+8\nu-8\nu^2)\nonumber\\
&-&
4\frac{4-7\nu}{e_r^2-1}(e_r\cosh v-3)+\frac{\nu^2-39\nu+60}{e_r\cosh v-1}\nonumber\\
&+&
\frac{\nu(15-\nu)(e_r^2-1)}{(e_r\cosh v-1)^2}
\bigg\}
\,.
\eea
The computation is done by using the integral representation of the Hankel functions of the first kind of order $p$ and argument $q$ (with Eqs. \eq{defupq})
\beq
H_p^{(1)}(q)=\frac{1}{ i\pi }\int_{-\infty}^\infty e^{q\sinh v -p v}dv\,.
\eeq
As the argument $q=iu$  of the Hankel function is purely imaginary, the Hankel function becomes converted into a 
Bessel K function, according to the  relation
\beq
H_p^{(1)}(iu)=\frac{2}{\pi}e^{-i \frac{\pi}{2}(p+1)}K_p(u)\,.
\eeq
Note that the order $p=i u/e_r$ of the Bessel functions is purely imaginary, and proportional to the (frequency-dependent)
argument $u = \omega e_r \bar a_r^{3/2}$. However, the order $p$ tends to zero when $e_r \to \infty$, which allows
some integrals to be explicitly computed when performing a large-eccentricity expansion.

Actually, the computation gives rise to several Bessel functions having the same argument $u$ but various orders differing by integers.
However,  standard identities valid for Bessel functions allows one to reduce the orders to either $p$ or $p+1$. 
When taking the large-eccentricity expansion, one then expands with respect to the order of the Bessel functions.
This gives rise, at LO, to $K_0(u)$, and $K_1(u)$,  and at NLO and NNLO to derivatives of $K_0(u)$, and $K_1(u)$
with respect to their orders. Such an expansion is explicitly shown below in Eqs. \eqref{K_exp},  while studying the Newtonian limit, and several useful relations are listed in Appendix \ref{A}.

Using ($\gamma=0.577215\ldots$)
\beq
{\rm Pf}_T \int_0^\infty d\tau \frac{\cos \omega \tau}{\tau}=-\ln (|\omega| T e^\gamma)\,,
\eeq
we find (see Section V of Ref. \cite{Bini:2017wfr} for details)
\begin{eqnarray}
\label{W1_final_exp}
W_1^{\rm tail,h}(E,j)&=& \frac{G^2H}{\pi c^5}2 \nu^2 \int_0^\infty d\omega {\mathcal K}(\omega) \ln \left(\omega \frac{2s}{c}e^\gamma  \right)
\,,\nonumber\\
\end{eqnarray}
where
\begin{eqnarray}
{\mathcal K}(\omega)&=&\frac15 \omega^6 |\hat I_{ab}(\omega)|^2\nonumber\\
&+&\eta^2 \left[\frac{\omega^8}{189}|\hat I_{abc}(\omega)|^2+\frac{16}{45}\omega^6 |\hat J_{ab}(\omega)|^2  \right]\nonumber\\
&+&\eta^4 \left[\frac{\omega^{10}}{9072}|\hat I_{abcd}(\omega)|^2+\frac{\omega^8}{84}|\hat J_{abc}(\omega)|^2  \right]\,.\nonumber\\
\end{eqnarray}
Here, the frequency-domain multipole moments are also given in a PN-expanded form, e.g.,
\beq
\hat I_{ab}(\omega)= \hat I_{ab}^{\rm N}(\omega) + \eta^2\hat I_{ab}^{\rm 1PN}(\omega)  
+ \eta^4\hat I_{ab}^{\rm 2PN}(\omega) + O(\eta^6)\,.
\eeq
The expression, Eq. \eqref{W1_final_exp}, for $W_1^{\rm tail,h}(E,j)$ is closely related to the total energy flux emitted during the scattering process
\beq \label{EGWomega}
\Delta E_{\rm GW}=\frac{G^2H}{\pi c^5}\nu^2  \int_{0}^\infty  d\omega  {\mathcal K}(\omega)\,.
\eeq
However, it crucially differs from it by the presence of the logarithmic term  $\ln \left(\omega \frac{2s}{c}e^\gamma  \right)$,
which is characteristic of the tail in the frequency domain \cite{Blanchet:1993ec}.
 
It is convenient to replace the integration over the frequency $\omega$ by an integration over the variable $u$, using
\beq
\omega= \frac{u}{e_r \bar a_r^{3/2}} \,.
\eeq
 The result is the following
\begin{eqnarray}
\label{W1_fourier}
W_1^{\rm tail,h}(E,j)&=& \frac{G^2H}{\pi c^5}\frac{2 \nu^2}{e_r \bar a_r^{3/2}} \int_0^\infty du {\mathcal K}(u) \ln \left(u \alpha\right)\,,\nonumber\\
\end{eqnarray}
with
\beq
{\mathcal K}(u)={\mathcal K}(\omega)\big|_{\omega= u/(e_r \bar a_r^{3/2})}\,,
\eeq
and
\beq
\alpha =  \frac{2s}{ce_r \bar a_r^{3/2}}e^\gamma \,.
\eeq
The integral in Eq. \eqref{W1_fourier} requires special care to be performed, even in  the Newtonian limit\footnote{At the Newtonian level, $\eta \to 0$, all eccentricities agree: $e_t=e_r=e_\phi=e$. However, we will continue to denote the eccentricity as $e_r$ to avoid confusion with the exponentials.} where
\beq
\left[{\mathcal K}(\omega)\right]^{\eta=0}=\frac15 \omega^6 |\hat I_{ab}^{\rm N}(\omega)|^2\,.
\eeq

In the Newtonian limit,  $\eta \to 0$, we have (in units of $G=c=1$, but putting back the appropriate power of $M$)
\begin{eqnarray}
\label{W1_fourier_Newtonian}
W_1^{\rm tail,h,N}(E,j)
&=& \frac25 \frac{M^2}{\pi }\frac{ \nu^2}{e_r^7 \bar a_r^{21/2}} \int_0^\infty du {\mathcal I}_{\rm N}(u) \ln \left(u \alpha\right)\,,\nonumber\\
\end{eqnarray}
with
\beq
{\mathcal I}_{\rm N}(u) \equiv u^6|\hat I_{ab}^{\rm N}(\omega)|^2\bigg|_{\omega= u/(e_r \bar a_r^{3/2})}\,.
\eeq
At this Newtonian level,  ${\mathcal I}_{\rm N}(u)$ can be given a very compact form in terms of Bessel $K_\nu(u)$ functions
\begin{eqnarray}
\label{I_N_gen}
{\mathcal I}_{\rm N}(u)&=&\frac{64 \nu^2}{p^4}\bar a_r^7 u^2e^{-i p\pi}[A K^2_{p+1}(u)\nonumber\\
&&+BK_p(u)K_{p+1}(u)+CK_p^2(u)]\,,
\end{eqnarray}
where $p=iu/e_r$ and
\begin{eqnarray}
A&=&\frac{u^2}{2}(p^2+u^2)(p^2+u^2+1)\,, \nonumber\\
B&=& -u(p^2+u^2)\left[\left(p-\frac32\right)u^2+p(p-1)^2  \right]\,, \nonumber\\
C&=& \frac{u^6}{2}+\left(2p^2-\frac32 p+\frac16 \right)u^4 +\left(\frac52 p^2-\frac72 p +1 \right) p^2u^2\nonumber\\
&& +p^4(p-1)^2\,.
\end{eqnarray}
Indeed, Eq. \eqref{I_N_gen} implies that ${\mathcal I}_{\rm N}(u)$ is quadratic  in  $K_p(u)$, and $K_{p+1}(u)$.
Furthermore, $p$ is purely imaginary and enters both the coefficients $A$, $B$, $C$ and the order of the Bessel K functions.

However, even at this Newtonian order,  the integration variable $(u)$ appears both in the argument ($u$) and in the order ($p=iu/e_r$) 
of the Bessel functions.
The computation proceeds then by expanding the integrand in the large eccentricity limit, with the useful consequence of removing,
at leading order in $\frac1{e_r}$, the 
$u$-dependence from the orders of the Bessel functions, reducing them either to $0$ or to $1$.
At the NLO in $\frac1{e_r}$, there appears the first derivative of $K_\nu(u)$ with respect to the order $\nu$, around the
two values $\nu=0$ and  $\nu=1$. Luckily, these first derivatives can be explicitly computed, namely (see Eqs. 9.1.66-9.1-68 of Ref. \cite{AS})
\beq \label{dKdnu}
\frac{\partial K_\nu(u)}{\partial \nu}\Bigg|_{\nu=0}=0\,,\qquad
\frac{\partial K_\nu(u)}{\partial \nu}\Bigg|_{\nu=1}=\frac{1}{u} K_0(u)\,.
\eeq
However, at the NNLO in the $\frac1{e_r}$ expansion, there appears the second derivative of $K_\nu(u)$ with respect to the order $\nu$.
Though there exist explicit representations for the latter (see Appendix \ref{A}), they introduce a level of complexity which did not
allow us to fully compute the NNLO expansion of $W_1^{\rm tail,h,N}(E,j)$, even at the presently discussed Newtonian
level, $\eta \to 0$.

When going beyond the Newtonian level, the Fourier transforms
\beq
e^{q\sinh v -(p+k) v}\to2e^{-i \frac{\pi}{2}(p+k)}K_{p+k}(u)\,,
\eeq
 become replaced by Fourier transforms of 
 $v^ne^{q\sinh v -(p+k) v}$ and $e^{q\sinh v -(p+k) v}V(v)$. The Fourier transforms of $v^ne^{q\sinh v -(p+k) v}$ lead to integrands involving
\beq
v^ne^{q\sinh v -(p+k) v}\to2(-1)^n\frac{\partial^n}{\partial p^n}\left[e^{-i \frac{\pi}{2}(p+k)}K_{p+k}(u)\right]\,,
\eeq
while the Fourier transforms of the terms $e^{q\sinh v -(p+k) v}V(v)$ would require to work with
 the large-$e_r$ expansion of the $V$-term (see Eq. \eqref{Vdef}), i.e.,
\begin{eqnarray}
\label{arctan_exp}
V(v)&=&2\,{\rm arctan}\left(\tanh \frac{v}{2}\right)
\nonumber\\
&+&
\frac1{e_r}\tanh v + \frac{\sinh v}{e_r^2 \cosh^2 v}
+O(e_r^{-3})\,.
\end{eqnarray}
Unfortunately, we did not find a way to replace the first, ${\rm arctan}\left(\tanh \frac{v}{2}\right)$,  
term by some uniform expansion in $v$ that could be integrated term-by-term.
Keeping it as is  complicates matters.
In some cases, we could overcome this new difficulty by exchanging the order of the $v$ and $u$ integrations, i.e., by integrating 
with respect to $u$ first. This has allowed us to analytically compute some of the remaining integrals.

More details on our computations are given in Appendices \ref{A} and \ref{B}. From the practical point of view, the main outcome 
of the frequency-domain approach
has been the analytical results \eq{solc00} and \eq{solc21}. In addition, this allowed us to analytically compute the first two
terms in the large-eccentricity expansion of the 5.5PN integrated action $W_{\rm 5.5PN}$, as discussed below.

\section{Final results for the h-route tail contribution to the scattering angle}
\label{sec_five}

The results discussed in the previous section are intermediate results towards our real aim which is to compute the nonlocal
scattering angle as a function of energy and angular momentum: $\chi^{\rm nonloc, f}(\pinf,j;\nu)$. In view of Eq. \eq{chivsW}, 
the knowledge of $\chi^{\rm nonloc, f}(\pinf,j;\nu)$ is equivalent to the knowledge of the integrated nonlocal Hamiltonian, 
$W^{\rm nonloc,f}_{\rm hyp}$ as a function of energy and angular momentum. 
The total nonlocal potential $W^{\rm nonloc}_{\rm hyp}(\pinf,j;\nu)$ was decomposed
in Eq. \eq{Wnonloctot} into three terms,
\beq
W^{\rm nonloc}_{\rm hyp}(\pinf,j;\nu)= W^{\rm tail,h} + W^{\rm tail,f-h}+ W^{\rm 5.5PN}\,,
\eeq
 which were defined in Eq. \eq{decompWnonloc}.
In the present section we finish the discussion of the first contribution,  $ W^{\rm tail,h} =  W_1^{\rm tail, h}+W_2^{\rm tail, h}$.

The expression  of $ W^{\rm tail,h}$ as a function of energy and angular momentum is obtained (besides adding together 
$W_1^{\rm tail, h}$ and $W_2^{\rm tail, h}$) from the results described above
(which were expressed in terms of the quasi-Keplerian elements $\bar a_r$ and $e_r$) by re-expressing $\bar a_r$ and $e_r$ 
in terms of $\pinf$ and $j$, using the links given in Appendix \ref{PN}. Introducing (see Eq. \eqref{ecc_newt})
\beq
e_{\rm N} \equiv e_{\rm N}(p_\infty, j) \equiv\sqrt{1+\pinf^2j^2}\,,
\eeq
we have
\bea
e_r &=& e_{\rm N}+\frac12 \frac{\pinf^2}{e_{\rm N}} [(\pinf^2j^2+1)\nu-4\pinf^2j^2-6]\eta^2\nonumber\\
&&
-\frac{\pinf^2}{8 j^2e_{\rm N}^3} [\pinf^2 j^2 (\pinf^2 j^2+1)^2\nu^2\nonumber\\
&&+(\pinf^2 j^2+1) (9 \pinf^4 j^4-91 \pinf^2 j^2-112)\nu\nonumber\\
&&-32 \pinf^6 j^6-36 \pinf^4 j^4+64 \pinf^2 j^2+64]\eta^4
\,,\nonumber\\
\bar a_r &=& \frac{1}{p_\infty^2}+2\eta^2 -\frac{(7\nu-4)}{j^2}\eta^4\,.
\eea
 The use of the new variables $\pinf$ and $j$ (instead of $\bar a_r$ and $e_r$) 
leads to a reshuffling of the large-eccentricity expansion. Indeed, we are actually interested in expanding $ W^{\rm tail,h}(\pinf,j)$
in  powers of $\frac1j$. This changes the meaning of the decomposition in LO, NLO and NNLO terms. To clarify this
change of meaning we  write the expansion in powers of $\frac1j$ as
\bea 
\label{decompWtailj}
W^{\rm tail,h}(\pinf,j)&=& W^{\rm tail,h \, LO_j}(\pinf,j) + W^{\rm tail,h \, NLO_j}(\pinf,j)\nonumber\\
&+& W^{\rm tail,h \, NNLO_j}(\pinf,j)+ O\left(\frac1{j^6}\right)\,,
\eea
where we have added a subscript $j$ to the superscripts N$^n$LO, because we are now referring to an expansion of 
$W^{\rm tail,h \, LO_j}(\pinf,j)$ in powers of $\frac1j$.

For instance, at the leading order in $\frac1j$, the combination, and re-expression, of Eqs. \eq{W1LO0}--\eq{W2LO0} yields the result
\begin{widetext}
\begin{eqnarray}
W^{\rm tail,h \, LO_j}(\pinf,j)
&=& \frac{2}{15}\frac{M^2\nu^2 p_{\infty}^4\pi}{ j^3}\left[
\frac{315}{4}+37\ln \left(\frac{p_{\infty}}{2} \right)
+\left[\frac{2753}{224}-\frac{1071}{8}\nu+\left(\frac{1357}{56}-\frac{111}{2}\nu\right)\ln \left(\frac{p_{\infty}}{2} \right)\right]p_{\infty}^2\eta^2\right. \nonumber\\
&& \left.+\left[\frac{155473}{1792}-\frac{109559}{8064}\nu+\frac{186317}{1008}\nu^2
+\left(\frac{27953}{672}-\frac{2517}{112}\nu+\frac{555}{8}\nu^2\right)\ln \left(\frac{p_{\infty}}{2} \right)\right]p_{\infty}^4\eta^4
\right]\,.
\end{eqnarray}
\end{widetext}
As we see, the LO$_j$ term is $\propto \frac1{j^3}$. Correspondingly,  the NLO$_j$ one will be
 $\propto \frac1{j^4}$, and the NNLO$_j$ one  $\propto \frac1{j^5}$.
As per Eq. \eq{chivsW}, the corresponding contributions to the scattering angle will be, respectively, $\propto \frac1{j^4}$, 
$\propto \frac1{j^5}$ and $\propto \frac1{j^6}$. Remembering that $\frac1j = O(G)$ this will give us the value of the scattering
angle up to the sixth order in $G$.
We display below the explicit results for the 2PN-accurate scattering angles associated with the LO$_j$, NLO$_j$ and NNLO$_j$ values of $W^{\rm tail,h}$.

Inserting  Eq. \eq{decompWtailj} in Eq. \eq{chivsW} yields
\beq
\chi^{\rm nonloc,f}= \chi^{\rm tail,h} + \chi^{\rm f-h}+ \chi^{\rm 5.5PN}\,,
\eeq
where the first (h-route) contribution is given by
\begin{eqnarray}
\frac12 \chi^{\rm tail,h} &=& \frac{1}{2 M^2\nu}\frac{\partial}{\partial j}W^{\rm tail,h}\nonumber\\
&=&\frac{\chi_4^{\rm tail,h}(\g,\nu)}{j^4} +  \frac{\chi_5^{\rm tail,h}(\g,\nu)}{j^5}+\frac{\chi_6^{\rm tail,h}(\g,\nu)}{j^6}\nonumber\\
&& +O\left(\frac1{j^6}\right)\,,
\end{eqnarray}
with
\bea
\frac{\chi_4^{\rm tail,h}}{j^4}&=&  \frac{1}{2 M^2\nu}\frac{\partial}{\partial j}W^{\rm tail,h \, LO_j},\nonumber\\
\frac{\chi_5^{\rm tail,h}}{j^5}&=&  \frac{1}{2 M^2\nu}\frac{\partial}{\partial j}W^{\rm tail,h \, NLO_j},\nonumber\\
\frac{\chi_6^{\rm tail,h}}{j^6}&=&  \frac{1}{2M^2 \nu}\frac{\partial}{\partial j}W^{\rm tail,h \, NNLO_j}.
\eea
As already announced, this yields results for $ \chi^{\rm tail,h}$ that can be written as
\bea \label{chinonlocxpbis}
\chi^{\rm tail, h}(\pinf,j;\nu) &=& \nu \frac{\pinf^4}{j^4} \left( A_0^{\rm tail, h}(\pinf;\nu) + \frac{A_1^{\rm tail, h}(\pinf;\nu)}{\pinf j}\right. \nonumber\\
&&\left.+ \frac{A_2^{\rm tail, h}(\pinf;\nu)}{(\pinf j)^2} + \cdots \right),
\eea
where the $\frac1j$ coefficients $A_0^{\rm tail, h}(\pinf;\nu)$, $A_1^{\rm tail, h}(\pinf;\nu)$, $A_2^{\rm tail, h}(\pinf;\nu)$,  are themselves 
given by a 2PN-accurate expansion  in powers of $\pinf$, say
\bea \label{Axptail}
A_0(\pinf;\nu)&=&A_0^{\rm tail, h, N}+ A_0^{\rm tail, h,1PN}+ A_0^{\rm tail, h, 2PN}+\cdots,\nonumber\\
A_1(\pinf;\nu)&=&A_0^{\rm tail, h, N}+ A_1^{\rm tail, h, 1PN}+  A_1^{\rm tail, h, 2PN}+\cdots, \nonumber\\
A_2(\pinf;\nu)&=&A_2^{\rm tail, h, N}+ A_2^{\rm tail, h, 1PN}+ A_2^{\rm tail, h, 2PN}+\cdots.\nonumber\\
\eea
Note that, in absence of the 5.5PN contribution, we do not have here fractional 1.5PN contributions, as in Eq. \eq{Axp}.

The LO$_j$ expressions given above yield
\begin{widetext}
\begin{eqnarray}
A_0^{\rm tail, h, N}&=&
\pi\left[-\frac{37}{5}\ln \left(\frac{p_{\infty}}{2} \right)-\frac{63}{4}\right]
\,,\nonumber\\
A_0^{\rm tail, h,1PN}&=&
\pi\left[\left(-\frac{1357}{280}+\frac{111}{10}\nu\right)\ln \left(\frac{p_{\infty}}{2} \right)-\frac{2753}{1120}+\frac{1071}{40}\nu\right]p_{\infty}^2
\,,\nonumber\\
A_0^{\rm tail, h, 2PN}&=&
\pi\left[\left(-\frac{27953}{3360}+\frac{2517}{560}\nu-\frac{111}{8}\nu^2\right
)\ln \left(\frac{p_{\infty}}{2} \right)-\frac{155473}{8960}+\frac{109559}{40320}\nu-\frac{186317}{5040}\nu^2\right]p_{\infty}^4
\,.
\end{eqnarray}
Our final results for the coefficients of the NLO$_j$ and NNLO$_j$ contributions to the tail part of the scattering angle then read
\begin{eqnarray}
A_1^{\rm tail, h, N}&=&
 -\frac{6656}{45}-\frac{6272}{45}\ln \left(4\frac{p_{\infty}}{2} \right)
\,,\nonumber\\
A_1^{\rm tail, h,1PN}&=&
\left[\left(-\frac{74432}{525}+\frac{13952}{45}\nu\right)\ln \left(4\frac{p_{\infty}}{2} \right)+\frac{114368}{1125}+\frac{221504}{525}\nu\right]p_{\infty}^2
\,,\nonumber\\
A_1^{\rm tail, h,2PN}&=&
\left[\left(-\frac{881392}{11025}+\frac{288224}{1575}\nu-\frac{21632}{45}\nu^2\right)\ln \left(4\frac{p_{\infty}}{2} \right)
+\frac{48497312}{231525}-\frac{5134816}{23625}\nu-\frac{25465952}{33075}\nu^2\right]p_{\infty}^4
\,.\nonumber\\
\end{eqnarray}
and
\begin{eqnarray}
A_2^{\rm tail, h, N}&=&\pi \left[ -122 \ln \left(\frac{p_{\infty}}{2} \right)-\frac{1633}{24} +\frac52 c_{00} -\frac{843}{4} \ln(2)
\right]\,,\nonumber\\
A_2^{\rm tail, h, 1PN}&=&\pi 
\left[\left(\frac{44845}{336} +\frac{5199}{8} \ln(2)+\frac52  c_{21} +\frac{811}{2} \ln\left(\frac{p_{\infty}}{2} \right)
-5c_{00} \right)\nu\right.\nonumber\\
&&\left.-\frac{6379}{192} +\frac{5}{2}c_{20} -\frac{13831}{56} \ln\left(\frac{p_{\infty}}{2} \right)+\frac{15}{2}c_{00} -\frac{74625}{224} \ln(2)  \right]p_{\infty}^2
\,,\nonumber\\
A_2^{\rm tail, h,  2PN}&=&\pi 
\left[\left(\frac52  c_{42} -785 \ln\left(\frac{p_{\infty}}{2} \right)-\frac{97127}{6048} +\frac{15}{2} c_{00} -5 c_{21} -\frac{39345}{32} \ln(2)\right)\nu^2\right.\nonumber\\
&&+\left(-\frac{2989465}{48384} +\frac{5}{2} c_{41} +\frac{75595}{168} \ln\left(\frac{p_{\infty}}{2} \right)+\frac52 c_{21} +\frac{152459}{448} \ln(2)-5 c_{20} -\frac{55}{4}c_{00} \right)\nu\nonumber\\
&&\left. +\frac52 c_{40} +\frac{986233}{1536} +\frac{396481}{2688} \ln(2)+\frac52  c_{20} +\frac{15}{4} c_{00} +\frac{64579}{1008} \ln\left(\frac{p_{\infty}}{2} \right)  \right]p_{\infty}^4
\,,
\end{eqnarray}
\end{widetext}
respectively.
Introducing the new set of parameters
\begin{eqnarray} \label{dmn}
d_{00}&=&   \frac52 c_{00}-\frac{1633}{24}-\frac{843}{4}\ln(2) 
\,,\nonumber\\
d_{20}&=&   \frac52 c_{20}+\frac{32813}{192}+\frac{66999}{224}\ln(2)   
\,,\nonumber\\
d_{21}&=&   \frac52 c_{21}-\frac{293}{112}+\frac{1827}{8}\ln(2) 
\,,\nonumber\\
d_{40}&=&   \frac52 c_{40}+\frac{293499}{512} +\frac{442237}{2688}\ln(2)    
\,,\nonumber\\
d_{41}&=&   \frac52 c_{41}-\frac{4431841}{48384}-\frac{28735}{64}\ln(2)  
\,,\nonumber\\
d_{42}&=&   \frac52 c_{42}+\frac{1105777}{6048}-\frac{4497}{32}\ln(2)  
\,,
\end{eqnarray}
with numerical values listed in Table \ref{num_resd},
the latter expressions can be rewritten as
\begin{eqnarray}
A_2^{\rm tail, h, N}&=&\pi \left[ -122 \ln \left(\frac{p_{\infty}}{2} \right)+d_{00}
\right]\,,\nonumber\\
A_2^{\rm  tail, h,1PN}&=&\pi 
\left[\left(d_{21} -2 d_{00}  +\frac{811}{2} \ln\left(\frac{p_{\infty}}{2} \right)\right)\nu\right.\nonumber\\
&+&\left.
d_{20} +3 d_{00} -\frac{13831}{56} \ln\left(\frac{p_{\infty}}{2} \right) 
 \right]p_{\infty}^2
\,,\nonumber\\
 A_2^{\rm tail, h, 2PN}&=&\pi 
\left[\left(d_{42} +3 d_{00} -2 d_{21}-785 \ln\left(\frac{p_{\infty}}{2} \right)\right)\nu^2\right.\nonumber\\
&+&
\left(d_{41}  +d_{21} -2 d_{20} -\frac{11}{2} d_{00}\right.\nonumber\\
&+&\left.
+\frac{75595}{168} \ln\left(\frac{p_{\infty}}{2} \right) 
 \right)\nu\nonumber\\
&+&\left.
d_{40} +\frac32 d_{00} +d_{20} +\frac{64579}{1008} \ln\left(\frac{p_{\infty}}{2} \right)
\right]p_{\infty}^4
\,.\nonumber\\
\end{eqnarray}

Among the numerical coefficients entering the NNLO$_j$ quantity 
$A_2(\pinf;\nu) = A_2^{\rm tail, h, N}+ A_2^{\rm tail, h, 1PN}+ A_2^{\rm tail, h, 2PN}$
 two can be written down in analytical form (thanks to our frequency-domain computation), namely,
 using Eqs. \eqref{solc00} and \eqref{solc21},
 \beq \label{d00}
d_{00}=-\frac{99}{4}-\frac{2079}{8}\zeta(3)\,,
\eeq
and 
\beq \label{d21}
d_{21}=\frac{1541}{8}+306 \zeta(3)\,,
\eeq
while the other ones are (only) known numerically (see Table \ref{num_res}).
%
%
%
%
\begin{table}
\caption{\label{num_resd} Numerical values of the coefficients \eqref{dmn}.}
\begin{ruledtabular}
\begin{tabular}{ll}
coefficient&numerical value\\
\hline
$d_{00}$&$-337.13453770$ \\
$d_{20}$&$+668.10143447$ \\
$d_{21}$&$+560.45441238$ \\
$d_{40}$&$+743.05631726$ \\
$d_{41}$&$-694.94788994$ \\
$d_{42}$&$-439.10050487$\\
\end{tabular}
\end{ruledtabular}
\end{table}

\section{Second-order tail contribution to the integrated action and to the scattering angle: $W^{5.5{\rm PN}}$ and $\chi^{5.5{\rm PN}}$}

Before finishing our discussion of the h-route tail contribution, $\chi^{\rm tail, h}(\pinf, j ; \nu)$, to the scattering angle, 
and of its $\nu$-dependence,
let us recall that, at the 6PN accuracy where we are working, the total scattering angle is made of the following four contributions:
\beq
\chi^{\rm tot}(\pinf, j ; \nu)= \chi^{\rm loc, f}+ \chi^{\rm tail, h} + \chi^{\rm 5.5PN}+ \chi^{\rm f-h}\,.
\eeq
Among these contributions two of them are
directly linked with nonlocal effects computed in harmonic-coordinates: indeed, $\chi^{\rm tail,h}$ comes from the {\it first-order tail}
(linear in $G {\cal M}$), while $\chi^{\rm 5.5PN}$ comes from the  {\it second-order  tail} (quadratic in $G {\cal M}$). 
Before being able to discuss the constraint that must be satisfied by the flexibility factor $f(t)$ entering the last contribution, $\chi^{\rm f-h}$,
we must control the structure (and, notably, the $\nu$-dependence) of the quadratic-tail contribution $\chi^{\rm 5.5PN}$.

From Eq. \eq{H5.5}, denoting  $B=-\frac{107}{105}$ and 
\beq
{\mathcal H}^{\rm split}(t,\tau)=\frac{G}{5c^5}[I^{(3)}_{ij}(t)I^{(4)}_{ij}(t+\tau)-I^{(3)}_{ij}(t)I^{(4)}_{ij}(t-\tau)]\,,
\eeq
 the 5.5PN Hamiltonian reads
\beq
H_{5.5{\rm PN}}^{\rm nonloc}=\frac{B}{2}\left(\frac{G{\mathcal M}}{c^3}\right)^2\int_{-\infty}^\infty
\frac{d\tau}{\tau}{\mathcal H}^{\rm split}(t,\tau)\,.
\eeq
Note that the function ${\mathcal H}^{\rm split}(t,\tau)$ is odd in $\tau$, so that $\tau^{-1}{\mathcal H}^{\rm split}(t,\tau)$
is even in $\tau$ (and regular at $\tau=0$).

As usual the computation can be done either in the time domain  or in the Fourier domain.
Working in the Fourier domain we find
\begin{eqnarray}
W_{5.5{\rm PN}}^{\rm nonloc}&=&-B \left(\frac{G{\mathcal M}}{c^3}\right)^2 \frac{G}{5c^5}\int_0^\infty
d\omega \omega^7 |\hat I_{ij}(\omega)|^2\,.\nonumber\\
\end{eqnarray}
At our present level of accuracy, it is enough to use the Newtonian approximation to the Fourier transform $\hat I_{ij}(\omega)$ of the
quadrupole moment. Using the relations given in Appendix \ref{A}  we have then
\begin{eqnarray} \label{W55fourier}
W_{5.5{\rm PN}}^{\rm nonloc}&=&-B \left(\frac{G{\mathcal M}}{c^3}\right)^2 \frac{G}{5c^5}\int_0^\infty
d\omega \omega^7 |\hat I_{ij}(\omega)|^2\nonumber\\
&=& \frac{107}{105}   \left(\frac{\bar n}{e_r}\right)^8 \left(\frac{G{\mathcal M}}{c^3}\right)^2 \frac{G}{5c^5} \int _0^\infty du u  {\mathcal I}_{\rm N}(u)\,,\nonumber\\
\end{eqnarray}
where the function ${\mathcal I}_{\rm N}(u)$ was defined in Eq. \eq{I_N_gen}.

As explained above, though the function ${\mathcal I}_{\rm N}(u)$ is here evaluated at the Newtonian level, it is quadratic in
Bessel functions $K_\nu(u)$ whose order is $u$-dependent: either $\nu=p$ or $\nu=p+1$, with $p=iu/e_r$. This makes it impossible
to compute $W_{5.5{\rm PN}}^{\rm nonloc}$ in closed form. However, it is enough for our purpose to compute the first two terms
in the large-eccentricity expansion of the integral \eq{W55fourier}. Thanks to the relations   \eq{dKdnu},
this computation only involves integrals containing $K_0(u)$ and $K_1(u)$. We find
\bea
W_{5.5{\rm PN}}^{\rm LO+NLO} &=&
 \frac{107}{105}  \left(\frac{\bar n}{e_r}\right)^8 \left(\frac{G{\mathcal M}}{c^3}\right)^2 \frac{G}{5c^5} \nu^2 32 \bar a_r^7 e_r^4 \nonumber\\
&\times &
\int _0^\infty du u  {\mathcal F}(u) \left(1+\frac{u}{e_r} +O\left(\frac{1}{e_r^2} \right)\right)\,,\nonumber\\
\end{eqnarray}
where LO and NLO refer to the large-eccentricity 
expansion, and where  
\begin{eqnarray}
\label{calFslpash}
{\mathcal F}(u)&=&\left(\frac{u^2}{3}+u^4\right)K_0^2(u)+3u^3 K_0(u)K_1(u)\nonumber\\
&+&(u^2+u^4)K_1^2(u)\,,
\end{eqnarray}
denotes (as in  Ref. \cite{Bini:2017wfr}, and in Appendix \ref{A})
the gravitational-wave energy spectrum in the  Newtonian-level ``splash" approximation \cite{rr1,rr2}, i.e., 
at Newtonian order, and at leading order in the large-eccentricity limit. 

The NLO-accurate result $W_{5.5{\rm PN}}^{\rm LO+NLO}$ involves the two nontrivial integrals 
${\sf f}^u=\int _0^\infty du u  {\mathcal F}(u)$, and ${\sf f}^{u^2}= \int _0^\infty du u^2  {\mathcal F}(u)$.
These integrals are computed in Appendix \ref{A}. This leads to the following explicit NLO result for $W_{5.5{\rm PN}}^{\rm nonloc}$ (using $G=1=c$):
\begin{eqnarray}
W_{5.5{\rm PN}}^{\rm LO+NLO}&=&
\frac{32}{5}\frac{107}{105}  \frac{p_\infty^{10}}{e_r^4}  M^2 \nu^2  
 \left(\frac{49}{9} +\frac{1}{e_r}\frac{297}{256}\pi^2\right.\nonumber\\
&&\left.
+O\left(\frac{1}{e_r^2} \right)\right)\,.
\end{eqnarray}
Replacing the eccentricity in terms of $j$ finally leads to the following explicit NLO$_j$ result for $W_{5.5{\rm PN}}^{\rm nonloc}$
\bea
W_{5.5{\rm PN}}^{\rm LO+NLO}&=& M^2\nu^2\left( \frac{23968}{675}\frac{p_\infty^6}{j^4}+\frac{10593}{1400}\pi^2 \frac{p_\infty^5}{j^5}\right. \nonumber\\
&&\left. +O\left(\frac{1}{j^6}\right)\right)\,.
\eea
Using the formula \eq{chivsW} we finally get 
\begin{eqnarray} \label{chi55j}
\chi_{5.5{\rm PN}}^{\rm LO+NLO}&=& -\nu \left(\frac{95872}{675}\frac{p_\infty^6}{j^5} +\frac{10593}{280}\pi^2 \frac{p_\infty^5}{j^6}\right.\nonumber\\
&&\left.
+O\left(\frac{1}{j^7}\right)\right)\nonumber\\
&=& -\nu \frac{95872}{675}\frac{p_\infty^6}{j^5}\left(1 +\frac{13365}{50176} \frac{\pi^2}{p_\infty j}\right.\nonumber\\
&&\left.
+O\left(\frac{1}{j^2}\right)\right)\,.\nonumber\\
\end{eqnarray}

\section{Analysis of the $\nu$-dependence of the harmonic-coordinate nonlocal scattering angle $\chi^{\rm nonloc,h}$}

Let us recall that a crucial tool of our method is to exploit
the special $\nu$-dependence \cite{Damour:2019lcq} satisfied by the total scattering angle $\chi^{\rm tot}(\pinf, j ; \nu)$. 
This structure is embodied in a restricted $\nu$-polynomial dependence
of the energy-rescaled PM-expansion coefficients of $\chi^{\rm tot}(\pinf, j ; \nu)$.

The total scattering angle  $\chi^{\rm tot}(\pinf, j ; \nu)$ is obtained as a sum of partial contributions, namely
\beq
\chi^{\rm tot}(\pinf, j ; \nu)= \chi^{\rm loc, f}+ \chi^{\rm tail, h} + \chi^{\rm 5.5PN}+ \chi^{\rm f-h}\,.
\eeq
Some of these contributions can fail to satisfy the special $\nu$-dependence satisfied by $\chi^{\rm tot}(\pinf, j ; \nu)$.
One ingredient of our method is to assume that $ \chi^{\rm loc, f}$ does satisfy  the latter special $\nu$-dependence.
We must, then, constrain  the flexibility factor $f$ in such a way that the complementary, nonlocal-related, contribution
\beq
\chi^{\rm tail, h} + \chi^{\rm 5.5PN}+ \chi^{\rm f-h}\,,
\eeq
does satisfy  the special $\nu$-dependence satisfied by $\chi^{\rm tot}(\pinf, j ; \nu)$.
This will be the task of the present section. We will start by recalling what is the  special $\nu$-structure
we are talking about. Then we will measure the extent to which the sum of the two h-route
nonlocal contributions, say $\chi^{\rm nonloc,h} \equiv \chi^{\rm tail,h} + \chi^{\rm 5.5PN}$
fails to satisfy the latter special $\nu$-dependence. This will finally allow us to constrain $f(t)$.

\subsection{Reminder of the $\nu$-rule to be satisfied} 
 
Let us define precisely the $\nu$-rule to be satisfied. We expand  in powers of $\frac1j =O(G)$ 
any partial contribution $\chi^X(\pinf, j ; \nu)$ to the total scattering angle, $\chi^{\rm tot} = \sum_X \chi^X$, say
\beq
\frac12 \chi^X(\pinf, j ; \nu)=\sum_n  \frac{\chi^X_n(\pinf; \nu)}{j^n}\,,
\eeq
and then define the energy-rescaled coefficients
\beq
\widetilde \chi^X_n(\pinf; \nu) \equiv h^{n-1}(\g,\nu) \chi^X_n(\pinf; \nu)\,,
\eeq
where we recall that
\beq
h(\g,\nu)=\sqrt{1+2\nu(\gamma-1)}\,,\qquad \gamma =\sqrt{1+p_{\infty}^2 \eta^2}\,,
\eeq
that is
\begin{eqnarray}
h&=& 1+\frac12 \nu p_{\infty}^2 \eta^2-\frac18 \nu(1+\nu) p_{\infty}^4\eta^4+O(\eta^6)\,.
\end{eqnarray}
The special $\nu$-structure says that 
\beq \label{rule}
\widetilde \chi^X_n(\pinf; \nu)=P^{X\g}_{d(n)}(\nu)\,,
\eeq
where $P^{X\g}_{d(n)}(\nu)$ denotes a polynomial in $\nu$, of degree
\beq
d(n) \equiv \left[\frac{n-1}{2}\right]\,,
\eeq
with $\g$-dependent coefficients. [Here, $\left[ \cdots\right]$ denotes the integer part.]
Let us analyze the $\nu$-structure Eq. \eq{rule} for the case where the label $X$ is equal to nonloc,h, in the
sense of the following definition of the sum of the two harmonic-coordinate nonlocal contributions
\beq \label{chinonlochtot}
\chi^{\rm nonloc,h \, tot} \equiv \chi^{\rm tail,h} + \chi^{\rm 5.5PN}\,.
\eeq
Our results above have led to the determination of $\chi_4^{\rm nonloc,h \, tot}$, $\chi_5^{\rm nonloc,h\, tot}$ and $\chi_6^{\rm nonloc,h\, tot}$. More precisely, we must, according to the definition, Eq. \eq{chinonlochtot}, add the 5.5PN contribution Eq. \eq{chi55j} to the
corresponding results for the $\frac1j$ expansion of $\chi^{\rm tail,h}$ given in section \ref{sec_five}.

Let us first remark that, in fact, the 5.5PN contribution $\chi^{\rm 5.5PN}$ {\it separately satisfies} the rule \eq{rule}.
Indeed, as we are at the 5.5PN level, we can use $h \approx 1$ so that $\widetilde \chi^{\rm 5.5PN}_n(\pinf; \nu) \approx \chi^{\rm 5.5PN}_n(\pinf; \nu)$. Then, for the relevant exponents $n=5,6$ of $\frac1j$, the rule \eq{rule} says that $\widetilde \chi^{\rm 5.5PN}_n(\pinf; \nu)$ should
be at most quadratic in $\nu$. However, our explicit results Eq. \eq{chi55j} for the $\widetilde \chi^{\rm 5.5PN}_n$'s show that they are
actually {\it linear} in $\nu$. 

In view of this structure of $\chi^{\rm 5.5PN}$, we can henceforth focus only on the remaining h-route contribution to $\chi$, namely $\chi^{\rm tail,h}$.
In the following, we shall use the notation
\beq
\chi^{\rm nonloc,h } \equiv \chi^{\rm tail,h}\,,
\eeq
to emphasize that this is the crucial additional h-route contribution to  the local piece $\chi^{\rm loc,f }$, to be eventually modified
by a suitable $f$-dependent piece $\chi^{\rm f-h }$.

We transform the results  given in section \ref{sec_five}
for $\chi^{\rm tail,h}= \chi^{\rm nonloc,h }$
into corresponding results for their energy-rescaled versions
$\widetilde \chi_4^{\rm nonloc,h}= h^3 \chi_4^{\rm tail,h}$, $\widetilde \chi_5^{\rm nonloc,h}=h^4 \chi_5^{\rm tail,h}$ and 
$\widetilde \chi_6^{\rm nonloc,h}=h^5 \chi_6^{\rm tail,h}$. 
We find 
\begin{widetext}
\begin{eqnarray}
\label{tildechinl456}
\widetilde \chi_4^{\rm nonloc,h} &=& \left(-\frac{63}{4}  -\frac{37}{5}  \ln\left(\frac{p_{\infty}}{2} \right)\right)\pi \nu p_{\infty}^4\nonumber\\
&&+\left(-\frac{2753}{1120}  -\frac{1357}{280}\ln\left(\frac{p_{\infty}}{2} \right)+\frac{63}{20}\nu \right)\pi \nu p_{\infty}^6 \eta^2 \nonumber\\
&&+\left(-\frac{27331}{10080}\nu^2+\frac{199037}{40320}\nu -\frac{155473}{8960} -\frac{27953}{3360}\ln\left(\frac{p_{\infty}}{2} \right)\right)\pi \nu p_{\infty}^8\eta^4
\,,\nonumber\\
\widetilde \chi_5^{\rm nonloc,h} &=& \left(-\frac{6656}{45}-\frac{12544}{45}\ln(2)-\frac{6272}{45} \ln\left(\frac{p_{\infty}}{2} \right)\right)\nu p_{\infty}^3 \nonumber\\
&&+\left[\left(-\frac{148864}{525}\ln(2)+\frac{114368}{1125}-\frac{74432}{525} \ln\left(\frac{p_{\infty}}{2} \right)\right) 
+\left(\frac{2816}{45}\ln(2)+\frac{198592}{1575}+\frac{1408}{45} \ln\left(\frac{p_{\infty}}{2} \right)\right)\nu\right]\nu p_{\infty}^5 \eta^2\nonumber\\
&&+\left[\left(-\frac{13888}{225}\ln(2)-\frac{6944}{225} \ln\left(\frac{p_{\infty}}{2} \right)+\frac{283168}{4725}\right)\nu-\frac{2448608}{33075}\nu^2\right. \nonumber\\
&&\left. +\left(\frac{48497312}{231525}-\frac{881392}{11025} \ln\left(\frac{p_{\infty}}{2} \right)-\frac{1762784}{11025}\ln(2)\right)\right]\nu p_{\infty}^7 \eta^4
\,,\nonumber\\
\widetilde \chi_6^{\rm nonloc,h} &=& \left(-122 \ln\left(\frac{p_{\infty}}{2} \right)+d_{00} \right)\pi \nu p_{\infty}^2 \nonumber\\
&&+\left[\left(-\frac{13831}{56} \ln\left(\frac{p_{\infty}}{2} \right)+d_{20} +3 d_{00} \right) +\left(\frac{201}{2} \ln\left(\frac{p_{\infty}}{2} \right)+\frac12 d_{00} +d_{21} \right)\nu\right]\pi \nu p_{\infty}^4 \eta^2\nonumber\\
&& +\left[\left(-\frac{30655}{336} \ln\left(\frac{p_{\infty}}{2} \right)+d_{21} +\frac{11}{8} d_{00} +\frac12 d_{20} +d_{41} \right)\nu+\left(\frac12 d_{21} +d_{42} -\frac18 d_{00} \right)\nu^2\right. \nonumber\\
&& \left.+\left(\frac{64579}{1008}\ln\left(\frac{p_{\infty}}{2} \right)+\frac32 d_{00} 
+d_{20} +d_{40} \right) \right]\pi \nu p_{\infty}^6\eta^4\,.
\end{eqnarray}

\end{widetext}

We recall that the powers of $\eta$ in Eqs. \eqref{tildechinl456} denote fractional (rather than absolute) PN corrections.
Actually, the leading-order contributions to $\widetilde \chi_4^{\rm nonloc,h} $, $\widetilde \chi_5^{\rm nonloc,h} $,
and $\widetilde \chi_6^{\rm nonloc,h} $ are all at the 4PN level, so that the $\eta^2$ (and $\eta^4$) terms denote
5PN (and  6PN) corrections, respectively.

\subsection{On the $\nu$-structure of the logarithmic contributions, and of the gravitational-wave energy loss}

The rule Eq. \eq{rule} says that $\widetilde \chi_4^X$ should be at most linear in $\nu$, while $\widetilde \chi_5^X$ and $\widetilde \chi_6^X$
should be at most quadratic in $\nu$. 

Let us first note that this rule is satisfied by {\it all the logarithmic contributions}. This is a nontrivial
check of the validity of this rule because all the logarithmic contributions have a genuinely nonlocal origin, and could not be compensated by
additional (logarithmic-free) local terms. Let us also note that the simple $\nu$-polynomial structure of the logarithmic (tail) contributions to
$\widetilde \chi_n^{\rm nonloc, h}$ is rather hidden in the structure of the multipole moments and, thereby, in the structure of the total
gravitational-radiation energy loss. It is worth pausing a moment to comment more on this structure.

From Eq. \eqref{Hnonloch}, one sees that the logarithmic contributions to\footnote{As explained in the
previous subsection, we henceforth label as ``nonloc, h" the crucial $4+5+6$PN nonlocal contribution, because the second-order
tail contribution separately satisfies the constraint we are studying.} $H^{\rm nonloc, h}$ are proportional to
\beq
\frac{G H_{}}{c^{5}}{\mathcal F}^{\rm split}(t,t)= \frac{G H_{}}{c^{5}}{\mathcal F}^{\rm GW}(t)\,,
\eeq
where ${\mathcal F}^{\rm GW}(t)$ is the instantaneous flux of  gravitational-wave energy.
Therefore, the logarithmic contributions to $W^{\rm nonloc, h} = \int dt H^{\rm nonloc, h}$ are proportional to
\beq
\frac{G H_{}}{c^{5}}\int dt\,  {\mathcal F}^{\rm GW}(t)= \frac{G H_{}}{c^{5}} \Delta E^{\rm GW}\,,
\eeq
where $\Delta E^{\rm GW}$ denotes the total energy radiated\footnote{Actually, as we are considering a time-symmetric interaction,
\`a la Fokker-Wheeler-Feynman, this energy is first absorbed by the system in the form of advanced waves, before being radiated
in the form of retarded waves.} during an hyperbolic encounter. Let us consider the functional dependence of $\Delta E^{\rm GW}$
on $\g$ (or, equivalently, $\pinf$), $j$ and $\nu$, and the expansion of $\Delta E^{\rm GW}(\g,j,\nu)$ in powers of $\frac1j$:
\beq
\Delta E^{\rm GW}(\g,j ;\nu)= \sum_{n\geq3} \frac{\Delta E^{\rm GW}_n(\g ;\nu)}{j^n}\,.
\eeq
The logarithmic contributions to $W^{\rm nonloc, h}$ have a $\frac1{j}$ expansion proportional to 
$h \sum_{n\geq3} \frac{\Delta E^{\rm GW}_n(\g ;\nu)}{j^n}$, so that the logarithmic contributions to 
$\chi^{\rm nonloc, h}=\frac1{M^2 \nu} \partial W^{\rm nonloc, h}/\partial j$ are proportional to
\beq
-\frac{h}{\nu} \sum_{n \geq3} n \frac{\Delta E^{\rm GW}_n(\g ;\nu)}{j^{n+1}}\,.
\eeq
In order for the rule Eq. \eq{rule} to be separately satisfied by the logarithmic contributions to the scattering angle,
and taking into account both the factor $\frac1{\nu}$ in the previous equation, and the fact that $\Delta E^{\rm GW} \propto \nu^2$,
we  conclude that the coefficient of $\frac1{j^n}$ in the gravitational-radiation loss should satisfy the non-trivial rule
\beq \label{ruleEGW}
h^{n+1}(\g,\nu)\frac{ \Delta E^{\rm GW}_n(\g ;\nu)}{\nu^2}=  P^{\g}_{\left[ \frac{n-2}{2}\right]}(\nu) \; {\rm for}\; n\geq3\,.
\eeq
We have confirmed the validity of this rule in two different ways.

First, we note that the rule \eq{ruleEGW} states that the leading-order contribution to $\Delta E^{\rm GW}(\g ,j;\nu)$
in its expansion in powers of $\frac1j$, \ie, its leading-order PM contribution $\propto \frac1{j^3} =O(G^3)$ must depend on
$\nu$ as $\propto \nu^2/h^4(\g,\nu)$. In view of the relation \cite{Damour:2017zjx}
\beq
\frac{GM}{b} = \frac{\pinf}{hj}\,,
\eeq
between $j$ and the impact parameter $b$, this is equivalent to saying that
\bea \label{hEGW}
h(\g,\nu) \Delta E^{\rm GW}(\g ,b,m_1,m_2) &=& (m_1m_2)^2 \left(\frac{G}{b}\right)^3 {\cal E}(\g)\nonumber\\
&& + O(G^4)\,,
\eea
where the dimensionless factor ${\cal E}(\g)$ depends only on $\g$ and not on the mass ratio.
The validity of this statement to all orders in the PN expansion is a non trivial fact which follows from the structure
of the LO post-Minkowskian gravitational Bremsstrahlung results of Refs. \cite{Kovacs:1977uw,Kovacs:1978eu}. Indeed, the
latter references have proven that the LO PM gravitational waveform has three properties: (i) it depends on the masses only through
an overall factor $G^2 m_1 m_2$; (ii) it depends on time through two separate time scales of the form $b f_A(\g)$, $b f_B(\g)$;
and, (iii) it enjoys a forward-backward symmetry in the {\it center-of-velocity frame} $\widetilde {\cal S}$.
These properties imply that the four-momentum $ P^{\mu}_{\rm GW}$ radiated as gravitational 
waves\footnote{As we are discussing the time-symmetric dynamics, the system ``emits"
both advanced and retarded waves and therefore absorbs $-P^{\mu}_{\rm GW}$ before emitting $+P^{\mu}_{\rm GW}$.} is
of the form 
\beq \label{PGW}
P^{\mu}_{\rm GW}=  (m_1m_2)^2 \left(\frac{G}{b}\right)^3 {\cal E}(\g) \frac{u_1^\mu +u_2^\mu}{\g+1} + O(G^4)\,,
\eeq
where $u_1^\mu$, $u_2^\mu$ denote the incoming 4-velocities. Computing from Eq. \eq{PGW} the {\it center-of-mass} energy loss
$\Delta E^{\rm GW} = - P^{\mu}_{\rm GW} (p_{1 \mu}+ p_{2 \mu})/|p_1+p_2| $ (where $p_{a \mu}=m_a u_{a \mu}$
and $|p_1+p_2|= M h(\g,\nu)$) leads to Eq. \eq{hEGW}.

Second, we have computed  $\Delta E^{\rm GW}$ to the 2PN accuracy
(thereby generalizing the 1PN-accurate result  of Blanchet and Sch\"afer \cite{Blanchet:1989cu}).
We give in Appendix \ref{gw2pn}, Eqs. \eq{C1}, \eq{C2}, 
the 2PN-level contribution to  $\Delta E^{\rm GW}$ when  
(following Ref. \cite{Blanchet:1989cu}) it is expressed in terms of $e_r=e_r^h$ and $j$. 
However, expressing $\Delta E_{\rm GW}$ in terms of $e_r$ and $j$ does not help to reveal its hidden simple $\nu$-dependence
because $e_r^h$ is itself a rather involved function of $\bar E \equiv (E_{\rm tot}-Mc^2)/\mu=(h(\g,\nu)-1)/\nu$, $j$ and $\nu$
given by
\beq
e_r^h=\sqrt{1+2\bar E j^2}\left(1+e_2\eta^2 +e_4 \eta^4 \right)\,,
\eeq
where
\begin{eqnarray}
e_2&=& \frac{\bar E}{2(2 \bar E j^2+1)} [(5 \bar E j^2+2)\nu-15 \bar E j^2-12] \,, \nonumber\\
e_4&=&\frac{\bar E}{8(2\bar E j^2+1)^2 j^2}[
\bar E^2 j^4 (7 \bar E j^2+4)\nu^2\nonumber\\
&& +(-210 \bar E^3 j^6+224+792 \bar E j^2+592 \bar E^2 j^4)\nu\nonumber\\
&&
+415 \bar E^3 j^6+200 \bar E^2 j^4-280 \bar E j^2-128] \,.
\end{eqnarray}
 
It is better to reexpress $\Delta E_{\rm GW}$ in terms of $\g$ (or equivalently $\pinf$) and of the (gauge-invariant) 
eccentricity-like variable\footnote{The quantity $e_{hj}$ is a PN-acceptable measure of the eccentricity in the range $0<e_{hj}<\infty$,
and the value $e_{hj}=1$ does describe parabolic motions (with zero binding energy). However, 
  $e_{hj}^2$ does not vanish along circular orbits. }
\beq
e_{hj}^2 \equiv  1+ (\g^2-1)h^2 j^2 = 1+ \pinf^2h^2 j^2\,,
\eeq
and of the related quantities 
\bea \label{defbeps}
\bar\epsilon &\equiv& \frac{1}{\pinf h j}=\frac{1}{\sqrt{e_{hj}^2 -1}}\,, \nonumber\\
 B(\bar \epsilon)&\equiv& \frac{\pi}{2}+ \arctan \bar \epsilon = \arccos \left(-\frac1{e_{hj}}\right)\,.
\eea
Note that $j$ enters these quantities only in the combination $h(\g,\nu)j$.
This leads to a 2PN-accurate result of the form
\bea \label{bepsxp}
h(\g,\nu)\frac{ \Delta E^{\rm GW}(\pinf, \bar \epsilon;\nu)}{M\nu^2}&=& 
\bar \epsilon^3 \widehat E_3(\pinf, \bar \epsilon) 
 + \nu \bar \epsilon^4  \widehat E_4(\pinf, \bar \epsilon) \nonumber\\
&+& \nu^2 \bar \epsilon^6 \widehat E_6(\pinf, \bar \epsilon)+ O(\pinf^{13})\,,\nonumber\\
\eea 
where the functions $ \widehat E_n(\pinf, \bar \epsilon) $, with $n=3,4,6$, are given in Table \ref{cal_E_funct}.


 \begin{table*}
 \caption{\label{cal_E_funct} Functions $ \widehat E_n(\pinf, \bar \epsilon) $ (for $n=3,4,6$), Eq. \eq{bepsxp}, 
  in terms of $\bar \epsilon$
 and $B\equiv B(\bar \epsilon)$, Eqs. \eq{defbeps}. }
 \begin{ruledtabular}
 \begin{tabular}{ll}
$\widehat E_3(\pinf, \bar \epsilon)$ & $[(-\frac{64579}{2520}\bar \epsilon^2+\frac{11947909}{3240}\bar \epsilon^6+\frac{19319}{189}\bar \epsilon^4+\frac{5839651}{1008}\bar \epsilon^8+\frac{27953}{5040}) B $\\
&$+\frac{5839651}{1008}\bar \epsilon^7+\frac{79675961}{45360}\bar \epsilon^5+\frac{64}{5}\frac{\bar \epsilon}{(1+\bar \epsilon^2)^2}+\frac{4309531}{136080}\bar \epsilon^3-\frac{8807569}{58800}\bar \epsilon+\frac{1060}{7}\frac{\bar \epsilon}{(1+\bar \epsilon^2)}]\pinf^{11}$\\
&$
+[(\frac{13831}{140}\bar \epsilon^2+\frac{13447}{20}\bar \epsilon^6+\frac{2259}{4}\bar \epsilon^4+\frac{1357}{420}) B
+\frac{31509}{700}\bar \epsilon-\frac{64}{5}\frac{\bar \epsilon}{(1+\bar \epsilon^2)}+\frac{13447}{20}\bar \epsilon^5+\frac{10219}{30}\bar \epsilon^3] \pinf^9$\\
&$+[(\frac{244}{5}\bar \epsilon^2+\frac{74}{15}+\frac{170}{3}\bar \epsilon^4) B
+\frac{1346}{45}\bar \epsilon+\frac{170}{3}\bar \epsilon^3] \pinf^7$ \\
$\widehat E_4(\pinf, \bar \epsilon)$ & $  [(-\frac{62813}{168}\bar \epsilon^3-\frac{1628347}{360}\bar \epsilon^5-\frac{258051}{40}\bar \epsilon^7+\frac{6131}{168}\bar \epsilon) B 
-\frac{16546}{ 105 (1+\bar \epsilon^2)}-\frac{427097}{180}\bar \epsilon^4-\frac{5912419}{37800}\bar \epsilon^2+\frac{280502}{1575}-\frac{258051}{40}\bar \epsilon^6-\frac{64}{5(1+\bar \epsilon^2)^2}]\pinf^{11}$\\
&$
+[(-\frac{1127}{3}\bar \epsilon^5-\frac{910}{3}\bar \epsilon^3-\frac{201}{5}\bar \epsilon) B 
-\frac{128}{9}+\frac{32}{5 (1+\bar \epsilon^2)}-\frac{1127}{3}\bar \epsilon^4-\frac{1603}{9}\bar \epsilon^2] \pinf^9$ \\
$\widehat E_6(\pinf, \bar \epsilon)$ & $ [(\frac{5929}{6}\bar \epsilon^3+\frac{485}{4}\bar \epsilon+\frac{5481}{4}\bar \epsilon^5)B
+\frac{2966}{45}+\frac{5481}{4}\bar \epsilon^4-\frac{578}{15 (1+\bar \epsilon^2)}+\frac{6377}{12}\bar \epsilon^2-\frac{16}{5(1+\bar \epsilon^2)^2}] \pinf^{11}$ \\
 \end{tabular}
 \end{ruledtabular}
 \end{table*}

These functions have a smooth limit as $\bar \epsilon \to 0$ (equivalent to $e_{hj} \to \infty$, or $j \to \infty$), \ie,
\beq
\widehat E_n(\pinf, \bar \epsilon)= \widehat E_{n0}(\pinf)+ \bar \epsilon \, \widehat E_{n1}(\pinf)+ \bar \epsilon^2 \, \widehat E_{n2}(\pinf) + \cdots\,.
\eeq
The error term $O(\pinf^{13})$ in \eq{bepsxp} indicates a fractional 3PN error level. Indeed,  the leading PN contribution
to  $\widehat E_3(\pinf, \bar \epsilon)$ is $O(\pinf^7)$ (corresponding to the large-eccentricity
Newtonian-level energy loss $\sim \bar \epsilon^3 \pinf^7 \sim e_{hj}^4 j^{-7}$).
 It is then easily checked that the properties embodied in the expansion of the expression \eq{bepsxp} in powers of $\bar \epsilon=1/(\pinf hj)$
implies that the expansion coefficients of $h \Delta E^{\rm GW}(\g,j;\nu)/\nu^2$ in powers of $\frac1j$ satisfy the rule Eq. \eq{ruleEGW}.

Let us also note that there is a simple link between the total gravitational-wave energy loss along an hyperbolic motion, and the 
gravitational-wave energy loss during one radial period of an elliptic motion, namely
\beq \label{ellvshyp}
\Delta E_{\rm GW}^{\rm elliptic}(\g,j)= \Delta E_{\rm GW}^{\rm hyperbolic}(\g,j)- \Delta E_{\rm GW}^{\rm  hyperbolic}(\g, - j).
\eeq
This result is obtained by analytically continuing the quasi-Keplerian representation of the hyperbolic motion used above \cite{Cho:2018upo}
back to the elliptic-motion case (expressing all quantities in terms of $\g$ and $j$ and analytically continuing $\g$ from 
$\g^{\rm  hyperbolic}>1$ to $\g^{\rm  elliptic}<1$). The result \eq{ellvshyp} is consistent with the analytic-continuation link
between the scattering angle and the periastron precession \cite{Kalin:2019inp}, as is easily seen in view of the link \cite{Bini:2017wfr}
used above between the tail contribution to the scattering angle and the time integral of the gravitational-wave energy loss.
The functional structure of $\Delta E_{\rm GW}^{\rm elliptic}(\g,j)$ is much simpler than that of $\Delta E_{\rm GW}^{\rm hyperbolic}(\g,j)$.
In particular the arccos, or arctan, factors present in  $\Delta E_{\rm GW}^{\rm hyperbolic}(\g,j)$ are simply replaced by $\pi$ in 
$\Delta E_{\rm GW}^{\rm elliptic}(\g,j)$. Finally, $\Delta E_{\rm GW}^{\rm elliptic}(\g,j)$ has a polynomial structure in $\pinf$
and $\bar \epsilon$.

Note also that our rule \eq{ruleEGW} about the special $\nu$-dependence of the hyperbolic gravitational-wave energy loss
implies, via the link \eq{ellvshyp}, that the same property should be satisfied by the elliptic gravitational-wave energy loss.
In view of the odd dependence of $\Delta E_{\rm GW}^{\rm elliptic}(\g,j)$ on $j$ (and therefore $\bar \epsilon$) displayed
in Eq. \eq{ellvshyp}, this transforms the result \eq{bepsxp} into
\bea \label{ellbepsxp}
&&\frac{ h(\g,\nu)\Delta E_{\rm GW}^{\rm elliptic}(\pinf, \bar \epsilon;\nu)}{M\nu^2}= \nonumber\\
&&\qquad\qquad \bar \epsilon^3 \pinf^7 \left[P_2(\bar \epsilon^2) + \pinf^2 P_3(\bar \epsilon^2)+ \pinf^4 P_4(\bar \epsilon^2)\right] \nonumber\\
&&\qquad\qquad +\nu \,\bar \epsilon^5  \pinf^9  \left[P_2(\bar \epsilon^2) + \pinf^2 P_3(\bar \epsilon^2)\right] \nonumber\\
&&\qquad\qquad +\nu^2 \,\bar \epsilon^7  \pinf^{11}  P_2(\bar \epsilon^2) + O(\pinf^{13})\,,\nonumber\\
\eea 
where each $P_n(\bar \epsilon^2)$ denotes a different polynomial of order $n$ in $\bar \epsilon^2$.

We have checked (using the elliptic 3PN results of Refs. \cite{Arun:2007rg,Arun:2007sg})
that  the remarkable constraint on the $\nu$-dependence of $\Delta E_{\rm GW}^{\rm elliptic}(\g,j ;\nu)$
displayed in Eqs. \eq{ruleEGW} and  \eq{ellbepsxp} is  satisfied at the 3PN level (with the evident generalization of
the structure \eq{ellbepsxp}). In particular, the $O(\nu^3)$ contribution is of the form 
$ \nu^3 \,\bar \epsilon^9  \pinf^{13}  P_2(\bar \epsilon^2) $.

\subsection{Contributions to  $\widetilde \chi_n^{\rm nonloc,h}$ violating the special $\nu$-structure}

Let us now highlight the relatively small number of contributions to $\widetilde \chi_n^{\rm nonloc,h}= h^{n-1}\chi_n^{\rm nonloc,h}$
that do not satisfy the rule \eq{rule} by separating them from those that satisfy the rule: 
\begin{eqnarray} \label{ruleviolations}
\widetilde \chi_4^{\rm nonloc,h}&=&[\widetilde \chi_4^{\rm nonloc,h}]^{\nu}+ \frac{63}{20}\nu^2 \pi  p_{\infty}^6 \eta^2 \nonumber\\
&+&
\left(\frac{199037}{40320} \nu^2 -\frac{27331}{10080}\nu^3 \right)\pi  p_{\infty}^8\eta^4
\,, \nonumber\\
\widetilde \chi_5^{\rm nonloc,h}&=& [\widetilde \chi_5^{\rm nonloc,h}]^{\nu + \nu^2}-\frac{2448608}{33075}\nu^3 p_{\infty}^7 \eta^4 
\,,\nonumber\\
\widetilde \chi_6^{\rm nonloc,h}&=&  [\widetilde \chi_6^{\rm nonloc,h}]^{\nu + \nu^2} + D \nu^3 \pi  p_{\infty}^6\eta^4
\,.
\end{eqnarray}
In other words, there are {\it only five terms} violating the rule \eq{rule} in $\widetilde \chi_n^{\rm nonloc,h}$ ($n=4,5,6$):
(i) one term of fractional order $\eta^2$, \ie, at 5PN [$O(\nu^2)$ term in $\widetilde \chi_4^{\rm nonloc,h}$]; and (ii) four terms of
fractional order $\eta^4$, \ie, at 6PN [$O(\nu^2)$ term and $O(\nu^3)$ term in $\widetilde \chi_4^{\rm nonloc,h}$,
and $O(\nu^3)$ terms in $\widetilde \chi_5^{\rm nonloc,h}$ and $\widetilde \chi_6^{\rm nonloc,h}$]. The coefficients 
of all those terms have been analytically derived, apart from the last one which has been only partially analytically derived.
However, we have  evaluated it numerically:
\beq \label{Dnum}
D \equiv\frac{1}{2}d_{21} +d_{42}-\frac{1}{8} d_{00}\approx -116.73148147\,.
\eeq
Note that the contributions $d_{21}$ and $d_{00}$ to the coefficient $D$ are known analytically and contain $\zeta(3)$
(see Eqs. \eq{d00}, \eq{d21}). The only integral we could not analytically compute is $d_{42} \approx  -439.10050487 $ (see Table \ref{num_resd}). 
For completeness, we give the explicit integral form of $c_{42}$ (equivalent to $d_{42}$, see Eq. \eq{dmn})
 in the Supplemental Material of this paper.

\section{Determination of the flexibility factor $f(t)$} \label{determiningf}

Let us recall again the logic behind the introduction of the flexibility factor $f(t)$. 
The total (local-plus-nonlocal) scattering angle $\chi^{\rm tot}(\pinf, j ; \nu)$ has been shown \cite{Damour:2019lcq} to
have a special $\nu$-dependence at each PM order, \ie, at each order in $\frac1j$. However, if we were to decompose $\chi^{\rm tot}(\pinf, j ; \nu)$
in its harmonic-coordinate nonlocal contribution  $\chi^{\rm nonloc, h}(\pinf, j ; \nu)$ and the complementary harmonic-coordinate local contribution $\chi^{\rm loc, h}(\pinf, j ; \nu)$, each contribution would not separately satisfy the special $\nu$-dependence of their sum
$\chi^{\rm tot}(\pinf, j ; \nu)= \chi^{\rm nonloc, h}(\pinf, j ; \nu) + \chi^{\rm loc, h}(\pinf, j ; \nu)$. This situation is  improved
by slightly modifying the (conventional) definition of the nonlocal Hamiltonian, and thereby the separation of $\chi^{\rm tot}(\pinf, j ; \nu)$
into a flexed nonlocal piece,  $\chi^{\rm nonloc, f}(\pinf, j ; \nu)$, and a complementary flexed local piece, $\chi^{\rm loc, f}(\pinf, j ; \nu)$,
such that each contribution to $\chi^{\rm tot}(\pinf, j ; \nu)= \chi^{\rm nonloc, f}(\pinf, j ; \nu) + \chi^{\rm loc, f}(\pinf, j ; \nu)$
{\it separately} satisfy the simple $\nu$-dependence satisfied by $\chi^{\rm tot}(\pinf, j ; \nu)$.
We have already determined in Ref. \cite{Bini:2020nsb} the structure of the flexed 6PN-accurate local Hamiltonian by using the condition
that its corresponding local scattering angle $\chi^{\rm loc, f}(\pinf, j ; \nu)$ satisfies the special $\nu$-dependence of Ref. \cite{Damour:2019lcq}. In the present section, we shall use the results  derived in the previous Section for the harmonic-coordinate nonlocal contribution  
$\chi^{\rm nonloc, h}(\pinf, j ; \nu)$ as a tool for determining the value of the flexibility factor $f$. Specifically,  by writing that the sum
\beq
\widetilde \chi^{\rm nonloc, f}(\pinf, j ; \nu) = \widetilde \chi^{\rm nonloc, h}(\pinf, j ; \nu) + \widetilde \chi^{\rm f-h}(\pinf, j ; \nu)\,,
\eeq
satisfies the special $\nu$-dependence satisfied by $\chi^{\rm tot}(\pinf, j ; \nu)$ we are going to get some constraints on
the value of $f$. 

The determination of $f(t)$ is done by going through three successive steps: (i) explicit computation of the few coefficients measuring to what extent
the harmonic-coordinate angle $\chi^{\rm nonloc, h}(\pinf, j ; \nu)$ fails to satisfy the special $\nu$-dependence; (ii) computation of
the $f-h$ additional contribution, $\chi^{\rm f-h}(\pinf, j ; \nu)$, to the nonlocal scattering angle; and (iii) determination of $f(t)$ by the
condition that $\chi^{\rm f-h}(\pinf, j ; \nu)$ compensates the rule-violating contributions, Eq. \eq{ruleviolations},
present in $\widetilde \chi^{\rm nonloc, h}(\pinf, j ; \nu) $. The step (i) was already accomplished in the previous section. We
now go through steps (ii) and (iii).

\subsection{Determination of $W^{f-h}$ and  $\widetilde \chi^{\rm f-h}(\pinf, j ; \nu)$} 

The $f$-induced additional contribution  $W^{\rm f-h}$ to $W^{\rm nonloc}= \int dt H^{\rm nonloc}$ is defined as
\begin{eqnarray}
\label{Wf1}
W^{\rm f-h}&=& + 2\frac{G H_{\rm real}}{c^{5}}\int dt {\mathcal F}^{\rm split}_{\rm 2PN}(t,t)  \ln(f(t))\,.
\end{eqnarray}
A simplification is that, as it is enough to look for a flexibility factor of the type
\beq
f(t)=1+ \eta^2 f_1(t)+ \eta^4 f_2(t) + O(\eta^6)\,,
\eeq
we have
\beq
\ln(f(t))= \eta^2 f_1+ \eta^4 \left(f_2 - \frac12 f_1^2\right) +  O(\eta^6)\,,
\eeq
so that it is enough to work at the 1PN fractional accuracy. 
[Indeed,  the factor $ 2\frac{G H_{\rm real}}{c^{5}}{\mathcal F}^{\rm split}_{\rm 2PN}(t,t) $ in Eq. \eq{Wf1} starts
at the 4PN order, while $\ln(f(t))=O(\frac1{c^2})$, so that $W^{\rm f-h}$ starts at the 5PN order, as appropriate
to cancell the 5PN+6PN rule-violating terms delineated in Eqs. \eq{ruleviolations}.]
Namely, we can use in Eq. \eq{Wf1} the 1PN-accurate gravitational-wave flux 
${\mathcal F}^{\rm GW}_{\rm 1PN}(t)={\mathcal F}^{\rm split}_{\rm 1PN}(t,t)$, and we can compute the integral by using the
1PN-accurate quasi-Keplerian dynamics.

There are several possible ways to parametrize a general 1PN expression for the flexibility function $f(t)$.
One could use a direct parametrization in terms of  harmonic-coordinate positions and velocities. 
Here, we shall follow our previous (Newtonian-accurate) determination of the 1PN term $\eta^2 f_1$ \cite{Bini:2020wpo}
by parametrizing $f(t)$ in terms of (1PN-accurate) harmonic-coordinate relative positions ${\bf x}$ and momenta ${\bf p}$.
[As in \cite{Bini:2020wpo}, we work with rescaled, dimensionless positions and momenta.] We write
\bea
\label{f_funct_expr}
f_1&=& \nu\left( c_1 p_r^2 +c_2 p^2 +c_3\frac{1}{r}\right)\,,\nonumber\\
f_2 &=& \nu \left(d_1 p_r^4 +d_2 p^4 +d_3\frac{1}{r^2}+d_4 p^2p_r^2\right.\nonumber\\
&&\left. +d_5 \frac{p_r^2}{r}+d_6\frac{p^2}{r}\right)\,.
\eea
The coefficients $c_i$ used here differ from the corresponding quantities in Ref. \cite{Bini:2020wpo} by an overall factor $\nu$:
$ \nu c_i^{\rm here}= c_i^{\rm there}$. As a consequence,  the $c_i^{\rm here}$'s can be chosen to be pure
numbers (independent of the value of $\nu$). On the other hand, in spite of a similar $\nu$ overall rescaling,
the coefficients $d_i$ will be found to be linear functions of $\nu$:
\beq \label{dinu}
d_i= d_i^0 + \nu  d_i^1\,.
\eeq
Eq. \eqref{Wf1} becomes
\begin{eqnarray}
\label{Wf2}
W^{\rm f-h}
&=&2  G H_{\rm real}\int dt {\mathcal F}^{\rm GW}_{\rm 1PN}(t) \left[ f_1+ \eta^2 \left(f_2 - \frac12 f_1^2\right)\right]\,.\nonumber\\
\end{eqnarray}
One should insert in Eq. \eqref{Wf2} the expression of ${\mathcal F}^{\rm GW}_{\rm 1PN}(t)$ \cite{Blanchet:1989cu} in terms of 
the 1PN-accurate momenta. It is derived in Appendix \ref{PN} using classic results on the 1PN Lagrangian for the (harmonic-coordinate)
relative motion (see, {\it e.g.}, Ref. \cite{dd}). See Eqs. \eq{L1PN},\eq{FGW1PN}.

It is straightforward to compute the integral \eq{Wf2}, and then to differentiate it with respect to $j$ to obtain
the corresponding contribution to the scattering angle:
\beq
\frac12 \chi^{f-h} = \frac1{2 M^2 \nu} \frac{\partial W^{\rm f-h}(\pinf,j;\nu)}{\partial j}\,.
\eeq
The result of this computation is a contribution that starts at the 5PN level, and that is fractionally 1PN-accurate, say
\beq
\frac12 \chi^{f-h}=\frac12 \chi^{f-h}_0+ \frac12\eta^2 \chi^{f-h}_2\,.
\eeq
The large-$j$ expansion of the 5PN-level contribution, $\frac12 \chi^{f-h}_0$, reads
\begin{eqnarray}
\frac12 \chi^{f-h}_0&=& -\frac{1}{10}\nu^2 \frac{p_\infty^6}{j^4}\pi (74 c_2+13 c_1) \nonumber\\
&-&\frac{128}{225}\nu^2 \frac{p_\infty^5}{j^5} (343 c_2+49 c_3+51 c_1) \nonumber\\
&-&\frac12 \nu^2 \frac{p_\infty^4}{j^6}\pi (63 c_1+488 c_2+122 c_3) + O\left(\frac1{j^7} \right)\,, \nonumber\\
\end{eqnarray}
while that of the 6PN-level one, $\frac12 \chi^{f-h}_2$, reads
\beq
\frac12 \chi^{f-h}_2= \chi_{4,2}^{f-h}\frac{p_\infty^8}{j^4} +\chi_{5,2}^{f-h}\frac{p_\infty^7}{j^5}+\chi_{6,2}^{f-h}\frac{p_\infty^6}{j^6}+ O\left(\frac1{j^7} \right)\,,
\eeq
with
\begin{eqnarray}
\chi_{4,2}^{f-h}&=& \frac{1}{2240}\nu^2 \pi (-10856 c_2+41440\nu c_2\nonumber\\
&&+2383 c_1+3572 c_1\nu-2912 d_4+2912\nu c_2 c_1\nonumber\\
&&-16576 d_2+574\nu c_1^2+8288\nu c_2^2-1148 d_1)
\,,\nonumber\\
\chi_{5,2}^{f-h}&=& +\frac{64}{11025}\nu^2  (107024\nu c_2-51669 c_2\nonumber\\
&&+7864 c_1\nu+439 c_1+10357\nu c_3-4019 c_3\nonumber\\
&&+6426\nu c_2 c_1-2226 d_1+714\nu c_1 c_3\nonumber\\
&&+21609\nu c_2^2-4802 d_6+4802\nu c_2 c_3\nonumber\\
&&-714 d_5+1113\nu c_1^2-6426 d_4-43218 d_2) 
\,,\nonumber\\
\chi_{6,2}^{f-h}&=& +\frac{1}{336}\nu^2  \pi (-12139 c_1-5460 d_1+17640\nu c_2 c_1\nonumber\\
&&+3528\nu c_1 c_3-136640 d_2+2730\nu c_1^2+21580 c_1\nu\nonumber\\
&&+58740\nu  c_3-246374 c_2+1708\nu c_3^2+68320\nu c_2^2\nonumber\\
&&+335600\nu c_2-30454 c_3-3416 d_3-17640 d_4\nonumber\\
&&-3528 d_5-27328 d_6+27328\nu c_2 c_3) 
\,.
\end{eqnarray}
The quantities of most interest are the corresponding  energy-rescaled coefficients of $\frac1{j^n}$ in the scattering angle
$\frac12 \chi^{f-h}=\sum_n \chi^{f-h}_n/j^n$, \ie,
\beq
\widetilde \chi^{f-h}_n=h^{n-1} \chi^{f-h}_n = \widetilde \chi^{f-h}_{n,0}+ \eta^2  \widetilde \chi^{f-h}_{n,2} +\cdots\,.
\eeq
They read (at the fractional 1PN accuracy, and setting $\eta=1$)
\begin{widetext}
\begin{eqnarray} \label{tildechifh456}
\pi^{-1} \widetilde\chi_4^{f-h}&=& \left(-\frac{13}{10} c_1-\frac{37}{5} c_2\right)\nu^2 p_\infty^6  \nonumber\\
&+&\left[\frac{2383}{2240}  c_1-\frac{1357}{280}  c_2-\frac{41}{80} d_1-\frac{37}{5}  d_2-\frac{13}{10}  d_4\right. \nonumber\\
&+& \left.\left(-\frac{199}{560} c_1+\frac{37}{5}  c_2+\frac{13}{10}  c_2c_1 +\frac{41}{160}  c_1^2+\frac{37}{10}  c_2^2\right) \nu \right]\nu^2 p_\infty^8 
\,,\nonumber\\
\widetilde\chi_5^{f-h}&=&\left(-\frac{2176}{75}  c_1-\frac{43904}{225} c_2-\frac{6272}{225}  c_3\right)\nu^2 p_\infty^5\nonumber\\  
&+&\left[\frac{28096}{11025}  c_1-\frac{367424}{1225}  c_2-\frac{257216}{11025}  c_3-\frac{6784}{525}  d_1-\frac{6272}{25} d_2 -\frac{6528}{175}  d_4-\frac{2176}{525}  d_5-\frac{6272}{225} d_6\right. \nonumber\\
&+&\left(-\frac{136448}{11025}  c_1+\frac{2546944}{11025}   c_2+\frac{16064}{3675}   c_3+\frac{6528}{175}  c_2 c_1+\frac{2176}{525}   c_1 c_3+\frac{3136}{25}   c_2^2\right.\nonumber\\
&&\left.\left.+\frac{6272}{225}   c_2 c_3+\frac{3392}{525}  c_1^2\right)\nu\right]\nu^2 p_\infty^7
\,,\nonumber\\
\pi^{-1} \widetilde \chi_6^{f-h}&=& \left(-\frac{63}{2} c_1-244 c_2-61 c_3\right)\nu^2 p_\infty^4   \nonumber\\
&+& \left[-\frac{12139}{336}  c_1-\frac{123187}{168}  c_2-\frac{15227}{168} c_3-\frac{65}{4}  d_1-\frac{1220}{3}  d_2-\frac{61}{6}  d_3-\frac{105}{2} d_4-\frac{21}{2}  d_5-\frac{244}{3}  d_6
\right. \nonumber\\
&+& \left.\left(-\frac{305}{21}   c_1+\frac{8165}{21}   c_2+\frac{625}{28} c_3+\frac{21}{2}   c_1 c_3+\frac{244}{3}   c_2 c_3+\frac{105}{2}   c_2 c_1+\frac{65}{8}  c_1^2+\frac{61}{12}   c_3^2+\frac{610}{3}   c_2^2\right)\nu 
\right]\nu^2 p_\infty^6 
\,.\nonumber\\
\end{eqnarray}


\subsection{Reparametrization of the flexibility factor $f(t)$, and constraints on its parameters}  

Combining the results \eq{ruleviolations} and \eq{tildechifh456}, we can now write the condition that the sums
$\widetilde \chi_n^{\rm nonloc,f}= \widetilde \chi_n^{\rm nonloc,h} + \widetilde \chi_n^{\rm f-h}$ satisfy
the  $\nu$-dependence of $\widetilde \chi_n^{\rm tot}$, \ie, 
\beq
[ \widetilde\chi_4^{\rm nonloc,f}]\sim \nu\,,\qquad [\widetilde \chi_5^{\rm nonloc,f}]\sim \nu+\nu^2\,,\qquad
[\widetilde \chi_6^{\rm nonloc,f}]\sim \nu+\nu^2\,.
\eeq
Remembering the $\nu$-independence of the $c_i$'s,
and the $\nu$-linearity of the $d_i$'s, Eq. \eq{dinu}, these conditions yield five equations.
One equation (already discussed  in Ref. \cite{Bini:2020wpo}) comes from the 5PN level and reads
\beq
\label{constr_5pn}
\frac{13}{10} c_1+ \frac{37}{5} c_2=\frac{63}{20}\,.
\eeq
The 6PN level yields four additional constraints, namely
\bea \label{constr_6pn}
(a)\qquad 0&=&-\frac{1357}{56} c_2+\frac{2383}{448} c_1-37 d_2^0-\frac{13}{2} d_4^0-\frac{41}{16} d_1^0+\frac{199037}{8064}
\,,\nonumber\\
(b)\qquad 0&=&37 c_2-\frac{13}{2} d_4^1+\frac{13}{2} c_2 c_1-37 d_2^1+\frac{41}{32} c_1^2+\frac{37}{2} c_2^2-\frac{199}{112}c_1-\frac{41}{16} d_1^1-\frac{27331}{2016}
\,,\nonumber\\
(c)\qquad 0&=&\frac{2546944}{441} c_2-\frac{6784}{21} d_1^1+\frac{2176}{21} c_1 c_3-\frac{136448}{441} c_1-\frac{6272}{9} d_6^1+\frac{6272}{9} c_2 c_3+3136 c_2^2-\frac{2176}{21} d_5^1+\frac{3392}{21} c_1^2\nonumber\\
&&+\frac{16064}{147} c_3-\frac{6528}{7} d_4^1+\frac{6528}{7} c_2 c_1-6272 d_2^1-\frac{2448608}{1323}
\,,\nonumber\\
(d)\qquad 0&=&\frac{65}{8} c_1^2+\frac{610}{3} c_2^2+\frac{625}{28} c_3+\frac{105}{2} c_2c_1-\frac{61}{6} d_3^1+\frac{21}{2} c_1c_3-\frac{65}{4} d_1^1+\frac{244}{3} c_2 c_3-\frac{1220}{3} d_2^1-\frac{21}{2} d_5^1+\frac{8165}{21} c_2\nonumber\\
&&-\frac{244}{3} d_6^1+\frac{61}{12} c_3^2-\frac{105}{2} d_4^1-\frac{305}{21} c_1+D
\,,
\eea
where the constant $D=\frac{1}{2}d_{21} +d_{42}-\frac{1}{8} d_{00}\approx -116.73148147$,
was already discussed above, see Eq. \eq{Dnum}.

There are many ways to satisfy these constraints. Indeed, at the 5PN level, we have one constraint, Eq. \eq{constr_5pn},
for three coefficients, $c_1, c_2, c_3$, while at the 6PN level we have four constraints, Eqs. \eq{constr_6pn}, for the twelve
coefficients  $d_1^0, d_1^1, d_2^0, d_2^1,d_3^0, d_3^1, d_4^0, d_4^1, d_5^0, d_5^1, d_6^0, d_6^1$.
We can, however, streamline the discussion of these constraints by defining a convenient reparametrization of the gauge-invariant
content of the Hamiltonian contribution associated with the flexibility factor $f(t)$, namely
\bea \label{Hfmenh6PN}
\Delta^{\rm f-h} H_{\rm 6PN}&=&2H{\mathcal F}^{\rm GW}_{\rm 1PN}\ln(f) \nonumber\\
&=&2H{\mathcal F}^{\rm GW}_{\rm 1PN}(p,p_r,r) \left[ f_1+ \eta^2 \left(f_2 - \frac12 f_1^2\right)\right]\,.
\eea
The latter flexibility-related Hamiltonian contains the three 5PN parameters $c_i$, and the four 6PN
parameters $d_i$ entering the flexibility factor $f(t)$, Eqs. \eq{f_funct_expr} . [Here, we count
for simplicity each $d_i$, $i=1,\ldots,4$, as one parameter, though one must remember that each $d_i(\nu)=d_i^0 + \nu d_i^1$
actually contains two numerical parameters.]
Let us, however, show that the flexibility described by $f(t)$ can be parametrized by three other 5PN parameters, $C_1, C_2, C_3$,
and {\it only four} 6PN parameters $D_1, D_2,D_3, D_4$. [Each new 6PN parameter $D_i$ will be again a linear function of $\nu$,
$D_i(\nu)=D_i^0 + \nu D_i^1$, and actually contain two numerical parameters.]

Indeed, it is shown in Appendix \ref{f} that  the 6PN flexibility contribution to the Hamiltonian, Eq. \eq{Hfmenh6PN},
 is canonically equivalent to the following ($p_r$-gauge-type) Hamiltonian 
 \beq \label{Hfmenh6PNnew}
\Delta^{\rm f-h} H'_{\rm 6PN}=\frac{M \nu^3}{r^4} \left[C_1 p_r^4+C_2 \frac{p_r^2}{r}+C_3\frac{1}{r^2}
+\eta^2\left(D_1 p_r^6+D_2 \frac{p_r^4}{r}+D_3 \frac{p_r^2}{r^2}+D_4\frac{1}{r^3}\right)\right]\,.
\eeq
The seven new parameters  $C_1, C_2, C_3$ and $D_1, D_2, D_3, D_4$ entering Eq. \eq{Hfmenh6PNnew} are defined
by the following explicit functions
of the original nine parameters $c_i, d_i$:
\begin{eqnarray} \label{Civsci}
C_1&=& \frac{16}{15}(13 c_1+74 c_2)\,,\nonumber\\
C_2&=& \frac{16}{15}(49 c_3+121 c_2+12 c_1)\,, \nonumber\\
C_3&=& \frac{64}{5}(c_2+c_3)\,,
\end{eqnarray}
and
\begin{eqnarray}
\label{Divsdi}
D_1&=&
\left(-\frac{328 c_1^2}{75}-\frac{1664 c_1
   c_2}{75}+\frac{38128 c_1}{525}-\frac{4736
   c_2^2}{75}+\frac{18944 c_2}{75}\right)\nu  
	-\frac{15356
   c_1}{525}+\frac{3424 c_2}{175}+\frac{656
   d_1}{75}+\frac{9472 d_2}{75}+\frac{1664 d_4}{75}
\,,\nonumber\\
D_2&=&
\left(-\frac{32 c_1^2}{5}-\frac{784 c_1
   c_2}{15}-\frac{272 c_1 c_3}{15}+\frac{1576
   c_1}{45}-\frac{3496 c_2^2}{15}-\frac{5488 c_2
   c_3}{45}-\frac{5584 c_2}{105}+\frac{70808
   c_3}{315}\right)\nu  \nonumber\\
	&&
	-\frac{11212 c_1}{63}-\frac{28496
   c_2}{45}+\frac{12944 c_3}{315}+\frac{64
   d_1}{5}+\frac{6992 d_2}{15}+\frac{784
   d_4}{15}+\frac{272 d_5}{15}+\frac{5488 d_6}{45}
\,,\nonumber\\
D_3&=&
\left(-\frac{64 c_1 c_2}{5}-\frac{64 c_1
   c_3}{5}-\frac{112 c_1}{5}-\frac{1928
   c_2^2}{15}-\frac{464 c_2 c_3}{3}-\frac{8440
   c_2}{21}-\frac{488 c_3^2}{15}-\frac{3048
   c_3}{35}\right)\nu  \nonumber\\
	&&
	-\frac{11708 c_1}{105}-\frac{18884
   c_2}{21}-\frac{1724 c_3}{3}+\frac{3856
   d_2}{15}+\frac{976 d_3}{15}+\frac{64 d_4}{5}+\frac{64
   d_5}{5}+\frac{464 d_6}{3}
\,,\nonumber\\
D_4&=&
\left(-\frac{32 c_2^2}{5}-\frac{64 c_2
   c_3}{5}-\frac{112 c_2}{5}-\frac{32
   c_3^2}{5}-\frac{112 c_3}{5}\right)\nu  
	-\frac{6332
   c_2}{105}-\frac{11708 c_3}{105}+\frac{64
   d_2}{5}+\frac{64 d_3}{5}+\frac{64 d_6}{5}
\,.  
\end{eqnarray}
\end{widetext}
The three $C_i$'s are in one-to-one correspondence with the three $c_i$'s, with the inverse relations $c_i=f_i(C_j)$
given in Eqs. \eq{civsCi}. On the other hand, the four $D_i$'s capture the full gauge-invariant content of the six $d_i$'s. 
[Two of the $d_i$'s being pure gauge parameters; see Eqs. \eq{divsDi}.]

The five constraints discussed in the previous subsection can be entirely re-expressed in terms of the
parameters  $C_i$ ($i=1\ldots3$), and  $D_i=D_i^0+\nu D_i^1$  ($i=1\ldots4$). Indeed, the scattering angle
only depends on the the time-integral (along an hyperbolic motion) of $\Delta^{\rm f-h} H_{\rm 6PN}$,
which is equal to the time-integral of $\Delta^{\rm f-h} H'_{\rm 6PN}$. This ensures that the scattering angle
 only depends on the $C_i$'s and $D_i$'s. Alternatively, 
using Eqs. \eq{civsCi}, \eq{divsDi}, we could reexpress the energy-rescaled scattering-angle coefficients 
\eq{tildechifh456} in terms the $C_i$'s and $D_i$'s. The results read
\begin{widetext}
\begin{eqnarray}
\pi^{-1} \widetilde\chi_4^{f-h}&=&
-\frac{3}{32}C_1\nu^2 p_\infty^6
-\frac{3}{32}\left[\left(\frac12-3\nu\right)C_1+\frac58D_1\right]\nu^2 p_\infty^8
\,,\nonumber\\
\widetilde\chi_5^{f-h}&=&
-\frac{8}{5}\left(C_1+\frac13C_2\right)\nu^2 p_\infty^5
-\frac{8}{5}\left[
\left(\frac{43}{14}-\frac{41}{14}\nu\right)C_1+\left(\frac16-\frac23\nu\right)C_2+\frac57D_1+\frac17D_2
\right]\nu^2 p_\infty^7
\,,\nonumber\\
\pi^{-1} \widetilde \chi_6^{f-h}&=&
-\frac{15}{16}\left(\frac32C_1+C_2+C_3\right)\nu^2 p_\infty^4\nonumber\\
&&
-\frac{15}{16}\left[
\left(\frac{41}{4}-\frac92\nu\right)C_1+\left(\frac{19}{6}-\frac{25}{12}\nu\right)C_2+\left(\frac12-\nu\right)C_3
+\frac54D_1+\frac12D_2+\frac16D_3
\right]\nu^2 p_\infty^6
\,.
\end{eqnarray}
\end{widetext}
Comparing these (simplified) expressions with the five contributions to $\widetilde \chi_n^{\rm nonloc,h}$
that do not satisfy the rule \eq{rule} (which were written down in Eqs. \eq{ruleviolations}), we now get the following
simplified versions of the five constraints \eq{constr_5pn}, \eq{constr_6pn}.

At 5PN we have only one constraint, Eq. \eqref{constr_5pn}, which now reads
\beq \label{C1constr}
C_1 =\frac{168}{5}\,.
\eeq
At 6PN, the four constraints, Eq. \eqref{constr_6pn}, now imply
\begin{eqnarray} \label{Dconstr}
D_1^0&=&\frac{398074}{4725}-\frac{4}{5}C_1\nonumber\\
&=&\frac{271066}{4725}
\,,\nonumber\\
D_1^1&=&-\frac{218648}{4725}+\frac{24}{5}C_1\nonumber\\
&=&\frac{21736}{189}
\,,\nonumber\\
D_2^1&=&-\frac{87428}{945}-\frac{7}{2}C_1+\frac{14}{3}C_2\nonumber\\
&=&-\frac{39712}{189}+\frac{14}{3}C_2
\,,\nonumber\\
D_3^1&=&\frac{65584}{105}+\frac{3}{2} C_1-\frac{3}{2} C_2+6 C_3+\frac{32}{5}D\nonumber\\
&=&\frac{70876}{105}-\frac{3}{2} C_2+6 C_3+\frac{32}{5} D
\,.
\end{eqnarray}
At the 5PN level, we have three flexibility parameters, $C_1, C_2, C_3$, and only one of them is determined, namely $C_1$, Eq. \eq{C1constr}.
It was pointed out in Ref. \cite{Bini:2020wpo} that the presence of two unconstrained 5PN
flexibility parameters (namely $C_2$ and $C_3$) is in one-to-one correspondence with the existence
of two 5PN-level undetermined coefficients in the local Hamiltonian (namely ${\bar d}_5^{\nu^2}$ and ${a}_6^{\nu^2}$).
More precisely, changing the values of $C_2$ and $C_3$ was shown to be equivalent to shifting the values of 
${\bar d}_5^{\nu^2}$ and ${a}_6^{\nu^2}$ (see Eqs. (8.21)--(8.22) of Ref. \cite{Bini:2020wpo}).
Alternatively, one could uniquely fix $C_2$ and $C_3$, i.e., uniquely fix the flexibility factor $f$,
so as to reduce $\Delta^{\rm f-h} H$ to be {\it minimal},  in a $p_r$-type-gauge, i.e., to contain the minimum
number of terms needed to satisfy the scattering constraints. This was formulated there in terms of the EOB
parametrization of the Hamiltonian. The result was that by choosing  (see Eqs. (8.24) of Ref. \cite{Bini:2020wpo}, 
here rescaled by $\nu$ as we recall)
\bea \label{c123min}
c_1^{\rm min} &=& \frac{189}{4} \,, \nonumber\\
c_2^{\rm min}&=&  -\frac{63}{8} \,,\nonumber\\
c_3^{\rm min}&=&\frac{63}{8} \,,
\eea
the $f-h$-piece of  the EOB effective Hamiltonian was reduced to be fully contained in the
following specific (minimal) $Q$ term
\beq
\Delta^{\rm f}Q^{\rm min}=\frac{336}{5} \nu^2 \frac{p_r^4}{r^4}\,.
\eeq
Let us now show how these results can be generalized  to the 6PN level\footnote{It can be shown that a similar result 
holds at higher PN orders.}. Let us first note that, when transcribing the 5PN-level minimal constraints \eq{c123min} 
in terms of the new parameters $C_i$, they are
 easily seen to simply correspond to completing the constraint \eq{C1constr} by the additional simple constraints
\bea \label{C230}
C_2^{\rm min}&=& 0 ,\nonumber\\
C_3^{\rm min}&=&0 \,.
\eea
If we then insert the latter results in the four 6PN-level constraints \eq{Dconstr}, we find
that, among the eight 6PN coefficients $D_i^0, D_i^1$, $i=1,\ldots,4$, four of them,
namely $D_1^0$, $D_1^1$, $D_2^1$ and $D_3^1$ are  completely fixed by combining the 5PN minimal choice \eqref{C230}
with the general 6PN constraints. 
This lead us to define the following {\it minimal} solution of the 5+6PN constraints
\bea
\label{CDmin}
C_1^{\rm min}&=&\frac{168}{5}\,,\nonumber\\
 C_2^{\rm min}&=&0\,,\nonumber\\
  C_3^{\rm min}&=&0
\,,\nonumber\\
D_1^{\rm min}&=&\frac{271066}{4725}+\frac{21736}{189}\nu
\,,\nonumber\\
D_2^{\rm min}&=&-\frac{39712}{189}\nu
\,,\nonumber\\
D_3^{\rm min}&=&\left(\frac{70876}{105}+\frac{32}{5} D\right)\nu
\,,\nonumber\\
D_4^{\rm min}&=&0\,.
\eea
Starting from this minimal solution of the flexibility constraints, we
can decompose $\Delta^{\rm f-h} H'_{\rm 6PN}$ into two parts, say
\beq \label{decompHf-h}
\Delta^{\rm f-h} H'_{\rm 6PN}=\Delta^{\rm f-h}{ H'}^{\rm min}_{\rm 6PN}+ \Delta^{\rm f-h} {H'}^{C D}_{\rm 6PN}\,.
\eeq
Here,  $\Delta^{\rm f-h}{ H'}^{\rm min}_{\rm 6PN}$ denotes the part that is  built with the minimal solution \eq{CDmin}, namely
   \bea \label{Hfmenh6PNnewmin}
\frac{\Delta^{\rm f-h}{ H'}^{\rm min}_{\rm 6PN}}{M} &=&\nu^3\frac{168}{5} \frac{p_r^4}{r^4}
+\nu^3\left(\frac{271066}{4725} + \frac{21736}{189}\nu \right) \frac{p_r^6}{r^4} \nonumber\\
 &-& \nu^4 \frac{39712}{189} \frac{p_r^4}{r^5}+
 \nu^4 \left(\frac{70876}{105}+\frac{32}{5}D \right) \frac{p_r^2}{r^6}\,.\nonumber\\
\eea
On the other hand, $\Delta^{\rm f-h} {H'}^{C D}_{\rm 6PN}$ denotes the part that involves the six flexibility parameters
that are left unconstrained by the general constraints \eq{C1constr}, \eq{Dconstr}, namely:  
 $C_2$, $C_3$, $D_2^0$, $D_3^0$, and $D_4= D_4^0+ \nu D_4^1$. Explicitly, we have
\bea \label{Hfmenh6PNunconstr}
&&\frac{\Delta^{\rm f-h} {H'}^{C D}_{\rm 6PN}}{M}= C_2 \frac{\nu^3 p_r^2}{r^5} +  C_3\frac{\nu^3}{r^6}\nonumber\\
&&
+\left(D_2^0+ \frac{14}{3} \nu C_2\right)\frac{\nu^3 p_r^4}{r^5}
\nonumber\\
&&
+\left[ D_3^0+ \nu\left(-\frac{3}{2} C_2+6C_3\right)\right]\frac{\nu^3 p_r^2}{r^6}+(D_4^0+\nu D_4^1)\frac{\nu^3}{r^7}
\,.\nonumber\\
\eea
By using a suitable canonical transformation to transform into standard EOB gauge
the harmonic-type gauge to which $\Delta^{\rm f-h} {H'}^{C D}_{\rm 6PN}$ belongs\footnote{Indeed, 
$\Delta^{\rm f-h} {H'}^{C D}_{\rm 6PN}$ is a contribution to the total nonlocal Hamiltonian $H^{\rm nonloc, f}$ which is
expressed in terms of harmonic coordinates.}, we can then transcribe
the unconstrained $f$-dependent Hamiltonian contribution $\Delta^{\rm f-h} {H'}^{C D}_{\rm 6PN}$
in EOB format, i.e., in terms of the potentials $A$, $\bar D$ and $Q$ parametrizing a general effective Hamiltonian
in $p_r$-gauge, as in Eqs. (4.1), (4.2) of Ref. \cite{Bini:2020nsb}). One then finds that adding the Hamiltonian
contribution $\Delta^{\rm f-h} {H'}^{C D}_{\rm 6PN}$, Eq. \eq{Hfmenh6PNunconstr},  is equivalent to adding to
the EOB potentials entering the f-route local Hamiltonian $H^{\rm nonloc, f}$ the
following supplementary (5PN and 6PN) contributions
\bea \label{ashifted}
A^{C D}&=&a_6^{C D} u^6 + a_7^{C D} u^7\,, \nonumber\\
{\bar D}^{C D}&=& {\bar d}_5^{C D} u^5 +  {\bar d}_6^{C D} u^6\,, \nonumber\\
{\widehat Q}^{CD} &=& q_{45}^{CD} p_r^4 u^5 \,,
\eea
with 5PN-level terms,
\bea \label{5pnshifts}
a_6^{C D} &=& 2 \nu^2 C_3\,, \nonumber\\
{\bar d}_{5}^{CD} &=& 2 \nu^2 C_2\,,
\eea
and 6PN-level ones:
\bea \label{6pnshifts}
a_7^{C D} &=& 2\nu^2 (D_4^0+\nu D_4^1) +\nu^2 (9-\nu) C_3\,, \nonumber\\
{\bar d}_{6}^{CD} &=& \nu^2 (2 D_3^0 + 17 C_2-8C_3) -\nu^3 (2 C_2 + 30 C_3)\,,\nonumber\\
 q_{45}^{CD} &=&  \nu^2 \left(2 D_2^0 +\frac{7}{3} C_2\right) - \frac{28}{3} \nu^3 C_2 \,.
\eea
By comparing the expressions \eq{5pnshifts}, \eq{6pnshifts} to the explicit form of the EOB potentials of the 6PN f-route
local Hamiltonian $H^{\rm loc, f}$, as displayed in Table X of \cite{Bini:2020nsb}, it is easily checked that the addition
of the contributions \eq{5pnshifts}, \eq{6pnshifts} (including their explicit $O(\nu^3)$ terms)
to $H^{\rm loc, f}$ is equivalent to replacing
the undetermined EOB coefficients $a_6^{\nu^2 \rm loc, f},  {\bar d}_{5}^{\nu^2 \rm loc,f}, \ldots$ appearing in 
$H^{\rm loc, f}(a_6^{\nu^2 \rm loc, f},  {\bar d}_{5}^{\nu^2 \rm loc,f}, \ldots)$ by the following shifted values
\bea \label{shifting}
a_6^{\nu^2 \rm shifted}&=& a_6^{\nu^2 \rm loc, f}+ 2C_3
\,,\nonumber\\
{\bar d}_{5}^{\nu^2 \rm shifted} &=& {\bar d}_{5}^{\nu^2 \rm loc,f} +2C_2
\,, \nonumber\\
a_7^{\nu^2 \rm shifted}&=& a_7^{\nu^2 \rm loc, f}+2 D_4^0 + 9 C_3
\,,\nonumber\\
a_7^{\nu^3 \rm shifted}&=& a_7^{\nu^3 \rm loc, f}+2 D_4^1 - C_3
\,,\nonumber\\
{\bar d}_{6}^{\nu^2 \rm shifted} &=& {\bar d}_{6}^{\nu^2 \rm loc,f}+2 D_3^0+17C_2-8C_3
\,,\nonumber\\
q_{45}^{\nu^2 \rm shifted} &=& q_{45}^{\nu^2 \rm loc,f} +2 D_2^0+ \frac73 C_2 
\,.
\eea
The first two (5PN-level) equations are equivalent to Eqs. (8.20)--(8.21) of Ref. \cite{Bini:2020wpo} (taking into account the fact that we
separated here the term $\nu^3\frac{168}{5} \frac{p_r^4}{r^4}$). 

In Eqs. \eq{shifting} the undetermined parameters $ a_6^{\nu^2 \rm loc, f}, \ldots$, 
 appearing on the right-hand sides of the definitions
 of the various shifted parameters depend on the choice of $f$
 (i.e., on the choice of the unconstrained $C_i$'s and $D_i$'s),  while the shifted parameters $a_6^{\nu^2 \rm shifted}, \ldots$, 
  on the left-hand sides do not depend on the choice of $f$ (because they parametrize the Hamiltonian 
 $H^{\rm tot} - H^{\rm loc,h} - \Delta^{\rm f-h}{ H'}^{\rm min}_{\rm 6PN}$).
Therefore, the choice of the values of the unconstrained flexibility parameters $C_2, C_3, D_2^0, \ldots$
is a kind of gauge-freedom that has no effect on the physical consequences of the total Hamiltonian (which 
only depends on the gauge-invariant shifted parameters defined in Eqs. \eq{shifting}). In other words, imposing 
the  simple  additional constraints
 \bea \label{CD0}
C_2&=&0\,,\nonumber\\
C_3&=&0\,,\nonumber\\
D_2^0&=& 0\,,\nonumber\\
D_3^0&=&0\,,\nonumber\\
D_4^0&=&0\,,\nonumber\\
D_4^1&=&0\,,
\eea
 which leads to the {\it minimal} values \eq{CDmin} of the flexibility parameters, is a ``gauge choice"  such that the 
 corresponding {\it minimal} values of the undetermined EOB parameters, say $a_6^{\nu^2 \rm min}, \ldots$, 
simply coincide with the general gauge-invariant shifted values defined in Eqs. \eq{shifting}:
 \bea \label{shifting3}
a_6^{\nu^2 \rm min}&=& a_6^{\nu^2 \rm shifted}
\,,\nonumber\\
{\bar d}_{5}^{\nu^2 \rm min} &=& {\bar d}_{5}^{\nu^2 \rm shifted}
\,, \nonumber\\
a_7^{\nu^2 \rm min}&=& a_7^{\nu^2 \rm shifted}
\,,\nonumber\\
a_7^{\nu^3 \rm min}&=& a_7^{\nu^3 \rm shifted}
\,,\nonumber\\
{\bar d}_{6}^{\nu^2 \rm min} &=& {\bar d}_{6}^{\nu^2 \rm shifted}
\,,\nonumber\\
q_{45}^{\nu^2 \rm min} &=& q_{45}^{\nu^2 \rm shifted}
\,.
\eea

In the following, we shall often use by default the minimal fixing of the flexibility factor, and of the
associated Hamiltonians, defined by using Eqs. \eq{CDmin} (i.e.,  satisfying Eqs. \eq{CD0}). This leads, in particular, to
the specific value of $\Delta^{\rm f-h}{ H'}_{\rm 6PN}$ given by Eq. \eq{Hfmenh6PNnewmin}.
The corresponding specific values of the original flexibility parameters $c_i$, $d_i$ defining the flexibility
factor $f(t)$ are discussed in Appendix \ref{f}.

\section{Nonlocal Delaunay Hamiltonian, ${\bar H}^{\rm nonloc, f}_{6 \rm PN}(I_R, I_\phi)$,
radial action, $I_{ R \,6 \rm PN}^{\rm nonloc, f}(E,J)$, and periastron precession}

As said in the Introduction,  besides the scattering angle, a second gauge-invariant characterization of the 
f-route nonlocal dynamics can be given. It consists in presenting the explicit form of the f-route nonlocal
contribution to the averaged (Delaunay) Hamiltonian, ${\bar H}^{\rm nonloc, f}(I_R, I_\phi)$, or equivalently
the corresponding contribution, $I_{ R \,6 \rm PN}^{\rm nonloc, f}(E,J)$, to the radial action. 
The (gauge-invariant) information contained in  ${\bar H}^{\rm nonloc, f}(I_R, I_\phi)$ or $I_{ R \,6 \rm PN}^{\rm nonloc, f}(E,J)$
is also nearly fully encoded in the corresponding contribution to the periastron advance. Indeed, we have the general identity
\beq
d{\bar H}(I_R, I_\phi)= \Omega_R d I_R +  \Omega_\phi d I_\phi= \Omega_R d I_R +  K \Omega_R d I_\phi\,,
\eeq
where
\beq
\Omega_R=\frac{\partial {\bar H}(I_R, I_\phi)}{\partial I_R}=  \left[\frac{\partial I_R(E,J)}{\partial E} \right]^{-1}\,,
\eeq
denotes the radial frequency $2\pi/T_R$, while
\bea
K&\equiv&\frac{\Phi(E,J)}{2\pi}
=\frac{\Omega_\phi }{\Omega_R}\nonumber\\
&=&- \frac{\partial I_R(E,J)}{\partial J}= +\frac1{\Omega_R} \frac{\partial {\bar H}(I_R, I_\phi)}{\partial I_\phi}\,,
\eea
denotes the periastron advance $K=1+k$ (where the value 1 would correspond to the absence of periastron advance).

We have given
in Table XI of  Ref. \cite{Bini:2020wpo} the explicit, 5PN-accurate, expression of the f-route {\it local} Delaunay Hamiltonian,
${\bar H}^{\rm loc, f}(I_R, I_\phi)$. We gave also the explicit value of the function 
$I_R^{\rm loc, f}(E,J)$ at the 5PN accuracy in Ref. \cite{Bini:2020wpo}. Concerning the 6PN-accurate f-route {\it local} dynamics,
we gave in Ref. \cite{Bini:2020nsb} the explicit expression of the radial action as a function
of the EOB effective energy $I_R^{\rm loc, f}(E_{\rm eff},J)$. We proved there that it had a remarkably
simple structure. Namely, it reads
\begin{eqnarray} \label{Irxp0}
\frac{I_r^{\rm loc,f}(\g,j)}{GM\mu} &=& -j+I^S_0(\g) +\frac{I_1^S(\g)}{hj}+\frac{I_3(\g; \nu)}{(hj)^3}\nonumber\\
&&+ \frac{I_5(\g; \nu)}{(hj)^5}+\frac{I_7(\g; \nu)}{(hj)^7}\nonumber\\
&&+\frac{I_9(\g; \nu)}{(hj)^9}+\frac{I_{11}(\g; \nu)}{(hj)^{11}}\,.
\end{eqnarray}
where $h=h(\g,\nu)=E^{\rm tot}/M$ as above; where the first two coefficients, $I^S_0(\gamma)$, $I^S_1(\gamma)$,
 only depend on $\g$ and have the following very simple exact expressions
\bea \label{IS0}
I^S_0(\gamma) &=& \frac{ 2\gamma^2-1 }{\sqrt{1-\gamma^2}}
\,,\nonumber\\
I_1^S(\gamma) &=& \frac{3}{4} (5 \g^2-1)\,,
\eea
and where all the other coefficients $ I_{2n+1}(\g; \nu)$ are polynomials in $\nu$ of order $n$:
\beq \label{Cn3}
I_{2n+1}(\g; \nu)=  I^S_{2n+1}(\g) +\sum_{k=1}^{n}  I_{2n+1}^{\nu^k}(\g) \nu^k\,.
\eeq
The explicit values of the coefficients $ I_{2n+1}(\g; \nu)$ were given (at the 6PN accuracy) in Table XIV of Ref. \cite{Bini:2020nsb},
while the exact (``Schwarzschild") values, $I^S_{2n+1}(\g)$,  of their test-mass limit, $\nu\to 0$, were given in Eq. (9.5) there.

In view of the existence of efficient algebraic-manipulation programs, there is no need to write down here the 
6PN-accurate f-route local effective Delaunay
Hamiltonian, ${\bar H}_{\rm eff}^{\rm loc, f}(I_R, I_\phi)$ corresponding to the inversion of the explicit expression
for $I_R^{\rm loc, f}(E_{\rm eff},J)$ given in Ref. \cite{Bini:2020nsb}. It might, however, be useful to
 emphasize again the relation between
the effective energy $E_{\rm eff}= \mu c^2 + \cdots$ and the total energy $E_{\rm tot}= M c^2 + \cdots$ (see Eq. \eqref{eobmap}):
\bea 
E_{\rm tot} &=& M c^2\sqrt{1 + 2 \nu \left( \frac{{\mathcal E}_{\rm eff}}{\mu c^2}-1\right)} \nonumber\\
&\equiv& M c^2 \sqrt{1 + 2 \nu(\e-1)} \equiv M c^2 h(\g,\nu)\,,
\eea
where
\beq
\e \equiv \frac{{\mathcal E}_{\rm eff}  }{\mu c^2} \equiv \g\,.
\eeq
Let us now complete the results of Ref. \cite{Bini:2020nsb} by explaining in detail how the results derived above
allow one to explicitly write down the complementary nonlocal contribution
${\bar H}^{\rm nonloc, f}_{6 \rm PN}(I_R, I_\phi)$ to the total Delaunay Hamiltonian
\beq
{\bar H}^{\rm tot}_{6 \rm PN}(I_R, I_\phi)={\bar H}^{\rm loc, f}_{6 \rm PN}(I_R, I_\phi)+ {\bar H}^{\rm nonloc, f}_{6 \rm PN}(I_R, I_\phi)\,.
\eeq
It is the sum of three contributions
\bea \label{H3terms}
{\bar H}^{\rm nonloc, f}_{6 \rm PN}(I_R, I_\phi)&=& {\bar H}^{\rm nonloc, h}_{4+5+6 \rm PN}(I_R, I_\phi)\nonumber\\ 
&+&{\bar H}^{\rm nonloc, h}_{5.5 \rm PN}(I_R, I_\phi) \nonumber\\
&+&  \Delta^{\rm f-h} {\bar H}(I_R, I_\phi)\,.
\eea
The first contribution was computed in Ref. \cite{Bini:2020nsb} (see Eq. (3.31) there) in terms of the harmonic coordinate semi-major axis $a_r^h$ and eccentricity\footnote{Here, we are talking about ellipticlike orbital elements.}  $e_t^h$ (as a power series expansion up to the order $O((e_t^h)^{10})$ included) and reads
\begin{eqnarray} \label{H6delaunay}
\frac{{\bar H}^{\rm nonloc, h}_{4+5+6 \rm PN}}{M}&=&\frac{\nu^2}{(a_r^h)^5}\left[{\mathcal A}^{\rm 4PN}(e_t^h)+{\mathcal B}^{\rm 4PN}(e_t^h)\ln a_r^h\right]\nonumber\\
&+&
\frac{\nu^2}{(a_r^h)^6}\left[{\mathcal A}^{\rm 5PN}(e_t^h)+{\mathcal B}^{\rm 5PN}(e_t^h)\ln a_r^h\right]
\nonumber\\
&+&
\frac{\nu^2}{(a_r^h)^7}\left[{\mathcal A}^{\rm 6PN}(e_t^h)+{\mathcal B}^{\rm 6PN}(e_t^h)\ln a_r^h\right]\,.\nonumber\\
\end{eqnarray} 
The explicit expressions of the 4PN and 5PN coefficients ${\mathcal A}^{\rm 4PN}$, $ {\mathcal B}^{\rm 4PN}$, $ {\mathcal A}^{\rm 5PN}$,
$ {\mathcal B}^{\rm 5PN}$ are written down in Table I of Ref. \cite{Bini:2020wpo}, while the explicit expressions of the 6PN coefficients
${\mathcal A}^{\rm 6PN}$, ${\mathcal B}^{\rm 6PN}$, have been written down in Table V or Ref. \cite{Bini:2020nsb}.

The second contribution was computed in Ref. \cite{Bini:2020wpo} and reads\footnote{After
correcting a sign error on the right-hand side of  Eq. (12.6) in Ref. \cite{Bini:2020wpo}.}
\beq \label{H55delaunay}
{\bar H}^{\rm nonloc, h}_{5.5 \rm PN} = +\frac{\mu^2}{M} c^2  \frac{6848}{525} \frac{\pi}{(a_r^h)^{13/2}} \varphi(e_t^h)\,,
\eeq
where the expansion of the function $\varphi(e)$ in powers of $e$ (up to the 16th order) is given in Eq. (12.7) there.

Let us clarify that  the intermediate (ellipticlike) orbital elements $a_r^h$ and $e_t^h$ used as arguments in these expressions
acquire a  gauge-invariant meaning when they are reexpressed
 as functions of $\bar E\equiv \frac{E_{\rm tot}-Mc^2}{\mu}$ and $j\equiv \frac{J}{G M \mu}$. The
 corresponding expressions are given in Eqs. \eq{aretvsEj} (see also Table III in Ref. \cite{Bini:2020nsb}).

Note that the replacement of the latter functions\footnote{For brevity, we henceforth omit the superscript $h$ on 
$a_r$, and $e_t$.} $a_r(E,J)$, $e_t(E,J)$ in the expressions \eq{H6delaunay}, \eq{H55delaunay},
would be appropriate for computing the corresponding values of the radial action, namely
\bea
I^{\rm nonloc, h}_{R \, 4+5+6 \rm PN}(E,J)&=& -\frac1{\Omega_R}{\bar H}^{\rm nonloc, h}_{4+5+6 \rm PN}(I_R, I_\phi)\,,\nonumber\\
I^{\rm nonloc}_{R \, 5.5 \rm PN}(E,J)&=& -\frac1{\Omega_R}{\bar H}^{\rm nonloc}_{5.5 \rm PN}(I_R, I_\phi)\,,
\eea
where $\Omega_R=2\pi/T_R$ denotes the radial frequency. The 2PN-accurate expression of $n \equiv GM \Omega_R$
in terms of $\bar E$ and $j$ is given in Eq. \eq{n2PN}.

Indeed, $E$ and $J$ are the natural arguments for the radial action. On the other hand, the natural variables for the Delaunay Hamiltonian are,
by definition,  $I_R$ and $I_\phi \equiv J$. Therefore we must use the (2PN-accurate) transformation between  $E$, $J$ and  $I_R$, $I_\phi$.
This transformation (first derived in \cite{Damour:1988mr})
is given (in both directions), at the 2PN accuracy, in Appendix \ref{PN} in terms of the rescaled action variables
\bea
i_r &\equiv& \frac{I_R}{GM\mu}\,, \nonumber\\
i_\phi &\equiv& \frac{I_\phi}{GM\mu} \equiv j\,,\nonumber\\
i_{r\phi} &\equiv& i_r+i_\phi \equiv i_r+j\,,
\eea
Note the important point that the function $e_t^2(i_r,i_\phi)$, given in Eq. \eq{aretvsiriphi},
contains $i_r$ as an overall factor. In other words,
$e_t^2$ vanishes like $i_r$ when $i_r \to 0$, keeping fixed $i_\phi$. This expresses the fact that the ellipticlike
eccentricity\footnote{Beware that it does not coincide with the analytic continuation of its hyperboliclike 
counterpart.} $e_t$ is a good
quasi-Keplerian eccentricity that vanishes along circular motions (the latter being intrinsically defined by the property $i_r=0$).
This property also ensures that the expression we computed for the nonlocal Delaunay Hamiltonian as a truncated expansion
in powers of $e_t$ (up to $e_t^{10}$ included) becomes transformed, when expressed as a function of $i_r$ and $i_\phi=j$,
as a truncated expansion in powers of $i_r$ (up to $i_r^5$ included). In turn, this ensures that, for example, the corresponding
contribution to the periastron advance is obtained as an expansion in powers of $i_r$ (up to $i_r^5$ included).

So far we have discussed the explicit expressions of the first two contributions to the nonlocal Delaunay Hamiltonian, Eq. \eq{H3terms}.
It remains to discuss the third contribution, namely $\Delta^{\rm f-h} {\bar H}(I_R, I_\phi)$.

In view of Eq. \eq{Wf1}, it is given by
\beq
\Delta^{\rm f-h} {\bar H}(I_R, I_\phi)= \frac{2\pi}{\Omega_R} W^{\rm f-h}_{\rm ell}\,,
\eeq
where
\begin{eqnarray}
\label{Wfell}
W^{\rm f-h}_{\rm ell}&=& + 2\frac{G H_{\rm tot}}{c^{3}}\oint dt {\mathcal F}^{\rm split}_{\rm 2PN}(t,t)  \ln(f(t)) \nonumber\\
&=&2  \frac{G H_{\rm tot}}{c^5}\oint dt {\mathcal F}^{\rm GW}_{\rm 1PN}(t) \left[ f_1+ \eta^2 \left(f_2 - \frac12 f_1^2\right)\right]\nonumber\\
&=& \oint dt \Delta^{\rm f-h} H'_{\rm 6PN}\,,
\end{eqnarray}
where $\Delta^{\rm f-h} H'_{\rm 6PN}$ is given by Eq. \eq{Hfmenh6PNnew}.
Using the 2PN-accurate quasi-Keplerian representation of {\it elliptic} motions in harmonic coordinates 
(see, e.g., section III of \cite{Bini:2020nsb}),
it is a straightforward matter to compute the elliptic integral $W^{\rm f-h}_{\rm ell}$. Its {\it exact expression} in terms of 
$a_r$ and $e_t$ reads
\beq
W^{\rm f-h}_{\rm ell}(a_r,e_t)=W^{\rm f-h}_{\rm ell\,0}+\eta^2 W^{\rm f-h}_{\rm ell\,2}\,,
\eeq
where
\begin{eqnarray}
W^{\rm f-h}_{\rm ell\,0}&=&2\pi M^2\nu^3\frac{w_0}{[a_r(1-e_t^2)]^{9/2}}\,, \nonumber\\
W^{\rm f-h}_{\rm ell\,2}&=&2\pi M^2 \nu^3\frac{w_2^{\nu^0}+\nu w_2^{\nu^1}}{[a_r(1-e_t^2)]^{11/2}}\,,
\end{eqnarray}
with
\begin{widetext}
\begin{eqnarray}
w_0&=&
C_3+\left(3C_3+\frac12C_2\right)e_t^2+\frac38(C_3+C_2+C_1)e_t^4
+\frac{1}{16}C_1e_t^6
\,,\nonumber\\
w_2^{\nu^0}&=& 
\frac92C_3+D_4^0+\left(\frac12D_3^0+81C_3+\frac{21}{4}C_2+5D_4^0\right)e_t^2
+\left(\frac{371}{16}C_2+\frac{3}{4}D_3^0+\frac{3}{8}D_2^0+\frac{99}{16}C_1+\frac{1539}{16}C_3+\frac{15}{8}D_4^0\right)e_t^4\nonumber\\
&&
+\left(\frac{117}{16}C_3+\frac{133}{16}C_2+\frac{199037}{7560}+\frac{339}{32}C_1+\frac{1}{16}D_3^0+\frac{3}{16}D_2^0\right)e_t^6
+\left(\frac{15}{16}C_1+\frac{199037}{60480}\right)e_t^8
\,,\nonumber\\
w_2^{\nu^1}&=& 
-\frac{1}{2}C_3+D_4^1+\left(-\frac{3}{2}C_2+\frac{32792}{105}+5D_4^1+\frac{16}{5}D+\frac{3}{4}C_1-25C_3\right)e_t^2\nonumber\\
&&
+\left(-\frac{555}{16}C_3-\frac{15}{2}C_2-\frac{9}{8}C_1+\frac{15}{8}D_4^1+\frac{24}{5}D+\frac{273271}{630}\right)e_t^4\nonumber\\
&&
+\left(-\frac{105}{32}C_1-\frac{45}{16}C_3+\frac{27331}{3780}-\frac{45}{16}C_2+\frac{2}{5}D\right)e_t^6
+\left(-\frac{27331}{15120}-\frac{9}{32}C_1\right)e_t^8
\,.
\end{eqnarray}
\end{widetext}
When using the minimal values, Eqs. \eq{CDmin}, of the flexibility parameters, this result takes
the following explicit form
\begin{eqnarray}
w_{0\,\rm min}&=&\frac{63}{5}e_t^4+\frac{21}{10}e_t^6
\,,\nonumber\\
w_{2\,\rm min}^{\nu^0}&=&\frac{2079}{10}e_t^4+\frac{2890019}{7560}e_t^6+\frac{2104157}{60480}e_t^8
\,,\nonumber\\
w_{2\,\rm min}^{\nu^1}&=&\left(\frac{35438}{105}+\frac{16}{5}D\right)e_t^2
+\left(\frac{249457}{630}+\frac{24}{5}D\right)e_t^4\nonumber\\
&&
+\left(-\frac{194707}{1890}+\frac{2}{5}D\right)e_t^6
-\frac{34043}{3024}e_t^8
\,.
\end{eqnarray}

Similarly to the treatment above of ${\bar H}^{\rm nonloc, h}_{4+5+6 \rm PN}$ and ${\bar H}^{\rm nonloc, h}_{5.5 \rm PN}$
we can then re-express $W^{\rm f-h}_{\rm ell}$ as a function of $E$ and $J$, and $\Delta^{\rm f-h} {\bar H}$ 
as a function of $I_R$,  and $I_\phi$, by using the 2PN-accurate transformations explicitly given above. 

As already mentioned, in view of the existence of efficient algebraic-manipulation programmes there is no need to write down
here the long expressions obtained after these transformations. Let us, instead, cite the explicit forms of two of the simplest
gauge-invariant quantities one can derive from our results: the value of the nonlocal contribution to the total energy {\it along
circular orbits}, and the value of the nonlocal contribution to the periastron advance, also computed {\it along
circular orbits}. They are both obtained by taking the limit $I_R \to 0$, namely
\bea
E^{\rm nonloc, X, circ}(J) &=& \left[ {\bar H}^{\rm nonloc, X}(I_R,I_\phi)\right]_{I_R=0}\,,\nonumber\\
K^{\rm nonloc, X, circ}(J) &=& \left[\frac1{\Omega_R} \frac{\partial {\bar H}^{\rm nonloc, X}(I_R, I_\phi)}{\partial I_\phi}\right]_{I_R=0}\,.\nonumber\\
\eea
Here, X, is a label distinguishing the various contributions to the nonlocal action. Following the decomposition \eq{H3terms} we have
\bea
E^{\rm nonloc,f, circ}(J)&=& E^{\rm nonloc, h, circ}_{4+5+6 \rm PN}(J)\nonumber\\
&+ &E^{\rm nonloc, h,circ}_{5.5 \rm PN}(J)\nonumber\\
& +& E^{\rm f-h, circ}(J)\,.
\eea
These three nonlocal contributions must be added to the f-route local contribution, $E^{\rm loc,f, circ}(J)$, to obtain the total
circular energy
\beq
E^{\rm tot,  circ}(J)= E^{\rm loc,f, circ}(J)+ E^{\rm nonloc,f, circ}(J)\,.
\eeq
Similarly, the total periastron advance along circular orbits can be decomposed as
\beq
K^{\rm tot,  circ}(J)= K^{\rm loc,f, circ}(J)+ K^{\rm nonloc,f, circ}(J)\,,
\eeq
where
\bea
K^{\rm nonloc,f, circ}(J)&=& K^{\rm nonloc, h, circ}_{4+5+6 \rm PN}(J)\nonumber\\
&+& K^{\rm nonloc, h,circ}_{5.5 \rm PN}(J)\nonumber\\
&+& K^{\rm f-h, circ}(J)\,.
\eea
Using rescaled variables, we find the following results for these quantities
\begin{widetext}
\bea
\frac{E^{\rm loc,f, circ}_{\leq6 \rm PN}(j)}{M}&=&1-\frac{\nu}{2}\frac{\eta^2}{j^2}
+\left(-\frac{\nu ^2}{8}-\frac{9 \nu }{8}\right)\frac{\eta^4}{j^4}
+\left(-\frac{\nu ^3}{16}+\frac{7 \nu ^2}{16}-\frac{81 \nu }{16}\right)\frac{\eta^6}{j^6}\nonumber\\
&&
+\left[-\frac{5 \nu ^4}{128}+\frac{5 \nu ^3}{64}+\left(\frac{8833}{384}-\frac{41
   \pi ^2}{64}\right) \nu ^2-\frac{3861 \nu }{128}\right]\frac{\eta^8}{j^8}\nonumber\\
&&
+\left[-\frac{7 \nu ^5}{256}+\frac{3 \nu ^4}{128}+\left(\frac{41 \pi
   ^2}{128}-\frac{8875}{768}\right) \nu
   ^3+\left(\frac{989911}{3840}-\frac{6581 \pi ^2}{1024}\right) \nu
   ^2-\frac{53703 \nu }{256}\right]\frac{\eta^{10}}{j^{10}}\nonumber\\
&&
+\left[\left(\frac{a_6^{\nu^2}}{2}+\frac{29335 \pi
   ^2}{2048}-\frac{1679647}{3840}\right) \nu ^3-\frac{21 \nu
   ^6}{1024}+\frac{5 \nu ^5}{1024}+\left(\frac{41 \pi
   ^2}{512}-\frac{3769}{3072}\right) \nu
   ^4\right.\nonumber\\
	&&\left.
	+\left(\frac{3747183493}{1612800}-\frac{31547 \pi ^2}{1536}\right)
   \nu ^2-\frac{1648269 \nu }{1024}\right]\frac{\eta^{12}}{j^{12}}\nonumber\\
&&
+\left[\nu ^3 \left(\frac{39
   a_6^{\nu^2}}{4}+\frac{a_7^{\nu^2}}{2}-\frac{1681 \pi
   ^4}{512}+\frac{10605841 \pi
   ^2}{24576}-\frac{10727952929}{1075200}\right)\right.\nonumber\\
	&&
	+\nu ^4
   \left(\frac{a_6^{\nu^2}}{4}+\frac{a_7^{\nu^3}}{2}-\frac{21383
   \pi ^2}{8192}+\frac{1007737}{7680}\right)-\frac{33 \nu
   ^7}{2048}-\frac{7 \nu ^6}{2048}+\left(\frac{41 \pi
   ^2}{1024}-\frac{2537}{3072}\right) \nu
   ^5\nonumber\\
	&&\left.
	+\left(\frac{576215112401}{29030400}+\frac{1322752463 \pi
   ^2}{3538944}-\frac{2800873 \pi ^4}{524288}\right) \nu ^2-\frac{27078705
   \nu }{2048}
	\right]\frac{\eta^{14}}{j^{14}}\,,
 \eea
\bea
\frac{E^{\rm nonloc, h, circ}_{4+5+6 \rm PN}(j)}{M}&=&
\frac{64}{5}\nu^2\frac{\eta^{10}}{j^{10}}\left\{
\ln\left(4\frac{e^{\gamma}}{j}\right)
+\left[
\frac{1}{2}+\frac{3793}{336}\ln\left(4\frac{e^{\gamma}}{j}\right)-\frac{155}{12}\ln(2)+\frac{1215}{896}\ln(3)\right.\right.\nonumber\\
&&\left.
+\left(\frac{1}{2}-\frac{7}{4}\ln\left(4\frac{e^{\gamma}}{j}\right)+\frac{155}{28}\ln(2)-\frac{1215}{224}\ln(3)\right)\nu
\right]\frac{\eta^2}{j^2}\nonumber\\
&&
+\left[
\frac{982207}{9072}\ln\left(4\frac{e^{\gamma}}{j}\right)-\frac{106783}{9072}\ln(2)+\frac{6075}{448}\ln(3)+\frac{5977}{672}\right.\nonumber\\
&&
+\left(-\frac{79727}{2016}\ln\left(4\frac{e^{\gamma}}{j}\right)+\frac{211849}{6048}\ln(2)+\frac{5977}{672}-\frac{83835}{1792}\ln(3)\right)\nu\nonumber\\
&&\left.\left.
+\left(\frac{76319}{1512}\ln(2)-\frac{5}{8}+\frac{1}{2}\ln\left(4\frac{e^{\gamma}}{j}\right)-\frac{13365}{448}\ln(3)\right)\nu^2
\right]\frac{\eta^4}{j^4}
\right\}
\,,\nonumber\\
\frac{E^{\rm nonloc, h,circ}_{5.5 \rm PN}(j)}{M}&=& \frac{6848}{525}\nu^2\pi\frac{\eta^{13}}{j^{13}}
\,,\nonumber\\
\frac{E^{\rm f-h, circ}(j)}{M}&=& 
\nu^3\frac{\eta^{12}}{j^{12}}\left[
C_3+(24C_3+D_4)\frac{\eta^2}{j^2}
\right]\,.
\eea
Note that the minimal version of  $\Delta^{\rm f-h} H'_{\rm 6PN}$, Eq. \eq{Hfmenh6PNnewmin},
 leads to a vanishing value of  $E^{\rm f-h, circ}(j)$
\beq
E^{\rm f-h, circ}_{\rm min}(j)=0\,.
\eeq
On the other hand,  if one does not use the minimal version of $\Delta^{\rm f-h} H'_{\rm 6PN}$,
the total energy is easily checked to depend only on the shifted versions of the undetermined parameters $a_6^{\nu^2}$,
  $a_7^{\nu^2}$ and  $a_7^{\nu^3}$
defined in Eqs. \eq{5pnshifts},  \eq{6pnshifts}, \eq{shifting}.

Similarly for the periastron advance
\bea
\label{Klocfcirc}
K^{\rm loc,f, circ}_{\leq6 \rm PN}(j)&=&1+3\frac{\eta^2}{j^2} +\left(\frac{45}{2}-6\nu\right)\frac{\eta^4}{j^4}
+\left[\frac{405}{2}+\left(-202+\frac{123}{32}\pi^2\right)\nu+3\nu^2\right]\frac{\eta^6}{j^6}\nonumber\\
&+&
\left[\frac{15795}{8}+\left(\frac{185767}{3072}\pi^2-\frac{105991}{36}\right)\nu+\left(-\frac{41}{4}\pi^2+\frac{2479}{6}\right)\nu^2\right]\frac{\eta^8}{j^8}  \nonumber\\
&+&
\left[\frac{161109}{8}+\left(-\frac{18144676}{525}+\frac{488373}{2048}\pi^2\right)\nu+\left(-\frac12 \bar d_5^{\nu^2}-\frac{15}{2} a_6^{\nu^2} -\frac{9225}{32}\pi^2+\frac{21399}{2}\right)\nu^2\right. \nonumber\\
&&\left.
+\left(-\frac{1627}{6}+\frac{205}{32}\pi^2\right)\nu^3\right]\frac{\eta^{10}}{j^{10}} \nonumber\\
&+&
\left[\frac{3383289}{16}
+\left(-\frac{2299413173213}{6350400}-\frac{10107671003}{1179648} \pi
   ^2+\frac{7335303}{65536} \pi ^4\right) \nu\right. \nonumber\\
&&
	+\left(-\frac{361}{2}a_6^{\nu^2}-\frac{21}{2}a_7^{\nu^2}-9
   \bar d_5^{\nu^2}-\frac{1}{2}\bar d_6^{\nu^2}+\frac{85731}{2048} \pi
   ^4-\frac{8043499}{1024} \pi ^2+\frac{1859633}{8}\right)\nu ^2 \nonumber\\
&&\left.
	+\left(\frac{15}{2}a_6^{\nu^2}+\frac{1}{2}\bar d_5^{\nu^2}-\frac{21}{2}a_7^{\nu^3}+\frac{1290233}{3072} \pi
   ^2-\frac{2190437}{144}\right)\nu ^3
	+\frac{75}{2}\nu^4
\right]\frac{\eta^{12}}{j^{12}}
\,,\nonumber\\
K^{\rm nonloc, h, circ}_{4+5+6 \rm PN}(j)&=&
-\frac{64}{10}\nu\frac{\eta^8}{j^8}\left\{
-11+\frac{157}{6}\ln\left(4\frac{e^{\gamma}}{j}\right)-\frac{277}{6}\ln(2)+\frac{729}{16}\ln(3)\right.\nonumber\\
&&
+\left[
-\frac{59723}{336}+\frac{9421}{28}\ln\left(4\frac{e^{\gamma}}{j}\right)-\frac{11237}{28}\ln(2)+\frac{112995}{224}\ln(3)\right.\nonumber\\
&&\left.
+\left(\frac{2227}{42}-\frac{617}{6}\ln\left(4\frac{e^{\gamma}}{j}\right)-\frac{1957}{2}\ln(2)+\frac{54675}{112}\ln(3)\right)\nu
\right]\frac{\eta^2}{j^2}\nonumber\\
&&
+\left[
-\frac{4446899}{2016}+\frac{11076725}{3024}\ln\left(4\frac{e^{\gamma}}{j}\right)-\frac{5347151}{1008}\ln(2)+\frac{10528947}{1792}\ln(3)+\frac{48828125}{145152}\ln(5)\right.\nonumber\\
&&
+\left(\frac{358987}{252}-\frac{363851}{168}\ln\left(4\frac{e^{\gamma}}{j}\right)-\frac{10931765}{1512}\ln(2)+\frac{4626963}{896}\ln(3)-\frac{48828125}{24192}\ln(5)\right)\nu\nonumber\\
&&\left.\left.
+\left(-\frac{136369}{1512}+\frac{775}{6}\ln\left(4\frac{e^{\gamma}}{j}\right)-\frac{1315051}{126}\ln(2)+\frac{4333905}{1792}\ln(3)+\frac{48828125}{16128}\ln(5)\right)\nu^2
\right]\frac{\eta^4}{j^4}
\right\}
\,,\nonumber\\
K^{\rm nonloc, h,circ}_{5.5 \rm PN}(j)&=& -\frac{99938}{315}\nu\pi\frac{\eta^{11}}{j^{11}}
\,,\nonumber\\
K^{\rm f-h, circ}(j)&=&
-\nu^2\frac{\eta^{10}}{j^{10}}\left\{
15C_3+C_2
+\left[
\frac{903}{2}C_3+\frac{53}{2}C_2+21D_4^0+D_3^0\right.\right.\nonumber\\
&&\left.\left.
+\left(-\frac{51}{2}C_3+\frac{70876}{105}-C_2+21D_4^1+\frac{32}{5}D\right)\nu
\right]\frac{\eta^2}{j^2}
\right\}\,.
\eea
The minimal version of  $\Delta^{\rm f-h} H'_{\rm 6PN}$, \eq{Hfmenh6PNnewmin},
 leads to the following simple value for $K^{\rm f-h, circ}(j)$
 \beq
 K^{\rm f-h, circ}_{\rm min}(j)= -\frac{32}{5}\nu^3\frac{\eta^{12}}{j^{12}}\left(\frac{17719}{168}+D\right)\,.
 \eeq
Again,  if one does not use the minimal version of $\Delta^{\rm f-h} H'_{\rm 6PN}$,
the total periastron advance is easily checked to depend only on the shifted versions of the undetermined parameters $\bar d_5^{\nu^2}, \ldots $ defined in Eqs. \eq{5pnshifts},  \eq{6pnshifts}, \eq{shifting}.

It is useful to express both the binding energy and the periastron advance along circular orbits in terms of the dimensionless frequency variable $x=(GM\Omega_\phi/c^3)^{2/3}$ by replacing $j$ as a function of $x$. For simplicity,
we henceforth use the minimal version, Eqs.\eq{CDmin}, of the flexibility factor (corresponding to
the explicit minimal Hamiltonian contribution  \eq{Hfmenh6PNnewmin}). [Accordingly, we replace the undetermined
parameters by their minimal values.]

We then find the following explicit  relation between $j$ and $x$
\bea
j&=&\frac{1}{\sqrt{x}}\left\{ 1+\left(\frac16 \nu+\frac32\right)x+\left(\frac{1}{24}\nu^2-\frac{19}{8}\nu+\frac{27}{8}\right) x^2
+\left[\frac{135}{16}+\frac{7}{1296}\nu^3+\frac{31}{24}\nu^2+\left(\frac{41}{24}\pi^2-\frac{6889}{144}\right)\nu\right] x^3\right.\nonumber\\
&&+\left[\frac{2835}{128}-\frac{55}{31104}\nu^4-\frac{215}{1728}\nu^3+\left(\frac{356035}{3456}-\frac{2255}{576}\pi^2\right)\nu^2
+\left(-\frac{128}{3}\gamma-\frac{6455}{1536}\pi^2\right.\right. \nonumber\\
&&\left.\left. -\frac{256}{3}\ln(2)-\frac{64}{3}\ln(x)+\frac{98869}{5760}\right)\nu\right] x^4
\nonumber\\
&&+\left[\frac{15309}{256}-\frac{1}{768}\nu^5-\frac{55}{768}\nu^4+\left(\frac{451}{128}\pi^2-\frac{25189}{256}\right)\nu^3\right. \nonumber\\
&&+\left(\frac{1312}{15}\ln(x)+\frac{1944}{7}\ln(3)+\frac{21337}{1536}\pi^2-2a_6^{\nu^2 \rm min}+\frac{6976}{105}\ln(2)+\frac{2624}{15}\gamma-\frac{341671}{1440}\right)\nu^2\nonumber\\
&&\left.+\left(\frac{59112343}{44800}+\frac{9976}{105}\ln(x)-\frac{486}{7}\ln(3)+\frac{47344}{105}\ln(2)+\frac{19952}{105}\gamma-\frac{126779}{768}\pi^2\right)\nu\right] x^5
\nonumber\\
&&
-\frac{89024}{1575}\pi\nu x^{11/2}
\nonumber\\
&&
+\left[\frac{168399}{1024}-\frac{1729}{6718464}\nu^6-\frac{3283}{248832}\nu^5+\left(\frac{18298567}{373248}-\frac{173635}{124416}\pi^2\right)\nu^4\right.\nonumber\\
&&
+\left(-\frac{41216}{135}\gamma-\frac73 a_7^{\nu^3\rm min} -\frac{4221791}{110592}\pi^2+\frac72 a_6^{\nu^2 \rm min} +1134\ln(3)-\frac{20608}{135}\ln(x)\right. \nonumber\\
&&\left.+\frac{49890383}{51840}-\frac{240112}{81}\ln(2)\right) \nu^3\nonumber\\
&&
+\left(\frac{99652}{81}\ln(2)-\frac73 a_7^{\nu^2 \rm min}-\frac{76581497731}{14515200}-\frac{11767}{2304}\pi^4\right. \nonumber\\
&&\left.+\frac{54738593}{110592}\pi^2-\frac{166324}{135}\gamma-\frac{5751}{2}\ln(3)-\frac72  a_6^{\nu^2 \rm min} 
-\frac{83162}{135}\ln(x)\right)\nu^2\nonumber\\
&&
+\left(\frac{247758680837}{43545600}+\frac{178844}{1215}\ln(x)+\frac{357688}{1215}\gamma-\frac{47656}{243}\ln(2)+\frac{19606111}{786432}\pi^4\right.\nonumber\\
&&\left.\left.\left.
+648\ln(3)-\frac{5802762665}{5308416}\pi^2\right)\nu\right]x^6 
\right\}\,.
\eea

Using the latter relation, the binding energy as a function of $x$ reads
\beq
E^{\rm tot,circ}_{\leq6 \rm PN}(x)=M c^2+ E^{\rm tot,circ}_{\leq4 \rm PN}(x)+E^{\rm tot, circ}_{5+5.5+6 \rm PN}(x)\,,
\eeq
where $E^{\rm tot,circ}_{\leq4 \rm PN}(x)$ is given by Eq. (5.5) of Ref. \cite{Damour:2014jta}, and where
\bea
E^{\rm tot,circ}_{5+5.5+6 \rm PN}(x)&=&-\frac{\mu}{2}x\bigg\{\left[-\frac{45927}{512}+ 
   \left(  -\frac{228916843}{115200}-\frac{23672}{35}\ln(2)-\frac{9976}{35}\gamma+\frac{729}{7}\ln(3)-\frac{4988}{35}\ln(x)+\frac{126779}{512}\pi^2\right)\nu\right. \nonumber\\
&& \left( \frac{1004021}{2880}+3a_6^{\nu^2 \rm min} -\frac{2916}{7}\ln(3)-\frac{3488}{35}\ln(2)-\frac{21337}{1024}\pi^2-\frac{656}{5}\ln(x)-\frac{1312}{5}\gamma  \right)\nu ^2\nonumber\\
&&
\left.+\left(\frac{75567}{512}-\frac{1353 \pi^2}{256}\right) \nu ^3
+\frac{55 }{512}\nu ^4
+\frac{1}{512}\nu^5
\right]x^5\nonumber\\
&&
+\frac{178048}{1575}\pi\nu x^{11/2}
\nonumber\\
&&
+\left[-\frac{264627}{1024}
+\left(\frac{9118627045}{5308416}\pi^2-\frac{389727504721}{43545600}-\frac{7128}{7}\ln(3)-\frac{30809603}{786432}\pi^4-\frac{3934568}{8505}\gamma\right.\right.\nonumber\\
&&\left.-\frac{1967284}{8505}\ln(x)+\frac{74888}{243}\ln(2)\right)\nu \nonumber\\
&&
+ \left(\frac{18491}{2304}\pi^4 +\frac{11}{2} a_6^{\nu^2\rm min}-\frac{86017789}{110592}\pi^2+\frac{914782}{945}\ln(x)+\frac{63261}{14}\ln(3)\right. \nonumber\\
&&\left.+\frac{120889797143}{14515200}+\frac{11}{3}a_7^{\nu^2\rm min} -\frac{156596}{81}\ln(2)+\frac{1829564}{945}\gamma\right)\nu ^2
\nonumber\\
&&+
\left(\frac{64768}{135}\gamma+\frac{32384}{135}\ln(x)-\frac{11}{2} a_6^{\nu^2\rm min}-1782\ln(3)+\frac{6634243}{110592}\pi^2\right. \nonumber\\
&&\left. +\frac{2641232}{567}\ln(2)-\frac{15582935}{10368}+\frac{11}{3}a_7^{\nu^3\rm min} \right)\nu ^3 
\nonumber\\
&&\left.
+\left(\frac{272855 \pi ^2}{124416}-\frac{28754891}{373248}\right)
   \nu ^4
+\frac{5159}{248832} \nu^5
+\frac{2717}{6718464} \nu ^6
\right]x^6\bigg\}\,.
\eea

Similarly, when using the minimal version of the flexibility factor, the periastron advance expressed in terms of $x$  reads
\bea
K^{\rm tot, circ}_{\le \rm 6PN}(x)&=&
1+3x+\left(\frac{27}{2}-7\nu\right) x^2+\left[\frac{135}{2}+7\nu^2+\left(-\frac{649}{4}+\frac{123}{32}\pi^2\right)\nu\right] x^3\nonumber\\
&&+\left[\frac{2835}{8} 
+\left(\frac{48007}{3072}\pi^2-\frac{1256}{15}\ln(x)-\frac{275941}{360}-\frac{1458}{5}\ln(3)-\frac{2512}{15}\gamma-\frac{592}{15}\ln(2)\right)\nu\right.\nonumber\\
&&\left.+\left(-\frac{451}{32}\pi^2+\frac{5861}{12}\right)\nu^2
-\frac{98}{27}\nu^3\right] x^4\nonumber\\
&&+\left[\frac{15309}{8} 
+\left(\frac{9477}{35}\ln(3)+\frac{3928}{35}\gamma+\frac{1964}{35}\ln(x)-\frac{1056789}{1600}-\frac{953995}{2048}\pi^2-\frac{26344}{35}\ln(2)\right)\nu\right. \nonumber\\
&&+\left(\frac{19832}{45}\ln(x)
-\frac12 \bar d_5^{\nu^2\rm min}-\frac{186997}{2304}\pi^2+\frac{42385559}{15120}
-\frac{15}{2}a_6^{\nu^2\rm min}\right.\nonumber\\
&&\left.\left.  +\frac{343408}{45}\ln(2)-\frac{95742}{35}\ln(3)+\frac{39664}{45}\gamma\right)\nu^2
+\left(-\frac{6512}{9}+\frac{1025}{48}\pi^2\right)\nu^3
+\frac{70}{81}\nu^4
\right]x^5\nonumber\\
&&
-\frac{99938}{315}\pi\nu x^{11/2}\nonumber\\
&&
+\left[\frac{168399}{16} 
+\left(\frac{343173600941}{12700800}-\frac{1261899}{280}\ln(3)-\frac{9765625}{4536}\ln(5)+\frac{12925966}{945}\ln(2)\right.\right. \nonumber\\
&&\left.-\frac{10245894299}{1179648}\pi^2+\frac{7335303}{65536}\pi^4+\frac{1287518}{945}\gamma+\frac{643759}{945}\ln(x)\right)\nu\nonumber\\
&&
+\left(-\frac{185881}{6144}\pi^2-\frac32 \bar d_5^{\nu^2\rm min} -\frac{21}{2}a_7^{\nu^2\rm min}
-56a_6^{\nu^2\rm min} +\frac{388640863}{20160}\right. \nonumber\\
&&  +\frac{943569}{140}\ln(3)+\frac{5043}{2048}\pi^4+\frac{9765625}{756}\ln(5)-\frac{38699404}{945}\ln(2)-\frac{86092}{105}\gamma \nonumber\\
&&\left.-\frac{43046}{105}\ln(x)-\frac12 \bar d_6^{\nu^2\rm min}\right)\nu^2\nonumber\\
&&+
\left(-\frac{10176}{5}\gamma-\frac{88248109}{15120}-\frac{9765625}{504}\ln(5)-\frac{2930337}{280}\ln(3)\right. \nonumber\\
&&+\frac{1499825}{9216}\pi^2+20a_6^{\nu^2\rm min}
+\frac{1836992}{35}\ln(2)+\frac43 \bar d_5^{\nu^2\rm min} \nonumber\\
&&\left.\left. 
-\frac{21}{2}a_7^{\nu^3\rm min}-\frac{5088}{5}\ln(x)-\frac{32}{5}D
\right)\nu^3
+\left(-\frac{1681}{96}\pi^2+\frac{22991}{36}\right)\nu^4
\right] x^6\,.
\eea
\end{widetext}
The 4PN-level periastron advance (along circular orbits) was first obtained in Refs. \cite{Damour:2015isa,Damour:2016abl}, and later 
rederived by a different approach in Ref. \cite{Bernard:2016wrg}. Ref. \cite{Damour:2015isa} also derived the 5.5PN periastron advance. 
The terms $O(x^5)$ and $O(x^6)$ corresponding to the 5PN and 6PN orders, respectively, are computed here for the first time, modulo the undetermined parameters $ \bar d_5^{\nu^2\rm min}, a_{6 }^{\nu^2 \rm min}, \ldots$, 
 that enter the  minimal version defined
in Eqs. \eq{CDmin}. We recall that, when using nonzero values of the unconstrained flexibility parameters, any physical
quantity will be given by the same expression as the minimal one, with the qualification that the parameters   
$a_{6 }^{\nu^2  \rm min}, \ldots$,  would be replaced by $a_6^{\nu^2 \rm shifted}$, etc, as defined
in Eqs. \eq{shifting}.
By contrast, the linear-in-$\nu$ part of these coefficients is  fully determined, reproducing the corresponding known terms 
\cite{Bini:2016qtx} in the EOB function $\rho(x)$ such that $K^{-2}(x)=1-6x+\nu\rho(x)+ O(\nu^2)$.

\section{Discussion}

The recent renewed interest in the gravitational scattering of a two-body system has led to further improvements in
the associated analytical modeling within PN-PM theory.
In this work we have raised the present knowledge of the nonlocal-in-time part of the scattering angle at the 6PN level, 
and at the next-to-next-to-leading order in the large eccentricity of the orbital dynamics. The intricacy of the
NNLO level in the scattering angle shows up in the appearance of $\zeta(3)$ in some of the integrals
making up the final result, see Eqs. \eq{solc00},  \eq{solc21}. It also shows up in the fact that we could not
compute analytically a third integral (namely $c_{42}$ or equivalently  $d_{42}$,  Eq. \eq{dmn}, Table \ref{num_resd})
entering the final result, though we did evaluate it numerically.  
Going beyond the NNLO in the large eccentricity expansion remains a challenge for future calculations.
 By considering the mass-ratio dependence of the scattering angle, we discovered in passing
 a hidden simplicity  in the mass-ratio dependence of the gravitational-wave energy loss of a two-body system (see
 subsection VIIB). The mass-ratio dependence of the nonlocal scattering angle allowed us to determine (in section \ref{determiningf}) the 
 contribution to the Hamiltonian linked to the flexibility factor $f(t)$. In particular, we discussed a minimal way
 to fix the residual gauge freedom present in the choice of $f(t)$, see Eqs. \eq{CDmin}, \eq{Hfmenh6PNnewmin}.

Besides our results on the scattering angle at the 6PN level, we gave several other gauge-invariant characterizations 
of the nonlocal-in-time  dynamics. We computed the nonlocal part of the averaged (Delaunay) Hamiltonian for ellipticlike motions
up to the tenth order in eccentricity, see section IX, and Appendix \ref{DelaunayHlocf}. 
We then extracted from the latter results two (partial but useful)
physical observables: the energy and the periastron precession along circular orbits. We expressed the latter quantities
both in terms of the angular momentum and in terms of the orbital frequency. 
Additional results and details are presented in several Appendices. In particular: (i) the details of our frequency-domain computations
are presented in Appendices \ref{A} and \ref{B}; (ii) 
Appendix \ref{q8} completes the information about the h-route nonlocal dynamics by giving the explicit value
of the $O(p_r^8)$ part of the corresponding EOB $Q$ potential; while (iii) Appendix \ref{Hlog} gives the elliptic-motion-average
of the  $\ln(r^h_{12}/s)$ part of the Hamiltonian.

Though our results for the nonlocal dynamics are complete, our method has allowed us to compute the complementary
local dynamics only modulo a small number of undetermined numerical parameters. Namely, two parameters at the
5PN level, and four at the 6PN level. Recent progress in the computer-aided evaluation of the 5PN-level dynamics
 of binary systems \cite{Foffa:2019hrb,Blumlein:2019zku,Blumlein:2020pog,Blumlein:2020znm} 
gives hope that it might become soon possible to extract the two missing 5PN coefficients 
(denoted $\bar d_5^{\nu^2}$ and $a_6^{\nu^2}$) by comparing the observables deducible from 
a 5PN-accurate Hamiltonian computed in (say) harmonic coordinates with the gauge-invariant functions
we presented above, thereby completing the knowledge of the 5PN dynamics. 
However several of the subtleties we had to cope with at 5PN might stand in the way.
We have particularly in mind the following two facts: (i) our method makes use of the knowledge of the 
multipole moments (including self-gravity contributions) acquired by the Multipolar Post-Minkowskian (MPM) formalism;
and (ii) our method uniquely determines all the terms quadratic in one mass in the action by a well-defined  matching
between the near zone (potential modes) and the wave zone (soft radiation modes) based on the use of a global
Green's function (computed by means of black-hole perturbation theory). By contrast, the EFT approach used in
Refs. \cite{Foffa:2019hrb,Blumlein:2019zku,Blumlein:2020pog,Blumlein:2020znm}: (i) has not yet succeeded in computing
self-gravity corrections to the multipole moments (Ref.\cite{Ross:2012fc} rederived only the linearized gravity limit
which had been known for many years \cite{Damour:1990gj}) ; and, (ii) has not yet presented a full {\it ab initio}
computation showing that the ``strategy of regions" \cite{Beneke:1997zp} does reproduce, at the 4PN level, 
the correct nearzone-wavezone matching, first obtained in Ref. \cite{Damour:2014jta} 
by combining the globally matched 4PN-level self-force result of Ref. \cite{Bini:2013zaa}, with the 4PN near-zone
computation of Ref. \cite{Jaranowski:2015lha}. [The EFT-based derivations of the full 4PN dynamics in 
Refs. \cite{Foffa:2019yfl,Blumlein:2020pog} have heuristically added together the results of two different EFT-like 
computations (namely a wavezone EFT computation \cite{Foffa:2011np,Galley:2015kus}) and a nearzone
EFT one \cite{Foffa:2019rdf,Blumlein:2020pog} without showing that this comes out automatically from
a decomposition of the original (non PN-expanded) action into the contributions coming from two different regions
of momentum space.]
We therefore expect that it will be difficult for a direct  EFT computation of the action to unambiguously apply,
in a technically complete way, the strategy of regions at the more intricate 5PN level.

Having this potential difficulty in mind, we therefore suggest  to use, within the EFT approach,  the
strategy that led to the first derivation \cite{Damour:2014jta} of the complete 4PN-level dynamics. Indeed, the basic
fact underlying the success (and completeness) of this strategy is that the  sole possible ambiguity
in combining the near-zone Hamiltonian with the wave-zone one
comes from combining the  {\it logarithmic} infrared divergence entering the former computation,
with the  {\it logarithmic} ultraviolet divergence enter the latter one. In other words, if we introduce
(like in the old-style computations of the Lamb shift) an intermediate scale $s$ (with $r_{12} \ll s \ll c/\Omega_{\phi}$),
the former computation contains a term $2\frac{G H_{}}{c^{5}}{\mathcal F}^{\rm GW}(t)  \ln \left( \frac{r_{12}^h(t)}{s_{\rm NZ}}\right)  $ while the latter one contains a term $2\frac{G H_{}}{c^{5}}{\mathcal F}^{\rm GW}(t)  \ln \left( \Omega_{\phi} s_{\rm WZ}/c\right)  $. 
Here, $s_{\rm NZ}$ denotes the intermediate scale $s$ when it is used as an infrared cutoff in a near-zone computation (involving
potential modes), while $s_{\rm WZ}$ denotes the intermediate scale $s$ when it is used as an ultraviolet cutoff in a wave-zone computation
(involving radiation modes).
In summing  the results of these two regions, the intermediate scale $s$ should disappear, but any 
ambiguity in the identification between  $s_{\rm NZ}$ and  $s_{\rm WZ}$ will introduce an ambiguity in the total Hamiltonian equal to 
\beq
H^C= 2 C \frac{G H_{}}{c^{5}}{\mathcal F}^{\rm GW}(t) \,,
\eeq
where
\beq \label{Cambig}
C =  \ln \left( \frac{ s_{\rm WZ}}{ s_{\rm NZ}} \right)\,,
\eeq
is some pure number. We emphasize here that the same result (presence of the single-parameter ambiguity \eq{Cambig}) 
holds also at the 5PN and 6PN levels because the only delicate divergences\footnote{We assume here that
the (unphysical \cite{DamourLH}) ultraviolet divergences due to the use of a point-mass description have been separately 
regularized; e.g. by using dimensional regularization.} entering the near-zone and wave-zone
computations are logarithmic, and have both the same, known coefficient $2\frac{G H_{}}{c^{5}}{\mathcal F}^{\rm GW}(t) $.

At the 4PN level, Ref. \cite{Damour:2014jta} had introduced such a single logarithmic ambiguity constant
and had shown how the sum of the near-zone (local\footnote{Note that in Ref. \cite{Damour:2014jta} and in
the present discussion the meaning of ``local" is different from the one used in our method.}) Hamiltonian and 
the wave-zone (tail-related) one,
together with the use of the globally-matched self-force 4PN Hamiltonian \cite{Bini:2013zaa}, led to
a unique answer for the full (local-plus-nonlocal) Hamiltonian. The advantage of this strategy is that it is
enough to know three partial results to apply it, namely: (i) a knowledge of the near-zone (potential-modes) Hamiltonian
restricted to the scales $r < s_{\rm NZ}$;
(ii) a knowledge of the wave-zone Hamiltonian, restricted to the scales $r > s_{\rm WZ}$; and (iii)
a knowledge of the globally-matched self-force result (which unambiguously determines the $O(\nu^2)$ 
part of the total Hamiltonian). Our method provides explicit (and complete) results for the items (ii) and (iii),
while it needs to be completed by a near-zone computation for determining the undetermined
parameters $a_6^{\nu^2}$, etc, entering our local Hamiltonian. 

From the practical point of view, 
 we are therefore suggesting to compare (say at the 6PN level) the gauge-invariant content of 
\bea
H^{\rm EFT, tot}_{\rm 6PN} &=& H^{\rm EFT, loc, s}_{\rm 6PN}+ H^C \nonumber\\
&-&
\frac{G H_{}}{c^{5}}{\rm Pf}_{2s/c}\int \frac{d\tau}{|\tau|}{\mathcal F}^{\rm split}_{\rm 2PN}(t,t+\tau)\,,
\eea
to that of our full Hamiltonian 
\bea
H^{\rm our, tot}_{\rm 6PN}&=& H^{\rm loc, f}_{\rm 6PN}+ H^{\rm nonloc, f}_{\rm 6PN}\nonumber\\
&=& H^{\rm loc, f}_{\rm 6PN}+ H^{\rm nonloc, h}_{\rm 6PN} + \Delta^{f-h} H_{\rm 6PN}\,.
\eea
As
\begin{eqnarray}  \label{Hnonloch2n}
 H^{\rm nonloc, h}_{\rm 6PN}(t)&=&
-\frac{G H_{}}{c^{5}}{\rm Pf}_{2s/c}\int \frac{d\tau}{|\tau|}{\mathcal F}^{\rm split}_{\rm 2PN}(t,t+\tau)\nonumber\\
&+&2\frac{G H_{}}{c^{5}}{\mathcal F}^{\rm GW}_{\rm 2PN}(t)  \ln \left( \frac{r_{12}^h(t)}{s}\right)
\,,
\end{eqnarray} 
we see that the identification between the two Hamiltonians boils down to identifying what one can call
their {\it near-zone parts}, namely, on the one hand,
\beq \label{HEFTNZ}
H^{\rm EFT, NZ}_{\rm 6PN} =H^{\rm EFT, loc, s}_{\rm 6PN}+ H^C \,,
\eeq
where $s$ denotes any scale used to regularize the infrared divergence of $H^{\rm EFT, loc}$,
and, on the other hand,
\bea \label{HBDGNZ}
H^{\rm our, NZ}_{\rm 6PN}&=&H^{\rm loc, f}_{\rm 6PN}+ \Delta^{f-h} H_{\rm 6PN}\nonumber\\
&+&
2\frac{G H_{}}{c^{5}}{\mathcal F}^{\rm GW}_{\rm 2PN}(t)  \ln \left( \frac{r_{12}^h(t)}{s}\right)\,.
\eea
There are various ways  to identify (in a gauge-invariant manner) these two near-zone Hamiltonians.
One can look for a canonical transformation mapping on into the other one, or one can identify gauge-invariant
observables. We have provided above (and in our previous papers \cite{Bini:2020wpo,Bini:2020nsb})
several gauge-invariant functions that can be used in this respect. However, as the last term on the right-hand side of 
Eq. \eq{HBDGNZ} has been incorporated in our recent developments into the nonlocal part of the Hamiltonian, 
and was not separately studied (in a gauge-invariant way), we decided to
 complete our gauge-invariant characterization of the near-zone dynamics by giving the
 value of its  Delaunay average, namely
 \beq
\langle  H_{\rm nonloc, \ln, h}^{4+5+6 \rm PN}\rangle =\frac{1}{\oint dt}\oint 2\frac{G H}{c^{5}}{\mathcal F}^{\rm split}_{\rm 2PN}(t,t)  \ln \left( \frac{r_{12}^h(t)}{s}\right)dt\,.
\eeq
 The explicit value of the latter 6PN-accurate  Delaunay average will be found in Appendix \ref{Hlog} as a function
 of $a_r^h$ and $e_t^h$ (up to the tenth order in $e_t^h$).
 
 Summarizing: the identification between Eq. \eq{HEFTNZ} and Eq. \eq{HBDGNZ} yields, in our opinion, an efficient way
 (avoiding a full use of the strategy of regions) to determine at once the values of our undetermined parameters
 $a_6^{\nu^2 }, \ldots$,  and the value of the single nearzone-wavezone separation ambiguity constant $C$ 
 (which we have incorporated here in Eq. \eq{HEFTNZ}).  Our determination of most of the $\nu$ dependence
 of the Hamiltonian will also provide many checks of the computation of $H^{\rm EFT, NZ}_{\rm 6PN}$.

\section*{Acknowledgments}
We thank Johannes Bl\"umlein and Gerhard Sch\"afer for useful exchanges, and Hyung Won Lee for a helpful suggestion. 
DB is grateful to the Institut des Hautes Etudes Scientifiques
 for warm hospitality at various stages during the development of the present project.

\appendix

\section{Compendium of useful PN results} \label{PN}

We collect in this Appendix some known results in PN theory. When working at the 6PN level we often need
only fractionally 2PN-accurate results on the dynamics. In some parts, we only need 1PN-level results such as the 
Einstein-Infeld-Hoffmann-Fichtenholz 1PN Lagrangian  for the relative motion (see, {\it e.g.}, Ref. \cite{dd})
\bea \label{L1PN}
\frac{{\mathcal L}_{\rm 1PN}^h}{\mu}&=&\frac12 v^2 +\frac{GM}{r}+ \eta^2 \left\{\frac18 (1-3\nu) v^4 \right.\nonumber\\
&+&
\left.\frac{GM}{2r }\left[(3+\nu)v^2+\nu ({\mathbf n}\cdot {\mathbf v})^2-\frac{GM}{r}\right]\right\},
\eea
where $v^2=\dot r^2+r^2 \dot \phi^2$. This determines the corresponding momenta
\beq
p_r=\frac{\partial {\mathcal L}_{\rm 1PN}^h}{\partial \dot r}=C_r \, \dot r
\,,\qquad 
p_\phi=\frac{\partial {\mathcal L}_{\rm 1PN}^h}{\partial \dot \phi}=C_\phi \, r^2 \dot \phi\,,
\eeq
with
\bea
C_r&=&1+\eta^2 \left(\frac{(1-3\nu)}{2}v^2 +\frac{GM}{r}(3+2\nu)\right)\,,\nonumber\\
C_\phi&=& 1+\eta^2 \left(\frac{1-3\nu}{2}v^2 +(3+\nu)\frac{GM}{r} \right)  \,,
\eea
so that $p^2=p_r^2+\frac{p_\phi^2}{r^2}=C_r^2 \dot r^2 +C_\phi^2 r^2\dot \phi^2$.

The corresponding 1PN-accurate Hamiltonian (expressed in terms of $p=p^{\rm phys}/\mu$; and using $c=1$) reads
\bea \label{H1PN}
&&\frac{H_{\rm 1PN}(r,p_r,j)- M }{\mu }=\left(\frac12 p^2-\frac{GM}{r}\right) \nonumber\\
&&\qquad
+ \eta^2\left[\frac18 (3\nu-1) p^4-\frac{GM}{2r} (\nu+3) p^2\right.\nonumber\\
&&\left.\qquad
-\frac{GM}{2 r}\nu p_r^2+\frac{(GM)^2}{2r^2}\right]. \nonumber\\
\eea
We often rescale $r$ according to $r^{\rm phys} = GM r$.

We will also need the expression of the 1PN-accurate gravitational wave energy flux \cite{Blanchet:1989cu}
in terms of $r=r^{\rm phys}/ GM$ and $p$:
\begin{widetext}
\begin{eqnarray} \label{FGW1PN}
{\mathcal F}^{\rm GW}_{\rm 1PN}(p,p_r,r)&=&\frac{8}{15}\nu^2\frac{(12 p^2-11 p_r^2)}{r^4}\nonumber\\
&&+\eta^2 \nu^2 \left\{
\frac{1}{r^4}\left[\left(\frac{1374}{35}-\frac{248}{7}\nu \right) p_r^4+\left(-\frac{5332}{105}+\frac{248}{7}\nu\right) p^2 p_r^2+\left(\frac{898}{105}+\frac{104}{35}\nu\right) p^4 \right]\right. \nonumber\\
&&\left.+\frac{1}{r^5}\left[\left(\frac{176}{21}\nu+\frac{9568}{105}\right) p_r^2+\left(-\frac{9472}{105}-\frac{1024}{105}\nu\right) p^2 \right]
+\frac{1}{r^6}\left( \frac{32}{105}-\frac{128}{105}\nu\right)  \right\}\,.
\end{eqnarray}
\end{widetext}
The parameters entering the quasi-Keplerian parametrization, Eq. \eq{hypQK2PN}, of the hyperbolic motion (in
harmonic coordinates) are listed
 in Table \ref{table_relations}, as functions of the  variables
\beq \label{defbarEj}
\bar E \equiv \frac{E_{\rm tot}-Mc^2}{\mu c^2}\; ; \; j \equiv \frac{c J}{GM\mu}\,.
\eeq
%
%
\begin{table*}
\caption{\label{table_relations} Quasi-Keplerian representation of the hyperbolic 2PN motion (in harmonic coordinates). We use the variables
$\nu={m_1m_2}/{(m_1+m_2)^2}$, $\bar E$, Eq. \eq{defbarE}, and $j$, Eq. \eq{defj2}.
}
\begin{ruledtabular}
\begin{tabular}{ll}
$\bar n$&$(2\bar E)^{3/2}\left[1+\frac{\bar E}{4}(15-\nu)\eta^2 +\frac{\bar E^2}{32}\left(555+30\nu+11\nu^2 \right)\eta^4 \right]$\\
$\bar a_r$&$\frac{1}{2\bar E}\left\{1+\frac{\bar E }{2}(7-\nu)\eta^2+\frac{\bar E^2}{4}\left[ 1+\nu^2-\frac{8}{\bar Ej^2}(7\nu-4) \right]\eta^4 \right\}$\\
$e_t^2$& $1+2\bar Ej^2 + \bar E\left[-\bar Ej^2(-17+7\nu) +4(1-\nu) \right] \eta^2+\bar E^2\left[2(3+18\nu+5\nu^2)+\bar Ej^2(112-47\nu+16\nu^2)+\frac{4}{\bar Ej^2}(-4+7\nu) \right]\eta^4 $\\
$e_r^2$& $1+2\bar Ej^2+ \bar E \left[-5\bar E j^2(3-\nu) +2(-6+\nu)\right] \eta^2+\bar E^2\left[30+74\nu+\nu^2+\bar Ej^2(80-45\nu+4\nu^2)+\frac{8}{\bar Ej^2}(-4+7 \nu)\right]\eta^4$\\
$e_\phi^2$&$ 1+2\bar Ej^2+ \bar E \left[ -\bar Ej^2(15-\nu)-12\right]\eta^2+\frac{\bar E^2}{4}\left[\frac{-416+91\nu+15\nu^2}{2\bar Ej^2}+2(-20+17\nu+9\nu^2) +2\bar Ej^2(160-31\nu+3\nu^2)\right]\eta^4 $\\
$f_t$& $\frac32 \frac{(2\bar E)^{3/2}}{j} \left(5-2\nu \right)\eta^4$\\
$g_t$&$ -\frac{(2\bar E)^{3/2}}{8j }\sqrt{1+2\bar Ej^2}\nu\left(-15+\nu \right)\eta^4$\\
$f_\phi$&$\frac{1+2\bar Ej^2}{8j^4} \left(1+19\nu -3\nu^2 \right)\eta^4 $\\
$g_\phi$&$\frac{1}{32}\frac{(1+2\bar Ej^2)^{3/2}}{j^4}    \nu\left(1-3\nu \right)\eta^4 $\\
$K$& $1+\frac{3}{j^2}\eta^2+\frac{3}{4j^4}\left[-2\bar Ej^2(-5+2\nu)+5(7-2\nu)\right] \eta^4$\\
\end{tabular}
\end{ruledtabular}
\end{table*}

Let us also recall (from Table II in Ref. \cite{Bini:2020nsb}) the expressions of the (harmonic-coordinates) rescaled semi-major
axis and time-eccentricity entering the 2PN-accurate quasi-Keplerian representation of elliptic motion in terms
of $\bar E$ and $j$:
\begin{widetext}
\bea \label{aretvsEj}
a_r &=&\frac{1}{(-2\bar E)}\left\{1+\frac{(-2\bar E )}{4}(-7+\nu)\eta^2
+\frac{(-2\bar E)^2}{16}\left[ 1+\nu^2+\frac{16}{(-2\bar E)j^2}(7\nu-4) \right]\eta^4 \right\}
\,, \nonumber\\
e_t^2&=&1+2\bar Ej^2 +\frac{(- 2\bar E)}{4}[-8(1-\nu)-(-2\bar E)j^2(-17+7\nu)]\eta^2 \nonumber\\
&&
+\frac{(-2\bar E)^2}{8}\left[4(3+18\nu+5\nu^2)-(-2\bar E)j^2(112-47\nu+16\nu^2)
\right. \nonumber\\
&& \left. -\frac{16}{(-2\bar E)j^2}(-4+7\nu)-24 \sqrt{-2\bar E}j(-5+2\nu)+\frac{24}{\sqrt{-2\bar E} j}(-5+2\nu)\right]\eta^4\,. 
\eea
\end{widetext}
Beware that the latter elliptic definition of $e_t^2$ is not equal to the analytic continuation in $\bar E$ of its
hyperbolic counterpart, listed in Table \ref{table_relations} (while $\bar a_r$ is the analytic continuation of $- a_r$).
Using the rescaled action variables
\bea \label{defiriphi}
i_r &\equiv& \frac{c I_R}{GM\mu}\,, \nonumber\\
i_\phi &\equiv& \frac{c I_\phi}{GM\mu} \equiv j\,,\nonumber\\
i_{r\phi} &\equiv& i_r+i_\phi \equiv i_r+j\,,
\eea
with 
\bea
i_r&=&-j-\frac{1}{\sqrt{-2{\bar E}}}+\left[\frac{3}{j}-\frac{1}{8} (\nu -15)\sqrt{-2{\bar E}}\right]\eta^2\nonumber\\
&&
+\left[-\frac{5 (2 \nu -7)}{4 j^3}+\frac{3 (2 \nu -5) (-2{\bar E})}{4j}\right.\nonumber\\
&&\left.
-\frac{1}{128} \left(3 \nu ^2+30 \nu +35\right) (-2{\bar E})^{3/2}\right]\eta^4 \,,
\eea
the 2PN-accurate Delaunay Hamiltonian reads \cite{Damour:1988mr} 
\begin{widetext}
\bea
{\bar E}(i_r,i_\phi) &=& -\frac{1}{2i_{r\phi}^2}\left[1 +\frac{1}{4} \frac{24 i_r+(9+\nu) i_\phi}{ i_\phi i_{r\phi} ^2}\eta^2\right.\nonumber\\
&&\left.
-\frac{1}{8}\frac{20 i_r^3 (2 \nu -7)+12 i_r^2 i_\phi (10 \nu -53)+72 i_r i_\phi^2
   (\nu -6)+i_\phi^3 \left(-\nu ^2+7 \nu -81\right)}{ i_\phi^3 i_{r\phi} ^4}\eta^4\right]
\,.
\eea
Using this transformation, we get the following explicit (2PN-accurate) expressions for 
the ellipticlike parameters $a_r$ and $e_r$ as functions
of the action variables $i_r$ and $i_\phi$:
\begin{eqnarray} \label{aretvsiriphi}
a_r &=&i_{r\phi}^2-2\frac{3 i_r+2 i_\phi}{ i_\phi}\eta^2+\frac12 \frac{5i_r^3 (2 \nu -7)+i_r^2i_\phi (44 \nu -95)+2i_r i_\phi^2 (26
   \nu -35)+18 i_\phi^3 (\nu -1)}{ i_{r\phi}^2 i_\phi^3}\eta^4
\,,\nonumber\\
e_t^2&=&\frac{i_r}{i_{r\phi}^2}\left[i_r+2i_\phi
+2\frac{i_r (\nu -1)+i_\phi (2 \nu -5)}{i_{r\phi}^2}\eta ^2\right.\nonumber\\
&&\left.
-\frac12\frac{4 i_r^3 (7 \nu -4)+i_r^2 i_\phi (66 \nu
   +25)-i_r i_\phi^2 \left(6 \nu ^2-28 \nu -207\right)-2 i_\phi^3 \left(6 \nu
   ^2-18 \nu -19\right)}{i_\phi^2 i_{r\phi}^4}\eta^4 
	\right]\,.
\end{eqnarray}
Another useful 2PN-accurate quantity is the (adimensionalized) radial frequency. It reads
\bea \label{n2PN}
n&=& \frac{GM\Omega_R}{c^3}\nonumber\\
&=&(-2\bar E)^{3/2}\left[1+\frac{(-2\bar E)}{8}(-15+\nu)\eta^2
+\frac{(-2\bar E)^2}{128}\left(555+30\nu+11\nu^2+
\frac{192(-5+2\nu)}{\sqrt{-2\bar E}j}\right)\eta^4 \right]\nonumber\\
&=&\frac{1}{i_{r\phi}^3}\left[1+\frac12\frac{(3+\nu)i_\phi+18i_r}{i_\phi i_{r\phi}^2}\eta^2\right.\nonumber\\
&&\left.
-\frac38\frac{-(9+5\nu+\nu^2)i_\phi^3+4i_ri_\phi^2(-37+5\nu)+6i_r^2i_\phi(-59+10\nu)+10i_r^3(2\nu-7)}{i_\phi^3i_{r\phi}^4}\eta^4 
	\right]\,.
\eea
\end{widetext}

\section{Large-eccentricity expansions of the frequency-domain, Newtonian-level, energy flux and integrated tail action} \label{A}

This Appendix discusses the frequency-domain computation of the Newtonian-level energy flux, Eq. \eq{EGWomega}, and the related integrated action,
Eq. \eq{W1_final_exp}, \eq{W1_fourier}.
The frequency-domain integrand \eqref{I_N_gen} is of lowest (Newtonian) order with respect to the PN expansion,
but  is exact in its eccentricity dependence. Let us consider its expansion in inverse powers of the eccentricity
at successive levels: LO, NLO, NNLO, etc. For simplicity, we use $1=GM=G=c$ in the following.

\subsubsection{Newtonian flux at the LO in the large-eccentricity expansion}

The expression \eqref{I_N_gen} can be easily evaluated at the LO in the large eccentricity expansion where 
$p= i \frac{u}{e_r } \to 0$ (see Eq. \eq{defupq}). 
This limit entails a big simplification (already studied in the literature, see e.g., \cite{rr1,rr2,Bini:2017wfr}) which leads to the following expression
\begin{eqnarray}
{\mathcal I}_{\rm N}^{\rm LO}(u)&=& 32 e_r^4 \nu^2\bar a_r^7 u^2 [ (u^2+1) K^2_{1}(u)+3u K_0(u)K_{1}(u)\nonumber\\
&+&\frac1{3}\left(3 u^2+1\right)K_0^2(u)]\,.
\end{eqnarray}
Using the notation introduced in Ref. \cite{Bini:2017wfr}
\begin{eqnarray}
\label{calF_def}
{\mathcal F}(u)&=&\left(\frac{u^2}{3}+u^4\right)K_0^2(u)+3u^3 K_0(u)K_1(u)\nonumber\\
&+&(u^2+u^4)K_1^2(u)\,,
\end{eqnarray}
we find
\beq
{\mathcal I}_{\rm N}^{\rm LO}(u)=  32 \nu^2 e_r^4 \bar a_r^7 {\mathcal F}(u)\,,
\eeq
so that
\begin{eqnarray}
\Delta E_{\rm GW}^{\rm LO}
&=&\frac{32}{5\pi} \frac{\bar n^7\bar a_r^7 }{e_r^3}  \nu^2  \int_{0}^\infty d u  {\mathcal F}(u)\,.
\end{eqnarray}
When using the Newtonian-level relations
\beq
\bar n=(\bar a_r)^{-3/2}=p_\infty^3\,,\qquad \bar a_r=p_\infty^{-2}\,,
\eeq
as well as
\beq
e_r=\sqrt{1+p_\infty^2 j^2}\qquad \to \qquad e_r^{\rm LO}=p_\infty j\,,
\eeq
one recovers the known result for the LO gravitational wave energy, or \lq\lq splash radiation,"
\begin{eqnarray}
\label{E_GW_LO}
\Delta E_{\rm GW\,N}^{\rm LO}
&=&\frac{32}{5\pi} \frac{p_\infty^4 }{j^3}  \nu^2  \int_{0}^\infty d u  {\mathcal F}(u)\nonumber\\
&=&\frac{37}{15}\pi \frac{p_\infty^4 }{j^3}  \nu^2 \,.
\end{eqnarray}
The tail potential $W_{1 \,\rm N}^{\rm (tail)\, LO}$ instead turns out to be
\begin{eqnarray}
\label{W1_LO}
W_{1 \,\rm N}^{\rm (tail)\, LO}
&=&\frac{64}{5\pi} \frac{p_\infty^4 }{j^3}  \nu^2  \int_{0}^\infty d u  {\mathcal F}(u)\ln(\bar\alpha u)\nonumber\\
&=&\frac{2}{15}\frac{\pi\nu^2 }{e_r^3\bar a_r^{7/2}}  \left[ 100 + 37 \ln \left(\frac{s}{4e_r\bar a_r^{3/2}}\right)\right]\,.\nonumber\\
\end{eqnarray}

Let us then pass to the extension of these results at the higher N$^n$LO levels of approximation in the large-eccentricity
expansion.

\subsubsection{Working at the NNLO accuracy in $\frac1{e_r}$}

Expanding the quantity \eqref{I_N_gen} for large $e_r$ up to the NNLO leads to
\beq
\label{IN_NNLO}
{\mathcal I}_{\rm N}(u)={\mathcal I}_{\rm N}^{\rm LO}+{\mathcal I}_{\rm N}^{\rm NLO}+{\mathcal I}_{\rm N}^{\rm NNLO}
+ O\left(\frac{{\mathcal I}_{\rm N}^{\rm LO}}{e_r^3} \right)\,,
\eeq
where
\begin{eqnarray}
\frac{{\mathcal I}_{\rm N}^{\rm LO}}{32\bar a_r^7 e_r^4 \nu^2  }&=& {\mathcal F}(u)\,,\qquad
\frac{{\mathcal I}_{\rm N}^{\rm NLO}}{ 32 \bar a_r^7 e_r^3 \nu^2 \pi}= u{\mathcal F}(u)\,, \nonumber\\
\frac{{\mathcal I}_{\rm N}^{\rm NNLO}}{16 \bar a_r^7 e_r^2 \nu^2}&=& C^{00}(u)+C^{20}(u)\frac{\partial^2 K_\nu(u)}{\partial \nu^2}\Bigg|_{\nu=0}\nonumber\\
&& +C^{21}(u)\frac{\partial^2 K_\nu(u)}{\partial \nu^2}\Bigg|_{\nu=1}\,,
\end{eqnarray}
where
\bea
C^{00}(u)&=& - 2u^2 \left[(3u^2+1)K_0^2(u)+7u K_0(u) K_1(u)\right. \nonumber\\
&&\left.+ (1+2u^2) K_1^2(u)\right]+\frac{\pi^2}{u^4}{\mathcal F}(u)\,,\nonumber\\
C^{20}(u)&=&-\frac{u^4}{3}[2(3u^2+1)K_0(u)+9uK_1(u)] \,,\nonumber\\
C^{21}(u)&=&-\frac{u^4}{3}[3u K_0(u)+2(u^2+1)K_1(u)] \,,
\eea
and where the Bessel functions $K_p(u)$ and $K_{p+1}(u)$ have been Taylor-expanded around $p=0$ to second order in $p$,
\begin{eqnarray}
\label{K_exp}
K_p(u)&=& K_0(u)+p \frac{\partial K_\nu(u)}{\partial \nu}\Bigg|_{\nu=0}\nonumber\\
&&+\frac12 p^2 \frac{\partial^2 K_\nu(u)}{\partial \nu^2}\Bigg|_{\nu=0}+O(p^3)\nonumber\\
&=&  K_0(u)+\frac12 p^2 \frac{\partial^2 K_\nu(u)}{\partial \nu^2}\Bigg|_{\nu=0}+O(p^3)
\,,\nonumber\\
K_{p+1}(u)&=& K_1(u)+p \frac{\partial K_\nu(u)}{\partial \nu}\Bigg|_{\nu=1}\nonumber\\
&&+\frac12 p^2 \frac{\partial^2 K_\nu(u)}{\partial \nu^2 }\Bigg|_{\nu=1}+O(p^3)\nonumber\\
&=& K_1(u)+\frac{p}{u} K_0(u) \nonumber\\
&&+\frac12 p^2 \frac{\partial^2 K_\nu(u)}{\partial \nu^2 }\Bigg|_{\nu=1}+O(p^3)\,.
\end{eqnarray}
In  Eqs. \eqref{K_exp} above we have used the known results (see Eqs. 9.1.66-9.1-68 of Ref. \cite{AS})
\beq
\frac{\partial K_\nu(u)}{\partial \nu}\Bigg|_{\nu=0}=0\,,\qquad
\frac{\partial K_\nu(u)}{\partial \nu}\Bigg|_{\nu=1}=\frac{1}{u} K_0(u)\,.
\eeq
Moreover, in what follows the derivatives of $K_\nu$ with respect to the order will only enter integrals of the type
\beq \label{Fint}
F(a,\mu)=\int_0^\infty du u^a K_\mu(u) \frac{\partial^2 K_\nu(u)}{\partial \nu^2 }\Bigg|_{\nu=0,1}\,.
\eeq
These integrals can be evaluated by considering the master integral
\bea \label{Gint}
G(a,\mu, \nu)&=&\int_0^\infty du u^a K_\mu(u) K_\nu(u)\nonumber\\
&=& \frac{2^{a-2} }{\Gamma (a+1)}\Gamma_1\Gamma_2\Gamma_3\Gamma_4\,,
\eea
where
\bea
\Gamma_1 &=&\Gamma \left(\frac{1}{2} (a-\mu -\nu +1)\right)\,,  \nonumber\\
\Gamma_2 &=&\Gamma \left(\frac{1}{2} (a+\mu -\nu +1)\right)\,,  \nonumber\\
\Gamma_3 &=&\Gamma \left(\frac{1}{2} (a-\mu +\nu +1)\right)\,, \nonumber\\
\Gamma_4 &=&\Gamma \left(\frac{1}{2}(a+\mu +\nu +1)\right)\,.
\eea
The resulting expression being valid when the four conditions ${\rm Re}[a \pm \mu \pm \nu] > -1$ are all satisfied.
Taking two derivatives of \eq{Gint} with respect to $\nu$ and evaluating the result at $\nu=0,1$
allows one to compute  the integral \eq{Fint}.

One can rewrite Eq. \eqref{IN_NNLO} in various ways.
For example, 
\begin{eqnarray}
\label{I_N_fin}
\frac{{\mathcal I}_{\rm N}(u)}{32 \bar a_r^7 e_r^4 \nu^2}
&=&{\mathcal F}(u)\left[1+\frac{\pi u}{e_r}  +\frac12 \left( \frac{\pi u}{e_r}\right)^2 \right]\nonumber\\
&& -\frac{{\mathcal G}(u)+{\mathcal H}(u)}{e_r^2}\,, 
\end{eqnarray}
where ${\mathcal F}(u)$ is defined in Eq. \eqref{calF_def},  and where we defined
\bea
{\mathcal G}(u)&\equiv&u^2 \left[(3u^2+1)K_0^2(u)+7u K_0(u) K_1(u)\right. \nonumber\\
&&\left.+ (1+2u^2) K_1^2(u)\right]
\,,\nonumber\\
{\mathcal H}(u)&\equiv&{\mathcal H}_0(u)\frac{\partial^2 K_\nu(u)}{\partial \nu^2}\Bigg|_{\nu=0}
+{\mathcal H}_1(u)\frac{\partial^2 K_\nu(u)}{\partial \nu^2}\Bigg|_{\nu=1}\,,\nonumber\\
\eea
with
\begin{eqnarray}
{\mathcal H}_0(u)&\equiv&u^4\left[\left(u^2+\frac13\right)K_0(u)+\frac32uK_1(u)\right]
\,,\nonumber\\
&=& -\frac{C^{20}(u)}{2}
\,,\nonumber\\
{\mathcal H}_1(u)&\equiv&u^4\left[\frac32u K_0(u)+(u^2+1)K_1(u)\right]\nonumber\\
&=& -\frac{C^{21}(u)}{2}\,.
\end{eqnarray}
The functions ${\mathcal G}(u), {\mathcal H}_0(u), {\mathcal H}_1(u)$ are such that
\bea
{\mathcal G}(u)&=&3{\mathcal F}(u)\nonumber\\
&-&u^2 [2u K_0(u)+(2+u^2)K_1(u)]K_1(u)\,,
\eea
and
\beq
{\mathcal H}_0(u)K_0(u)+{\mathcal H}_1(u)K_1(u)=u^2 {\mathcal F}(u)\,.
\eeq
We list  in Table \ref{integ_NNLO} the  integrals needed in order to compute both the gravitational wave energy $\Delta E_{\rm GW}$ and the tail potential $W_1^{\rm tail}$ at the NNLO level  in $e_r^{-1}$ (but still at the Newtonian level, $\eta^0$):


 \begin{table}
 \caption{\label{integ_NNLO} Integrals needed for the Newtonian-level gravitational wave energy $\Delta E_{\rm GW}$, and the tail potential $W_1^{\rm tail}$, at the NNLO level in $e_r^{-1}$.}
 \begin{ruledtabular}
 \begin{tabular}{lll}
 & expression & value\\
\hline
${\sf f}$ & $\int_{0}^\infty du  {\mathcal F}(u)$ & $\frac{37}{96}\pi^2$\\
${\sf f}^{\ln {}}$ &$\int_{0}^\infty du  {\mathcal F}(u)\ln u$ & $ \frac{1}{96}\pi^2 [100 - 37\gamma - 111 \ln(2)]$\\
${\sf f}^{u}$&$ \int_{0}^\infty du  u{\mathcal F}(u)$ & $ \frac{49}{9}$\\
${\sf f}^{u\ln{}}$ &$\int_{0}^\infty du u{\mathcal F}(u)\ln u$ & $ \frac{1}{54}[139 - 294\gamma + 294  \ln(2)]$\\
${\sf f}^{u^2}$&$ \int_{0}^\infty du u^2{\mathcal F}(u)$ & $ \frac{297}{256}\pi^2$\\
${\sf f}^{u^2\ln{}}$&$ \int_{0}^\infty du u^2{\mathcal F}(u)\ln u$ & $ \frac{3}{256}\pi^2 [350 - 99\gamma - 297\ln(2)]$\\
${\sf g}$&$ \int_{0}^\infty du  {\mathcal G}(u)$ & $ \frac{403}{512}\pi^2$\\
${\sf g}^{\ln} $&$\int_{0}^\infty du  {\mathcal G}(u)\ln u$ & $\frac{1}{2048}\pi^2 [4591 - 1612\gamma - 4836\ln(2)]$\\
 \end{tabular}
 \end{ruledtabular}
 \end{table}

The integral of ${\mathcal H}(u)$ can be written as the sum of the two pieces
\beq
{\sf h}=\int_0^\infty {\mathcal H}(u)du= \frac{\partial^2{\sf h}_0(\nu)}{\partial \nu^2}\Bigg|_{\nu=0}+ \frac{\partial^2{\sf h}_1(\nu)}{\partial \nu^2}\Bigg|_{\nu=1}\,,
\eeq
where
\begin{eqnarray}
{\sf h}_0(\nu)&=& \int_0^\infty du\, {\mathcal H}_0(u) K_\nu(u)\,,\nonumber\\
{\sf h}_1(\nu)&=& \int_0^\infty du\, {\mathcal H}_1(u) K_\nu(u)\,.
\end{eqnarray}
We find
\begin{eqnarray}
\frac{\partial^2{\sf h}_0(\nu)}{\partial \nu^2}\Bigg|_{\nu=0}&=&\frac{2007 \pi ^4}{8192}-\frac{179 \pi ^2}{80}
\,,\nonumber\\ 
\frac{\partial^2{\sf h}_1(\nu)}{\partial \nu^2}\Bigg|_{\nu=1}&=&\frac{2745 \pi ^4}{8192}-\frac{7527 \pi ^2}{2560}\,,
\end{eqnarray}
so that
\beq
{\sf h}=\frac{297 \pi ^4}{512}-\frac{2651 \pi ^2}{512}\,.
\eeq
The integral of ${\mathcal H}(u)\ln u$ can be computed in the same way
\bea
{\sf h}^{\ln}&=&\int_0^\infty {\mathcal H}(u)\ln udu\nonumber\\
&=&-\frac{\pi ^2 }{6144}\left[49896 \zeta (3)+53437+132 \gamma  \left(27 \pi ^2-241\right)\right. \nonumber\\
&&\left. -95436 \log (2)+36 \pi ^2 (297 \log (2)-350)\right]\,.
\eea
Finally, from
\begin{eqnarray}
\int_0^\infty {\mathcal I}_{\rm N}(u) du
&=&32 \bar a_r^7 e_r^4 \nu^2 \left\{\left[{\sf f}+\frac{\pi}{e_r}{\sf f}^{u}  +\frac12 \left( \frac{\pi }{e_r}\right)^2 {\sf f}^{u^2}\right]\right. \nonumber\\
&&\left. -\frac{1}{e_r^2}{\sf g}
-\frac{1}{e_r^2 }{\sf h}\right\}+O\left(\frac{1}{e_r^3}\right)\,, 
\end{eqnarray}
we have that the first and second-order eccentricity corrections to the (Newtonian-level) splash radiation energy \eqref{E_GW_LO} read 
\begin{widetext}
\beq
\label{delta_E_GW_NNLO}
\Delta E_{\rm GW \,N}^{\rm LO+NLO+NNLO}=
\frac{\nu^2 }{e_r^3\bar a_r^{7/2}}\left(\frac{37 \pi }{15}+\frac{1568}{45 e_r}+\frac{281 \pi }{10 e_r^2}
+ O\left(\frac1{e_r^3} \right)\right)\,.
\eeq
Similarly,
\begin{eqnarray}
W_{1 \rm N}^{\rm (tail)\, LO+NLO+NNLO}
&=&\frac{2}{15}\frac{\nu^2 }{e_r^3\bar a_r^{7/2}}  \left\{\pi\left[ 100 + 37 \ln \left(\frac{s}{4e_r\bar a_r^{3/2}}\right)\right]
+\frac1{e_r}\left[\frac{2224}{9}+\frac{1568}{3}\ln \left(\frac{4s}{e_r\bar a_r^{3/2}}\right)\right]\right.\nonumber\\
&&\left.
+\frac{\pi}{e_r^2}\left[\frac{2479}{4}+\frac{6237}{8}\zeta(3)+\frac{843}{2}\ln \left(\frac{s}{4e_r\bar a_r^{3/2}}\right)\right] + O\left(\frac1{e_r^3} \right)  \right\}\,.\nonumber\\
\end{eqnarray}
\end{widetext}
This result  allows one to fix the previously defined parameter $c_{00}$ (see Eq. \eqref{solc00}) which could not
be computed in the time domain.

\subsubsection{Going at N$^3$LO in the energy flux}

The next term in the $e_r^{-1}$ expansion of Eq.  \eqref{I_N_fin} is the following
\begin{widetext}
\begin{eqnarray}
\frac{{\mathcal I}_{\rm N}(u)}{32 \bar a_r^7 e_r^4 \nu^2 u^6}
&=&\ldots +\frac{1}{e_r^3}\left[
-\frac{\pi}{u^5}  {\mathcal H}_0(u) \frac{\partial^2 K_\nu(u)}{\partial \nu^2}\Bigg|_{\nu=0} -\frac{\pi}{u^5}  {\mathcal H}_1(u)  
\frac{\partial^2 K_\nu(u)}{\partial \nu^2}\Bigg|_{\nu=1}+ \frac{\pi^3}{6u^3}{\mathcal F}(u)-\frac{\pi}{u^5}{\mathcal G}(u)\right]\nonumber\\
&=&\ldots +\frac{1}{e_r^3}\left[ -\frac{\pi}{u^5}  {\mathcal H}(u)+ \frac{\pi^3}{6u^3}{\mathcal F}(u)-\frac{\pi}{u^5}{\mathcal G}(u)\right]\,.
\end{eqnarray}
\end{widetext}
Multiplying both sides by $u^6$ one has then  
\begin{eqnarray}
\frac{{\mathcal I}_{\rm N}(u)}{32 \bar a_r^7 e_r^4 \nu^2}
&=&{\mathcal F}(u)\left[1+\frac{\pi u}{e_r}  +\frac12 \left( \frac{\pi u}{e_r}\right)^2 +\frac16\left( \frac{\pi u}{e_r}\right)^3 \right] \nonumber\\
&&
-\frac{{\mathcal G}(u)+{\mathcal H}(u)}{e_r^2}\left(1+\frac{\pi u}{e_r}\right)\,,
\end{eqnarray}
that is the $O\left(\frac{1}{e_r^3} \right)$-accurate truncation of the compact expression
\begin{eqnarray}
\frac{{\mathcal I}_{\rm N}(u)}{32 \bar a_r^7 e_r^4 \nu^2}
&=&e^{\pi u/e_r}\left[{\mathcal F}(u)-\frac{{\mathcal G}(u)+{\mathcal H}(u)}{e_r^2}\right]\Bigg|_{O\left(\frac{1}{e_r^3} \right)}\,.
\end{eqnarray}
Eq. \eqref{delta_E_GW_NNLO} is then extended as
\bea
\label{delta_E_GW_NNNLO}
\Delta E_{\rm GW \,N}^{{\rm LO}+\ldots +{\rm N}^3{\rm LO} }(\bar a_r, e_r)&=&
\frac{\nu^2 }{e_r^3\bar a_r^{7/2}}\left(\frac{37 \pi }{15}+\frac{1568}{45 e_r}\right. \nonumber\\
&&+\frac{281 \pi }{10 e_r^2}+\frac{7808}{45 e_r^3}\nonumber\\
&&\left.+O\left(\frac{1}{e_r^4}\right)\right)\,.
\eea
Expressing $e_r$ and $a_r$ in terms of $\pinf$ and $j$ the above (Newtonian-level) expression becomes
\bea
\label{delta_E_GW_NNNLO_pinf_j}
\Delta E_{\rm GW \,N}^{{\rm LO}+\ldots +{\rm N}^3{\rm LO} }(\pinf, j)&=&\nu^2 \left[
\frac{37}{15}\pi \frac{\pinf^4}{j^3}+\frac{1568}{45}\frac{\pinf^3}{j^4}\right. \nonumber\\
&&+\frac{122}{5}\pi \frac{\pinf^2}{ j^5}+\frac{4672}{45}\frac{\pinf}{j^6}\nonumber\\
&&\left.+O\left(\frac{1}{j^7}\right) \right]\,.
\eea

No special, additional difficulties arise for the Newtonian energy flux when going up to higher orders in the large eccentricity expansion in the frequency domain.

\section{Eccentricity expansion of the 1PN-accurate  frequency-domain energy flux and tail action} \label{B}

Let us now consider the 1PN corrections ($\propto \eta^2$) to the (frequency-domain) energy flux, and tail action.
The corresponding integrands are linear in $\nu$ (after factoring out an overall factor). In this case Eq. \eqref{arctan_exp} should be used, and complications arise already at the NLO as we are going to show. We have been able to compute the NNLO level too, but the associated expressions are very long and will  not be displayed below.

Let us write ($1=GM=G=c$) the 1PN contribution to the energy flux as follows,
\beq
\Delta E^{\rm 1PN}=\frac{1}{\pi e_r \bar a_r^{3/2}}\, \int_0^\infty du {\mathcal F}_{\rm GW}^{\rm 1PN}(u)\,,
\eeq
with a subsequent large-eccentricity expansion:
\bea
{\mathcal F}_{\rm GW}^{\rm 1PN}&=&\frac{\nu^2}{e_r^2 \bar a_r^3}\left[{\mathcal F}_{\rm GW}^{\rm 1PN,LO}+\frac{1}{e_r}{\mathcal F}_{\rm GW}^{\rm 1PN,NLO}\right. \nonumber\\
&&\left. +\frac{1}{e_r^2}{\mathcal F}_{\rm GW}^{\rm 1PN,NNLO} +O\left(\frac{1}{e_r^3}\right)\right]\,.
\eea
We recall that, at 1PN, each term ${\mathcal F}_{\rm GW}^{{\rm 1PN}, {\rm N}^n{\rm LO}}$ depends on $\nu$ linearly, i.e.
\beq
{\mathcal F}_{\rm GW}^{{\rm 1PN}, {\rm N}^n{\rm LO}}={\mathcal F}_{{\rm GW}, \nu^0}^{{\rm 1PN}, {\rm N}^n{\rm LO}}+\nu
{\mathcal F}_{{\rm GW}, \nu^1}^{{\rm 1PN}, {\rm N}^n{\rm LO}}\,.
\eeq
We then find the explicit expressions
\begin{widetext}
\begin{eqnarray}
{\mathcal F}_{{\rm GW}, \nu^0}^{{\rm 1PN}, {\rm LO}}&=&-\frac{8}{105}u^2 \left[(-10u^4 +82u^2+156)K_0(u)^2 -4u(-43+61u^2)K_0(u)K_1(u)\right.\nonumber\\
&&\left.
 +(-10u^4-45u^2-300)K_1^2(u)\right] 
\,,\nonumber\\
{\mathcal F}_{{\rm GW}, \nu^1}^{{\rm 1PN}, {\rm LO}}&=&-\frac{8}{105}u^2 \left[ (-16+40u^4)K_0^2(u)+(-48u^3-64u)K_0(u)K_1(u)+(-4u^2+40u^4)K_1^2\right] 
\,,\nonumber\\
{\mathcal F}_{{\rm GW}, \nu^0}^{{\rm 1PN}, {\rm NLO}}&=&-\frac{64}{21}u^3 \left(-\frac{u^4}{4}+\frac{23}{2}u^2+\frac{141}{20} \right)K_0^2(u)+\frac{128}{35}u^4 \left(\frac{61}{12}u^2 -\frac{653}{24} \right) K_0(u)K_1(u)\nonumber\\
&&-\frac{64}{21}u^3 \left(-\frac{u^4}{4}+\frac{333}{40}u^2+\frac{39}{20} \right)K_1^2(u) \nonumber\\
&&-\frac{24}{5} iu^4 \left[(A(u)+uB(u))K_0(u)+(B(u)+2uA(u))K_1(u)\right]
\,,\nonumber\\
{\mathcal F}_{{\rm GW}, \nu^1}^{{\rm 1PN}, {\rm NLO}}&=& \pi \left[-\frac{64}{21}u^3 \left(u^4-\frac{21}{20}u^2 -\frac{3}{4} \right)K_0^2(u)+\frac{128}{35}u^4\left(u^2+\frac{95}{24} \right)K_0(u)K_1(u)\right.\nonumber\\
&&\left.
-\frac{64}{21}u^3 \left( u^4 -\frac{23}{20}u^2-\frac{21}{20}\right)K_1^2(u)\right]\,,
\end{eqnarray}
\end{widetext}
where
\begin{eqnarray}
A&=& -\frac{i}{2}(G_2(u)+G_2^*(u))+\frac{i}{2}(G_{-2}(u)+G_{-2}^*(u))\nonumber\\
&=& -i [G_2^{\rm S}(u)-G_{-2}^{\rm S}(u)]\nonumber\\
&=& -2iG_{[2]}^{\rm S}(u)
\,,\nonumber\\
B&=& -\frac{1}{4}(G_3(u)-G_3^*(u))+\frac{1}{4}(G_{-3}(u)-G_{-3}^*(u))\nonumber\\
&&
+\frac{5}{4}(G_1(u)-G_1^*(u))+\frac{5}{4}(G_{-1}(u)-G_{-1}^*(u))\nonumber\\
&=& -\frac12 G_3^{\rm A}(u)-\frac12 G_{-3}^{\rm A}(u)+\frac{5}{2}G_1^{\rm A}(u)+\frac52 G_{-1}^{\rm A}(u)\nonumber\\
&=& -G_{(3)}^{\rm A}(u)+5 G_{(1)}^{\rm A}(u)\,.
\end{eqnarray}
Here we introduced the notation
\beq
G_n(u) \equiv\int_{-\infty}^\infty \, dv\, {\rm arctan}\left(\tanh \frac{v}{2} \right)\, e^{iu\sinh v -nv}\,,
\eeq
as well as $G_n^{\rm S}(u)=\frac12( G_n(u)+G_n^*(u))$, $G_n^{\rm A}(u)=\frac12 (G_n(u)-G_n^*(u))$ and the symmetry-related 
expressions $G_{(n)}^{\rm S}(u)=\frac12(G_n^{\rm S}(u)+G_{-n}^{\rm S}(u))$, $G_{[n]}^{\rm S}(u)=\frac12(G_n^{\rm S}(u)-G_{-n}^{\rm S}(u))$, $G_{(n)}^{\rm A}(u)=\frac12(G_n^{\rm A}(u)+G_{-n}^{\rm A}(u))$, $G_{[n]}^{\rm A}(u)=\frac12(G_n^{\rm A}(u)-G_{-n}^{\rm A}(u))$.
Due to parity reasons
\beq
G_{(n)}^{\rm S}(u)\equiv 0 \equiv G_{[n]}^{\rm A}(u)\,.
\eeq
This, however, has no effect on $A$ and $B$ which only contain $G_{[n]}^{\rm S}(u)$ and $G_{(n)}^{\rm A}(u)$.

Going to the NNLO, the energy flux also contains terms involving the derivatives of the Bessel K functions with respect to the order.
Furthermore, integrals over $v$ enter the term ${\mathcal F}_{{\rm GW}, \nu^0}^{{\rm 1PN}, {\rm NNLO}}$.
All integrations can be done analytically, leading to 
\bea
\Delta E_{\rm GW}^{\rm 1PN}&=&\frac{\nu^2}{e_r^3 \bar a_r^{9/2}}\left[
\left(\frac{1143}{280}-\frac{37}{30}\nu\right)\pi\right.\nonumber\\
&&
+\frac1{e_r}\left(\frac{944}{1575}-\frac{1136}{45}\nu\right)\nonumber\\
&&\left.
+\frac1{e_r^2}\left(-\frac{22333}{560}-\frac{609}{20}\nu\right)\pi
\right]\,.
\eea

By contrast, $W_1$ can be  analytically computed (in the frequency-domain) only at the LO.
Indeed, consider for instance the NLO term $W_1^{{\rm 1PN}, {\rm NLO}}$, which we have succeeded to compute in the time domain (see Eq. \eqref{W1NLO0}), with the result
\bea \label{W1PNNLO}
W_1^{{\rm 1PN}, {\rm NLO}}&=&\frac{2}{15} \frac{\nu^2}{e_r^4\bar a_r^{9/2}}H_{\rm real}
\left[-\frac{28072}{225}-\frac{38872}{63}\nu\right.\nonumber\\
&+&\left.
\left(\frac{944}{105}-\frac{1136}{3}\nu\right)\ln \left(\frac{4s}{e_r\bar a_r^{3/2}}\right)\right]\,.
\eea
By contrast, the computation of $W_1^{{\rm 1PN}, {\rm NLO}}$ in the frequency-domain yields the expression
\bea
W_1^{{\rm 1PN}, {\rm NLO}}&=&\frac{2}{15} \frac{\nu^2}{e_r^4\bar a_r^{9/2}}H_{\rm real}
\left[-\frac{1768}{9}-\frac{38872}{63}\nu\right.\nonumber\\
&+&
\frac{6144}{5}\gamma-\frac{6144}{5}\ln(2)+\frac{15}{2}X\nonumber\\
&+&\left.
\left(\frac{944}{105}-\frac{1136}{3}\nu\right)\ln \left(\frac{4s}{e_r\bar a_r^{3/2}}\right)\right]
\,,\nonumber\\
\eea
where the quantity $X$ denotes the following double integral:
\beq
X=\frac{48}{\pi}\int_{0}^\infty du\int_{-\infty}^\infty dv\,u^4{\mathcal X}(u,v)\,{\rm arctan}\left(\tanh \frac{v}{2} \right)\ln(u)\,,
\eeq
with
\bea
{\mathcal X}(u,v)&=&\alpha(u)\left(-\frac15\cosh(3v)+\cosh(v)\right)S(u,v)\nonumber\\
&&
-\frac45\beta(u)\sinh(2v) C(u,v)\,.
\eea
Here, to shorten the expression, we denoted $\alpha(u)\equiv uK_0(u)+K_1(u)$, $\beta(u) \equiv \frac12 K_0(u)+uK_1(u)$,
as well as $[S(u,v),C(u,v)]=[\sin(u\sinh(v)),\cos(u\sinh(v))]$.

When attempting to compute $X$, one can first integrate over $u$ by replacing $\ln(u)\to u^a$, taking then a derivative with respect to $a$, before finally setting $a\to0$.
Unfortunately, this method of integration generates derivatives of hypergeometric functions with respect to the parameters, which did not allow us to compute the integral over $v$ in closed form. 
However, direct comparison with the time-domain result \eq{W1PNNLO} yields the following simple result for $X$:
\beq
X=\frac{3584}{375}-\frac{4096}{25}\gamma+\frac{4096}{25}\ln(2)\,.
\eeq

Going to the NNLO, one can similarly extract a Fourier space representation for the missing coefficients $c_{20}$ and $c_{21}$.
A straightforward calculation shows that
\bea
c_{20} &=& -\frac{599223}{560}\ln(2)+\frac{1637641}{3360}+\frac{99837}{160}\zeta(3)\nonumber\\
&&-\frac{1584}{5}\gamma-\frac{1}{\pi^2} (Y_1+Y_2)\,,
\eea
where
\bea
Y_1&=&48\int_{0}^\infty du\int_{-\infty}^\infty dv\,u^4\left[\frac{\pi}{2}u{\mathcal X}(u,v)+{\mathcal Y}(u,v)\right]\nonumber\\
&&\times
\,{\rm arctan}\left(\tanh \frac{v}{2} \right)\ln(u)
\,,\nonumber\\
Y_2&=&24\int_0^\infty du \int_{-\infty}^\infty dv\, u^4{\mathcal X}(u,v)\tanh v\ln(u)\,,
\eea
with
\beq
{\mathcal Y}(u,v)=A(u,v)C(u,v)+B(u,v)S(u,v)\,,
\eeq
and
\bea
A(u,v)&=& -\alpha(u) uv \left(-\frac15\cosh(3 v)+\cosh(v)\right)\nonumber\\
&&
+2\beta(u)\left(\frac15 \sinh(3v)+\sinh(v)\right)\,,\nonumber\\
B(u,v)&=& -2\alpha(u)\left(1+\frac15\cosh (2v)\right)\nonumber\\
&&
-\frac45\beta(u) uv  \sinh(2v)\,. 
\eea
These Fourier-domain expressions did not allow us to compute $c_{20}$.
The remaining coefficient $c_{21}$, instead, was determined by a time-domain computation (see Eq. \eqref{solc21}).

\section{Gravitational wave  energy emitted during a scattering process at the 2PN accuracy} \label{gw2pn}

The total gravitational-wave energy emitted during a scattering process
\beq
\Delta E_{\rm GW}=\Delta E_{\rm GW}^{\rm N}+ \Delta E_{\rm GW}^{\rm 1PN}+ \Delta E_{\rm GW}^{\rm 2PN}+ \cdots
\eeq
 has been computed long ago at the 1PN accuracy by Blanchet and Sch\"afer (see Eq. (5.7) of Ref. \cite{Blanchet:1989cu}).
Let us extend their result by giving here the 2PN term, $\Delta E_{\rm GW}^{\rm 2PN}$,
when  $\Delta E_{\rm GW}$ is expressed in terms of $e_r=e_r^h$ and $j$, as in Ref. \cite{Blanchet:1989cu}:
\begin{widetext}
\begin{eqnarray} \label{C1}
\Delta E_{\rm GW}^{\rm 2PN}(e_r,j)&=& \frac{2}{15}\frac{\nu^2}{j^{11}}\left[{\mathcal E}_1 {\rm arccos}\left(-\frac{1}{e_r}\right)
+{\mathcal E}_2\sqrt{e_r^2-1} \right]
\,,
\end{eqnarray}
with
\begin{eqnarray} \label{C2}
{\mathcal E}_1&=& \frac{1636769}{189}+\frac{2380852}{189}e_r^2+\frac{596996}{63}e_r^4+\frac{494977}{48}e_r^6+\frac{1615745}{672}e_r^8\nonumber\\
&&+\nu \left(-\frac{74435}{21}-\frac{23953}{3}e_r^2-\frac{527659}{28}e_r^4-\frac{1775713}{112}e_r^6-\frac{120745}{56}e_r^8\right)\nonumber\\
&&+\nu^2\left(48+\frac{1463}{2}e_r^2+\frac{31215}{8}e_r^4+\frac{10155}{2}e_r^6+518e_r^8\right)\nonumber\\
{\mathcal E}_2&=& \frac{307844062}{19845}+\frac{1280690597}{158760}e_r^2+\frac{1596923303}{158760}e_r^4+\frac{76924511}{7840}e_r^6\nonumber\\
&& +\nu\left(-\frac{281551}{45}-\frac{25157339}{2520}e_r^2-\frac{104242423}{5040}e_r^4-\frac{3209299}{280}e_r^6 \right)\nonumber\\
&& +\nu^2 \left(\frac{453}{4}+\frac{10777}{8}e_r^2+\frac{10765}{2}e_r^4+3434e_r^6 \right)\,.
\end{eqnarray}
In the parabolic orbit limit $e_r\to 1$ we find (see Ref. \cite{Blanchet:1989cu})
\begin{eqnarray}
\Delta E_{\rm GW}^{{\rm 2PN\, term}\, e_r\to 1}&=& \frac{2 \pi}{15}\frac{\nu^2}{j^{11}}\left[\frac{29198255}{672}-\frac{774153}{16}\nu+\frac{82215}{8}\nu^2\right]\,.
\end{eqnarray}
In the main text we study the 2PN-accurate expression of $\Delta E_{\rm GW}$ when it is expressed in terms of the energy
and the angular momentum (and more precisely in terms of $p_\infty$  and $ h j$, where $h=E_{\rm tot}/M$).

\section{Reparametrization and minimal value of the flexibility factor} \label{f}

The proof of the canonical equivalence of the two flexibility-related Hamiltonians $\Delta^{\rm f-h} H_{\rm 6PN}(r,p_r,j)$, Eq. \eq{Hfmenh6PN}, and $\Delta^{\rm f-h} H'_{\rm 6PN}(r',p_r',j)$, Eq. \eq{Hfmenh6PNnew}, (with $p_\phi'=j=p_\phi)$), i.e.,
\beq
\Delta^{\rm f-h} H'_{\rm 6PN} = \Delta^{\rm f-h} H_{\rm 6PN} - \{g, H_{\rm 1PN} \} \,,
\eeq
is obtained by a direct construction of the generating function $g(r,p_r',j)$ of the canonical transformation
\beq
r'=r+\frac{\partial g(r,p'_r,j)}{\partial p'_r}\,,\qquad
p_r=p'_r +\frac{\partial g(r,p'_r,j)}{\partial r}\,,
\eeq
with the 1PN (harmonic-coordinate) Hamiltonian (recalled in Eq. \eq{H1PN}).
Using  the 1PN-accurate gravitational wave energy flux given in Eq. \eq{FGW1PN},
one can solve for all the unknowns, i.e., the $C_i$'s and the $D_i$'s as functions of the $c_i$'s and $d_i$'s (see
Eqs. \eq{Civsci}, \eq{Divsdi}), as well as  the coefficients $g_i, n_i$ entering $g(r,p_r,j)$:
\beq
g(r, p_r,j) = \frac{\nu^3p_r}{r^3}\left[g_1\frac{1}{r}+g_2\frac{j^2}{r^2}+g_3 p_r^2
+\eta^{2}\left(n_1\frac{1}{r^2}+n_2 \frac{j^4}{r^4}+n_3 p_r^4+n_4 \frac{j^2}{r^3}+n_5 \frac{p_r^2}{r}+n_6 \frac{j^2 p_r^2}{r^2}\right)\right]\,.
\eeq
We found the explicit results
\begin{eqnarray}
g_1 &=& \frac{64}{5}(c_2+c_3)\,, \nonumber\\
g_2 &=& \frac{64}{5}c_2\,, \nonumber\\
g_3 &=& \frac{16}{45}(12 c_1+73 c_2) \,,
\end{eqnarray}
and
\begin{eqnarray}
n_1 &=&
\left(-\frac{32 c_2^2}{5}-\frac{64 c_2
   c_3}{5}-\frac{112 c_2}{5}-\frac{32
   c_3^2}{5}-\frac{2096 c_3}{105}\right)\nu  
	-\frac{4988
   c_2}{105}-\frac{3476 c_3}{35}+\frac{64
   d_2}{5}+\frac{64 d_3}{5}+\frac{64 d_6}{5}
\,,\nonumber\\
n_2 &=&
\left(-\frac{32 c_2^2}{5}-\frac{48 c_2}{7}\right) \nu
   +\frac{2468 c_2}{105}+\frac{64 d_2}{5}
\,,\nonumber\\
n_3 &=&
\left(-\frac{32 c_1^2}{25}-\frac{176 c_1
   c_2}{25}+\frac{8216 c_1}{525}-\frac{4696
   c_2^2}{225}+\frac{67888 c_2}{1575}\right)\nu  
	-\frac{1268
   c_1}{225}+\frac{33832 c_2}{1575}+\frac{64
   d_1}{25}+\frac{9392 d_2}{225}+\frac{176 d_4}{25}
\,,\nonumber\\
n_4 &=&
\left(-\frac{32 c_2^2}{5}-\frac{64 c_2
   c_3}{5}-\frac{4112 c_2}{105}-\frac{48
   c_3}{7}\right)\nu  
	-\frac{3252 c_2}{35}+\frac{2468
   c_3}{105}+\frac{64 d_2}{5}+\frac{64 d_6}{5}
\,,\nonumber\\
n_5 &=&
\left(-\frac{64 c_1 c_2}{15}-\frac{64 c_1
   c_3}{15}-\frac{2096 c_1}{315}-\frac{1448
   c_2^2}{45}-\frac{272 c_2 c_3}{9}-\frac{2968
   c_2}{45}+\frac{1592 c_3}{45}\right)\nu  \nonumber\\
	&&
	-\frac{3476
   c_1}{105}-\frac{15752 c_2}{105}+\frac{1144
   c_3}{63}+\frac{2896 d_2}{45}+\frac{64
   d_4}{15}+\frac{64 d_5}{15}+\frac{272 d_6}{9}
\,,\nonumber\\
n_6 &=&
\left(-\frac{64 c_1 c_2}{15}-\frac{16
   c_1}{7}-\frac{872 c_2^2}{45}+\frac{1472
   c_2}{315}\right)\nu  
	+\frac{2468 c_1}{315}+\frac{2680
   c_2}{63}+\frac{1744 d_2}{45}+\frac{64 d_4}{15}
\,.
\end{eqnarray}

Let us note that, while the three $C_i$'s are in one-to-one correspondence with the three $c_i$'s, with the inverse relations
\begin{eqnarray} \label{civsCi}
c_1 &=& \frac{456}{32}C_1+\frac{90656}{1536}C_3-\frac{1856}{128}C_2\,,\nonumber\\
c_2 &=& -\frac{31856}{3072}C_3+\frac{656}{256}C_2-\frac{156}{64}C_1\,,\nonumber\\
c_3 &=& -\frac{656}{256}C_2+\frac{34256}{3072}C_3+\frac{156}{64}C_1\,,
\end{eqnarray}
 one cannot express the the six $d_i$'s in terms of the four $D_i$'s.
However,  Eqs. \eqref{Divsdi} can be inverted to express the first four $d_i$'s, namely $d_1,\ldots,d_4$, in terms of $d_5$, $d_6$, 
and of the new parameters $C_i$ and $D_i$: 
\begin{eqnarray} \label{divsDi}
d_1 &=& 
\left(\frac{356325 C_1^2}{616448}-\frac{1409225
   C_2 C_1}{1232896}+\frac{67026425 C_3
   C_1}{14794752}-\frac{146435
   C_1}{8428}+\frac{16657175
   C_2^2}{29589504}+\frac{5345155025
   C_3^2}{608698368}+\frac{390545
   C_2}{28896}\right.\nonumber\\
	&&\left.
	-\frac{789897575 C_2
   C_3}{177537024}-\frac{161067325
   C_3}{4854528}\right) \nu +\frac{1235525
   C_1}{134848}-\frac{3774345
   C_2}{539392}+\frac{37456235
   C_3}{1078784}\nonumber\\
	&&
	-\frac{181445}{4816}
+\frac{11175 D_1}{4816}-\frac{1815 D_2}{1204}
+\frac{10275 D_3}{4816}-\frac{208925 D_4}{19264}	
+\frac{8 d_5}{301}-\frac{944
   d_6}{129}
	\,,\nonumber\\
d_2 &=& \left(\frac{10275 C_1^2}{2465792}-\frac{34925
   C_2 C_1}{4931584}+\frac{1251725 C_3
   C_1}{59179008}-\frac{476265
   C_1}{539392}+\frac{329225
   C_2^2}{118358016}+\frac{22825775
   C_3^2}{2434793472}+\frac{715625
   C_2}{924672}\right.\nonumber\\
	&&\left.
	-\frac{9696425 C_2
   C_3}{710148096}-\frac{174415025
   C_3}{77672448}\right) \nu +\frac{716985
   C_1}{2157568}-\frac{1500925
   C_2}{8630272}+\frac{303139985
   C_3}{310689792}\nonumber\\
	&&
	+\frac{225
   D_1}{2408}-\frac{615
   D_2}{9632}+\frac{3805
   D_3}{38528}-\frac{232105
   D_4}{462336}-\frac{32
   d_5}{301}-\frac{137 d_6}{129}
\,,\nonumber\\
d_3 &=& 
\left(-\frac{10275 C_1^2}{2465792}+\frac{34925
   C_2 C_1}{4931584}-\frac{1251725 C_3
   C_1}{59179008}+\frac{476265
   C_1}{539392}-\frac{329225
   C_2^2}{118358016}-\frac{15395375
   C_3^2}{2434793472}-\frac{715625
   C_2}{924672}\right.\nonumber\\
	&&\left.
	+\frac{9696425 C_2
   C_3}{710148096}+\frac{185034305
   C_3}{77672448}\right) \nu +\frac{1305735
   C_1}{2157568}-\frac{7264195
   C_2}{8630272}+\frac{1196779135
   C_3}{310689792}\nonumber\\
	&&
	-\frac{225
   D_1}{2408}+\frac{615
   D_2}{9632}-\frac{3805
   D_3}{38528}+\frac{268225
   D_4}{462336}+\frac{32 d_5}{301}+\frac{8
   d_6}{129}
\,,\nonumber\\
d_4 &=& 
\left(-\frac{675 C_1^2}{19264}+\frac{225 C_2
   C_1}{4816}-\frac{57075 C_3
   C_1}{616448}+\frac{5362055
   C_1}{539392}-\frac{2925
   C_2^2}{308224}+\frac{1229725
   C_3^2}{4227072}-\frac{7298675
   C_2}{924672}\right.\nonumber\\
	&&\left.
	-\frac{80125 C_2
   C_3}{2465792}+\frac{1425084095
   C_3}{77672448}\right) \nu -\frac{7428725
   C_1}{2157568}+\frac{13974505
   C_2}{8630272}-\frac{3276924925
   C_3}{310689792}\nonumber\\
	&&
	-\frac{3375
   D_1}{2408}+\frac{9225
   D_2}{9632}-\frac{54065
   D_3}{38528}+\frac{3297965
   D_4}{462336}+\frac{179
   d_5}{301}+\frac{384 d_6}{43}
\,.
\end{eqnarray}

We introduced in the text a minimal way, namely Eqs.  \eq{CDmin}, of fixing the values of  the gauge-invariant 
 parameters $C_i$ and $D_i$ associated with some flexibility factor $f(t)$. However, this unique choice of the
 $C_i$ and $D_i$ 
  still leave some gauge freedom in the choice of the flexibility factor $f(t)$ itself.
If ever one wants to have also a specific value for the flexibility factor $f(t)$ itself, (i.e., specific
values of the original flexibility parameters $c_i$, $d_i$ entering Eq. \eq{f_funct_expr}) one needs, in
addition to the explicit values \eq{c123min}, to insert the minimal values  Eqs.  \eq{CDmin} in the
relations \eq{divsDi} expressing the $d_i$'s in terms of the $C_i$'s, the $D_i$'s and of $d_5$ and $d_6$.
This yields
\bea
d_1^{\rm min} &=&
\left(\frac{4110 D}{301}+\frac{1269775907}{606816}\right) \nu +\frac{33448631}{75852}\nonumber\\
&&+\frac{8
   d_5}{301}-\frac{944 d_6}{129}
\,,\nonumber\\
d_2^{\rm min} &=&
\left(\frac{761 D}{1204}+\frac{159864493}{2427264}\right) \nu +\frac{13371067}{809088}\nonumber\\
&&-\frac{32
   d_5}{301}-\frac{137 d_6}{129}
\,,\nonumber\\
d_3^{\rm min} &=&
\left(-\frac{761 D}{1204}-\frac{159864493}{2427264}\right) \nu +\frac{12115205}{809088}\nonumber\\
&&+\frac{32
   d_5}{301}+\frac{8 d_6}{129}
\,,\nonumber\\
d_4^{\rm min} &=&
\left(-\frac{10813 D}{1204}-\frac{34223993}{33712}\right) \nu -\frac{52885925}{269696}\nonumber\\
&&+\frac{179
   d_5}{301}+\frac{384 d_6}{43}
\,.
\eea
In these expressions, $d_5$ and $d_6$ can be given arbitrary values.

\section{6PN-accurate f-route local Delaunay Hamiltonian} \label{DelaunayHlocf}

The 6PN-accurate f-route local effective Delaunay Hamiltonian (expressed in terms of $I_2 \equiv j$ and 
$I_3 \equiv i_r+j \equiv i_{r\phi}$, see Eqs. \eq{defiriphi}) is given by 
\beq 
\label{Hdelaunay_n}
\frac{ H_{\rm eff}^{\rm 6PN, loc,f}(I_2,I_3;\nu)}{\mu c^2}=\eta^{-2}+\sum_{k=0}^6\eta^{2k} \bar E_{\rm eff}^{2k}(I_2,I_3;\nu)+O(\eta^{14})\,.
\eeq
The coefficients up to the 5PN order (i.e., $O(\eta^{10})$) are listed in Table XI of Ref. \cite{Bini:2020nsb}.
We complete this result by adding the 6PN coefficient $\bar E_{\rm eff}^{12}$ (see Table \ref{table_barE12} below).


\begin{table*}[h]
\caption{\label{table_barE12} 6PN coefficient $\bar E_{\rm eff}^{12}$ entering the PN expansion of the Delaunay effective Hamiltonian \eqref{Hdelaunay_n}}
\begin{ruledtabular}
\begin{tabular}{ll}
$\bar E_{\rm eff}^{12}$&$
\Big[\nu ^2 \left(\frac{1911 a_6^{\nu^2}}{32}+\frac{63
   a_7^{\nu^2}}{16}+\frac{273 \bar d_5^{\nu^2}}{32}+\frac{21
   \bar d_6^{\nu^2}}{32}+\frac{9 q_{45}^{\nu^2}}{32}-\frac{529515 \pi
   ^4}{65536}+\frac{179354853 \pi
   ^2}{65536}-\frac{2062272503}{22400}\right)$\\
&$+\nu ^3 \left(-\frac{315
   a_6^{\nu^2}}{32}-\frac{63 \bar d_5^{\nu^2}}{32}+\frac{63
   a_7^{\nu^3}}{16}-\frac{24980025 \pi
   ^2}{65536}+\frac{978061}{64}\right)+\frac{819 \nu
   ^5}{256}+\left(\frac{38745 \pi ^2}{2048}-\frac{428085}{512}\right) \nu
   ^4$\\
&$
+\left(\frac{3236467169}{30240}+\frac{188085303629 \pi
   ^2}{50331648}-\frac{350055909 \pi ^4}{8388608}\right) \nu
   -\frac{14196819}{256}\Big]\frac{1}{I_2^{11} I_3^3}$\\ 
&$+\Big[\nu ^2 \left(\frac{315 a_6^{\nu^2}}{16}+\frac{63
   \bar d_5^{\nu^2}}{16}-\frac{45387 \pi ^4}{32768}+\frac{5648989 \pi
   ^2}{4096}-\frac{2795413}{48}\right)-\frac{117 \nu
   ^4}{2}+\left(\frac{48355}{8}-\frac{16113 \pi ^2}{128}\right) \nu
   ^3$\\
&$+\left(\frac{255513551}{1920}-\frac{55414387 \pi ^2}{32768}\right)
   \nu -\frac{16298667}{256}\Big]\frac{1}{I_2^{10} I_3^4}$\\
&$+\Big[\nu ^2 \left(-\frac{1225 a_6^{\nu^2}}{16}-\frac{35
   a_7^{\nu^2}}{8}-\frac{245 \bar d_5^{\nu^2}}{16}-\frac{35
   \bar d_6^{\nu^2}}{32}-\frac{21 q_{45}^{\nu^2}}{32}+\frac{176505 \pi
   ^4}{32768}-\frac{346823785 \pi
   ^2}{98304}+\frac{13914839443}{115200}\right)$\\
&$
+\nu ^3 \left(\frac{525
   a_6^{\nu^2}}{32}+\frac{133 \bar d_5^{\nu^2}}{32}-\frac{35
   a_7^{\nu^3}}{8}+\frac{1057777 \pi
   ^2}{1536}-\frac{16220123}{576}\right)-\frac{3465 \nu
   ^5}{256}+\left(\frac{3398185}{1536}-\frac{50225 \pi ^2}{1024}\right)
   \nu ^4$\\
&$
+\left(-\frac{34107960371}{345600}-\frac{590940624077 \pi
   ^2}{113246208}+\frac{387365405 \pi ^4}{8388608}\right) \nu
   +\frac{9066235}{256}\Big]\frac{1}{I_2^9 I_3^5}$\\
&$+\Big[\nu ^2 \left(-\frac{225 a_6^{\nu^2}}{8}-\frac{75
   \bar d_5^{\nu^2}}{8}+\frac{25215 \pi ^4}{16384}-\frac{12455065 \pi
   ^2}{4096}+\frac{3386395}{24}\right)+300 \nu ^4+\left(\frac{22755 \pi
   ^2}{64}-\frac{75595}{4}\right) \nu ^3$\\
&$
+\left(\frac{170705335 \pi
   ^2}{32768}-\frac{822324589}{2688}\right) \nu
   +\frac{33326145}{256}\Big]\frac{1}{I_2^8 I_3^6}$\\
&$+\Big[\nu ^2 \left(\frac{1095 a_6^{\nu^2}}{32}+\frac{15
   a_7^{\nu^2}}{16}+\frac{365 \bar d_5^{\nu^2}}{32}+\frac{15
   \bar d_6^{\nu^2}}{32}+\frac{15 q_{45}^{\nu^2}}{32}-\frac{25215 \pi
   ^4}{65536}+\frac{14759815 \pi
   ^2}{8192}-\frac{50590683}{1120}\right)$\\
&$
+\nu ^3 \left(-\frac{225
   a_6^{\nu^2}}{32}-\frac{85 \bar d_5^{\nu^2}}{32}+\frac{15
   a_7^{\nu^3}}{16}-\frac{49631575 \pi
   ^2}{98304}+\frac{23365741}{1152}\right)+\frac{2835 \nu
   ^5}{128}+\left(\frac{21525 \pi ^2}{512}-\frac{532105}{256}\right) \nu
   ^4$\\
&$
+\left(-\frac{22549379339}{423360}+\frac{161199909365 \pi
   ^2}{75497472}-\frac{81987555 \pi ^4}{8388608}\right) \nu
   +\frac{41670225}{512}\Big]\frac{1}{I_2^7 I_3^7}$\\
&$+\Big[\nu ^2 \left(\frac{63 a_6^{\nu^2}}{16}+\frac{63
   \bar d_5^{\nu^2}}{16}-\frac{11767 \pi ^4}{32768}+\frac{8802269 \pi
   ^2}{4096}-\frac{21668549}{180}\right)-462 \nu
   ^4+\left(\frac{67123}{4}-\frac{14637 \pi ^2}{64}\right) \nu
   ^3$\\
&$
+\left(\frac{8093748209}{28800}-\frac{590358503 \pi ^2}{98304}\right)
   \nu -\frac{11382315}{128}\Big]\frac{1}{I_2^6 I_3^8}$\\
&$+\Big[\nu ^2 \left(-\frac{25 a_6^{\nu^2}}{8}-\frac{25
   \bar d_5^{\nu^2}}{8}-\frac{\bar d_6^{\nu^2}}{32}-\frac{3
   q_{45}^{\nu^2}}{32}-\frac{26480553 \pi
   ^2}{32768}+\frac{130268403}{44800}\right)$\\
&$
+\nu ^3 \left(\frac{15
   a_6^{\nu^2}}{32}+\frac{15 \bar d_5^{\nu^2}}{32}+\frac{1432335 \pi
   ^2}{8192}-\frac{438383}{64}\right)-\frac{2205 \nu
   ^5}{128}+\left(\frac{219555}{256}-\frac{12915 \pi ^2}{1024}\right) \nu
   ^4$\\
&$
+\left(\frac{2754048127363}{25401600}+\frac{12817445435 \pi
   ^2}{12582912}-\frac{135909 \pi ^4}{8388608}\right) \nu
   -\frac{11393277}{64}\Big]\frac{1}{I_2^5 I_3^9}$\\
&$+\Big[270 \nu ^4+\left(\frac{3321 \pi ^2}{128}-\frac{12231}{2}\right) \nu
   ^3+\left(\frac{188409}{4}-\frac{1471401 \pi ^2}{4096}\right) \nu
   ^2+\left(\frac{57005721 \pi ^2}{32768}-\frac{3445375221}{22400}\right)
   \nu +\frac{3686445}{256}\Big]\frac{1}{I_2^4 I_3^{10}}$\\
&$+\Big[\frac{1575 \nu ^5}{256}+\left(\frac{1435 \pi
   ^2}{2048}-\frac{268555}{1536}\right) \nu
   ^4+\left(\frac{449845}{384}-\frac{915845 \pi ^2}{65536}\right) \nu
   ^3+\left(\frac{51899359}{67200}+\frac{7988539 \pi ^2}{65536}\right) \nu
   ^2$\\
&$
+\left(-\frac{8318335583}{176400}-\frac{27753064819 \pi
   ^2}{50331648}\right) \nu +\frac{76277895}{256}\Big]\frac{1}{I_2^3
   I_3^{11}}$\\
&$+\Big[-\frac{99 \nu ^4}{2}+\frac{8415 \nu ^3}{8}-10395 \nu
   ^2+\frac{4058505 \nu }{64}-\frac{34218855}{128}\Big]\frac{1}{I_2^2
   I_3^{12}}$\\
&$+\Big[-\frac{189 \nu ^5}{256}+\frac{9975 \nu ^4}{512}-\frac{15825 \nu
   ^3}{64}+\frac{1047627 \nu ^2}{512}-\frac{6699213 \nu
   }{512}+\frac{47435571}{512}\Big]\frac{1}{I_2
   I_3^{13}}
	-\frac{24188177}{2048}\frac{1}{I_3^{14}}$\\
\end{tabular}
\end{ruledtabular}
\end{table*}

In the case of circular motions, the f-route, local contribution to the 6PN-accurate effective energy 
$\bar E_{\rm eff} \equiv (E_{\rm eff}-\mu)/\mu$ is found to be
\begin{eqnarray}
\label{bar_E_delaunay_circ}
\bar E_{\rm eff}^{\rm circ} &=& -\frac{1}{2j^2} -\frac{9}{8}\frac{\eta^2}{j^4}+\left(-\frac{81}{16}+\nu\right)\frac{\eta^4}{j^6}
+\left[-\frac{3861}{128}+\left(-\frac{41}{64}\pi^2+\frac{157}{6}\right)\nu\right] \frac{\eta^6}{j^8}\nonumber\\
&+&\left[-\frac{53703}{256}+\left(-\frac{6581}{1024}\pi^2+\frac{8357}{30}\right)\nu+\left(-\frac{275}{12}+\frac{41}{64}\pi^2\right)\nu^2\right] \frac{\eta^8}{j^{10}}\nonumber\\
&+&\left[-\frac{1648269}{1024}+\left(\frac{15592753}{6300}-\frac{31547}{1536}\pi^2\right)\nu+\left(-\frac{4725}{8}+\frac12 a_6^{\nu^2}+\frac{2337}{128}\pi^2\right)\nu^2+2\nu^3\right] \frac{\eta^{10}}{j^{12}}\nonumber\\
&+&\left[-\frac{27078705}{2048}+\left(-\frac{2800873}{524288}\pi^4+\frac{298273237}{14175}+\frac{1322752463}{3538944}\pi^2\right)\nu
\right. \nonumber\\
&&+\left(-\frac{1681}{512}\pi^4+\frac{39}{4} a_6^{\nu^2}+\frac{1389451}{3072}\pi^2-\frac{3321439}{288}+\frac12 a_7^{\nu^2}\right)\nu^2\nonumber\\
&&\left.+\left(-\frac{615}{64}\pi^2+\frac{1369}{4}+\frac12 a_7^{\nu^3}\right)\nu^3\right]\frac{\eta^{12}}{j^{14}}\,.
\end{eqnarray}

The relation between $\bar E_{\rm eff}$ and the specific binding energy $\bar E$, Eq. \eq{defbarE}, is given by
\beq
\bar E=\frac{\sqrt{1+2\nu \bar E_{\rm eff} }-1}{\nu}\,,
\eeq
so that in the circular case we get
\begin{eqnarray}
\label{bar_E_h_circ}
\bar E^{\rm circ}_{\rm loc, f} &=& -\frac{1}{2j^2} 
+\left(-\frac{\nu }{8}-\frac{9}{8}\right)\frac{\eta^2}{j^4}
+\left(-\frac{\nu ^2}{16}+\frac{7 \nu }{16}-\frac{81}{16}\right)\frac{\eta^4}{j^6}
+\left[-\frac{5 \nu ^3}{128}+\frac{5 \nu ^2}{64}+\left(\frac{8833}{384}-\frac{41
   \pi ^2}{64}\right) \nu -\frac{3861}{128}
\right] \frac{\eta^6}{j^8}\nonumber\\
&+&
\left[-\frac{7 \nu ^4}{256}+\frac{3 \nu ^3}{128}+\left(\frac{41 \pi
   ^2}{128}-\frac{8875}{768}\right) \nu
   ^2+\left(\frac{989911}{3840}-\frac{6581 \pi ^2}{1024}\right) \nu
   -\frac{53703}{256}
\right] \frac{\eta^8}{j^{10}}\nonumber\\
&+&
\left[\left(\frac{a_6^{\nu^2}}{2}+\frac{29335 \pi
   ^2}{2048}-\frac{1679647}{3840}\right) \nu ^2-\frac{21 \nu
   ^5}{1024}+\frac{5 \nu ^4}{1024}+\left(\frac{41 \pi
   ^2}{512}-\frac{3769}{3072}\right) \nu
   ^3\right.\nonumber\\
	&&\left.
	+\left(\frac{3747183493}{1612800}-\frac{31547 \pi ^2}{1536}\right)
   \nu -\frac{1648269}{1024}
\right] \frac{\eta^{10}}{j^{12}}\nonumber\\
&+&
\left[\nu ^2 \left(\frac{39
   a_6^{\nu^2}}{4}+\frac{a_7^{\nu^2}}{2}-\frac{1681 \pi
   ^4}{512}+\frac{10605841 \pi
   ^2}{24576}-\frac{10727952929}{1075200}\right)+\nu ^3
   \left(\frac{a_6^{\nu^2}}{4}+\frac{a_7^{\nu^3}}{2}-\frac{21383
   \pi ^2}{8192}+\frac{1007737}{7680}\right)\right.\nonumber\\
	&&\left.
	-\frac{33 \nu
   ^6}{2048}-\frac{7 \nu ^5}{2048}+\left(\frac{41 \pi
   ^2}{1024}-\frac{2537}{3072}\right) \nu
   ^4+\left(\frac{576215112401}{29030400}+\frac{1322752463 \pi
   ^2}{3538944}-\frac{2800873 \pi ^4}{524288}\right) \nu
   -\frac{27078705}{2048}
\right]\frac{\eta^{12}}{j^{14}}\,.\nonumber\\
\end{eqnarray}

The f-route, local 6PN-accurate periastron advance (along arbitrary eccentric orbits),
expressed in terms of $\bar E$ and $j$, reads
\bea
K(\bar E,j)_{\rm loc, f}&=&1+\frac{3}{j^2} \eta ^2
+\left[\left(\frac{15}{2}-3 \nu\right)\frac{\bar E}{j^2}+\left(\frac{105}{4}-\frac{15 \nu
   }{2}\right)\frac{1}{j^4}\right]\eta ^4\nonumber\\
	&&
+\left[\left(3 \nu ^2-\frac{15 \nu
   }{4}+\frac{15}{4}\right)\frac{\bar E^2}{j^2}+
   \left(\frac{45 \nu ^2}{2}+\left(\frac{123 \pi ^2}{64}-218\right) \nu
   +\frac{315}{2}\right)\frac{\bar E}{j^4}\right.\nonumber\\
	&&\left.
	+\left(\frac{105 \nu ^2}{8}+\left(\frac{615
   \pi ^2}{128}-\frac{625}{2}\right) \nu +\frac{1155}{4}\right)\frac{1}{j^6}\right]\eta ^6\nonumber\\
	&&
+\left[
\left(\frac{15 \nu ^2}{4}-3 \nu
   ^3\right)\frac{\bar E^3}{j^2}+\left(-45 \nu
   ^3+\left(\frac{4045}{8}-\frac{615 \pi ^2}{128}\right) \nu
   ^2+\left(\frac{35569 \pi ^2}{2048}-\frac{20323}{24}\right) \nu
   +\frac{4725}{16}\right)\frac{\bar E^2}{j^4}\right.\nonumber\\
	&&
	+\left(-\frac{525 \nu ^3}{8}+\left(\frac{35065}{16}-\frac{615 \pi
   ^2}{16}\right) \nu ^2+\left(\frac{257195 \pi
   ^2}{2048}-\frac{293413}{48}\right) \nu
   +\frac{45045}{16}\right)\frac{\bar E}{j^6}\nonumber\\
	&&\left.	
	+\left(-\frac{315 \nu
   ^3}{16}+\left(\frac{132475}{96}-\frac{7175 \pi ^2}{256}\right) \nu
   ^2+\left(\frac{2975735 \pi ^2}{24576}-\frac{1736399}{288}\right) \nu
   +\frac{225225}{64}\right)\frac{1}{j^8}
\right]\eta ^8\nonumber\\
	&&
+\left[
\left(\nu ^2 \left(-\frac{15
   a_6^{\nu^2}}{4}-\frac{15 \bar d_5^{\nu^2}}{4}-\frac{1203065 \pi
   ^2}{2048}+\frac{310189}{12}\right)+\frac{1575 \nu
   ^4}{8}+\left(\frac{35055 \pi ^2}{256}-\frac{240585}{32}\right) \nu
   ^3\right.\right.\nonumber\\
	&&\left.
	+\left(\frac{4899565 \pi ^2}{4096}-\frac{33023719}{840}\right) \nu
   +\frac{315315}{32}\right)\frac{\bar E^2}{j^6}\nonumber\\
	&&
	+\left(\nu ^2 \left(-\frac{105 a_6^{\nu^2}}{4}-\frac{35
   \bar d_5^{\nu^2}}{4}-\frac{12964665 \pi
   ^2}{8192}+\frac{549451}{8}\right)+\frac{2205 \nu
   ^4}{16}+\left(\frac{121975 \pi ^2}{512}-\frac{271705}{24}\right) \nu
   ^3\right.\nonumber\\
	&&\left.
	+\left(\frac{16173395 \pi ^2}{8192}-\frac{30690127}{240}\right) \nu
   +\frac{765765}{16}\right)\frac{\bar E}{j^8}+
   \left(3 \nu ^4-\frac{15 \nu ^3}{4}+\frac{15 \nu
   ^2}{16}\right)\frac{\bar E^4}{j^2}\nonumber\\
	&&
	+\left(75
   \nu ^4+\left(\frac{1107 \pi ^2}{128}-\frac{7113}{8}\right) \nu
   ^3+\left(\frac{9689}{6}-\frac{35569 \pi ^2}{1024}\right) \nu
   ^2+\left(\frac{15829 \pi ^2}{256}-\frac{12160657}{8400}\right) \nu
   +\frac{3465}{16}\right)\frac{\bar E^3}{j^4}\nonumber\\
	&&
	+\left(\nu ^2 \left(-\frac{315
   a_6^{\nu^2}}{16}-\frac{63 \bar d_5^{\nu^2}}{16}-\frac{15796431
   \pi ^2}{16384}+\frac{5156991}{128}\right)+\frac{3465 \nu
   ^4}{128}+\left(\frac{90405 \pi ^2}{1024}-\frac{127995}{32}\right) \nu
   ^3\right.\nonumber\\
	&&\left.\left.
	+\left(\frac{1096263 \pi ^2}{1024}-\frac{61358067}{640}\right) \nu
   +\frac{2909907}{64}\right)\frac{1}{j^{10}}
\right]\eta ^{10}\nonumber
\eea
\bea
&&
+\left[
\left(-3 \nu ^5+\frac{15 \nu ^4}{4}-\frac{15 \nu ^3}{16}\right)\frac{\bar E^5}{j^2}
+\left(-\frac{225 \nu ^5}{2}+\left(\frac{2737}{2}-\frac{861 \pi ^2}{64}\right)
   \nu ^4+\left(\frac{462397 \pi ^2}{8192}-\frac{247189}{96}\right) \nu
   ^3\right.\right.\nonumber\\
	&&\left.\left.
	+\left(\frac{25148189}{11200}-\frac{47487 \pi ^2}{512}\right) \nu
   ^2+\left(\frac{104950259 \pi ^2}{1048576}-\frac{25669261}{29400}\right)
   \nu +\frac{3465}{64}\right)\frac{\bar E^4}{j^4}\right.\nonumber\\
	&&
+\left(\nu ^2 \left(-\frac{25 a_6^{\nu^2}}{2}-\frac{25
   \bar d_5^{\nu^2}}{2}-\frac{5 \bar d_6^{\nu^2}}{4}-\frac{15
   q_{45}^{\nu^2}}{4}-\frac{8648125 \pi
   ^2}{2048}+\frac{58338869}{480}\right)\right.\nonumber\\
	&&
	+\nu ^3 \left(15
   a_6^{\nu^2}+15 \bar d_5^{\nu^2}+\frac{6795375 \pi
   ^2}{4096}-\frac{2255935}{32}\right)-\frac{3675 \nu
   ^5}{8}+\left(\frac{601625}{32}-\frac{89175 \pi ^2}{256}\right) \nu
   ^4\nonumber\\
	&&\left.
	+\left(-\frac{39123984017}{635040}-\frac{3128795225 \pi
   ^2}{1572864}-\frac{679545 \pi ^4}{1048576}\right) \nu
   +\frac{525525}{32}\right)\frac{\bar E^3}{j^6}\nonumber\\
	&&
+\left(\nu ^2 \left(-\frac{735 a_6^{\nu^2}}{2}-\frac{105
   a_7^{\nu^2}}{4}-\frac{245 \bar d_5^{\nu^2}}{2}-\frac{105
   \bar d_6^{\nu^2}}{8}-\frac{105 q_{45}^{\nu^2}}{8}+\frac{176505 \pi
   ^4}{16384}-\frac{28607145 \pi
   ^2}{1024}+\frac{890209513}{960}\right)\right.\nonumber\\
	&&
	+\nu ^3 \left(\frac{735
   a_6^{\nu^2}}{4}+70 \bar d_5^{\nu^2}-\frac{105
   a_7^{\nu^3}}{4}+\frac{392482055 \pi
   ^2}{49152}-\frac{48508187}{144}\right)-\frac{2205 \nu
   ^5}{4}+\left(\frac{9492035}{192}-\frac{1083425 \pi ^2}{1024}\right) \nu
   ^4\nonumber\\
	&&\left.
	+\left(-\frac{36266340619}{60480}-\frac{749028566195 \pi
   ^2}{18874368}+\frac{573912885 \pi ^4}{2097152}\right) \nu
   +\frac{16081065}{64}\right)\frac{\bar E^2}{j^8}\nonumber\\
	&&
+\left(\nu ^2 \left(-\frac{17325 a_6^{\nu^2}}{16}-\frac{315
   a_7^{\nu^2}}{4}-\frac{3465 \bar d_5^{\nu^2}}{16}-\frac{315
   \bar d_6^{\nu^2}}{16}-\frac{189 q_{45}^{\nu^2}}{16}+\frac{1588545
   \pi ^4}{16384}-\frac{446396685 \pi
   ^2}{8192}+\frac{12054492193}{6400}\right)\right.\nonumber\\
	&&
	+\nu ^3 \left(\frac{4725
   a_6^{\nu^2}}{16}+\frac{1197 \bar d_5^{\nu^2}}{16}-\frac{315
   a_7^{\nu^3}}{4}+\frac{11337249 \pi
   ^2}{1024}-\frac{7270879}{16}\right)-\frac{31185 \nu
   ^5}{128}+\left(\frac{10090605}{256}-\frac{452025 \pi ^2}{512}\right)
   \nu ^4\nonumber\\
	&&\left.
	+\left(-\frac{31568079821}{19200}-\frac{551913398477 \pi
   ^2}{6291456}+\frac{3486288645 \pi ^4}{4194304}\right) \nu
   +\frac{101846745}{128}\right)\frac{\bar E}{j^{10}}\nonumber\\
	&&
+\left(\nu ^2 \left(-\frac{21021 a_6^{\nu^2}}{32}-\frac{693
   a_7^{\nu^2}}{16}-\frac{3003 \bar d_5^{\nu^2}}{32}-\frac{231
   \bar d_6^{\nu^2}}{32}-\frac{99 q_{45}^{\nu^2}}{32}+\frac{5824665
   \pi ^4}{65536}-\frac{1972903383 \pi
   ^2}{65536}+\frac{22684997533}{22400}\right)\right.\nonumber\\
	&&
	+\nu ^3 \left(\frac{3465
   a_6^{\nu^2}}{32}+\frac{693 \bar d_5^{\nu^2}}{32}-\frac{693
   a_7^{\nu^3}}{16}+\frac{274780275 \pi
   ^2}{65536}-\frac{10758671}{64}\right)-\frac{9009 \nu
   ^5}{256}+\left(\frac{4708935}{512}-\frac{426195 \pi ^2}{2048}\right)
   \nu ^4\nonumber\\
	&&\left.\left.
	+\left(-\frac{35601138859}{30240}-\frac{2068938339919 \pi
   ^2}{50331648}+\frac{3850614999 \pi ^4}{8388608}\right) \nu
   +\frac{156165009}{256}\right)\frac{1}{j^{12}}
\right]\eta ^{12}
\,.
\eea
 In the circular case this reduces to Eq. \eqref{Klocfcirc}.

\section{Completing the information on the h-route, nonlocal $q_8$ EOB potential} \label{q8}

One of the intermediate steps of our analysis is to transform the h-route (i.e., $r_{12}^h$-scaled) nonlocal
Hamiltonian, $ H_{\rm nonloc, h}^{4+5+6 \rm PN}(t)$, defined in Eq. \eq{Hnonloch}, into its gauge-equivalent
EOB potentials, $A^{\rm nonloc, h}(r)$, ${\bar D}^{\rm nonloc, h}(r)$ and ${\widehat Q}^{\rm nonloc, h}(r,p_r)$.
We have listed the PN-expansion coefficients of these potentials in Table  IV of  \cite{Bini:2020wpo} (for the
4+5PN-level contributions), and in Table VI of \cite{Bini:2020nsb} (for the 6PN-level contributions). 
However, we did not include in Table  IV of  \cite{Bini:2020wpo} the values of the 4+5PN-level coefficients
entering the $q_8$ EOB potential, i.e. the coefficients denoted $q_{81}^{\rm nonloc}$ and $q_{82}^{\rm nonloc}$
in the last line of Eq. (2.22) in \cite{Bini:2020wpo}. The aim of this Appendix is to remedy this gap by giving
the 6PN-accurate values of the PN expansion coefficients of the $O(p_r^8)$ part of ${\widehat Q}^{\rm nonloc, h}_{4+5+6 \rm PN}(r,p_r;\nu)$, namely
\beq
\left[{\widehat Q}^{\rm nonloc, h}_{4+5+6 \rm PN}(r,p_r;\nu)\right]_{p_r^8}=  p_r^8 \,q_8(u;\nu)=p_r^8 \left(q_{81}^{\rm nonloc, h}(\nu) u+  q_{82}^{\rm nonloc, h}(\nu)u^2+  q_{83}^{\rm nonloc, h}(\nu)u^3 \right).
\eeq
As indicated here, the coefficients $q_{81}^{\rm nonloc, h}$ (4PN level),  $q_{82}^{\rm nonloc, h}$(5PN level) and  
$q_{83}^{\rm nonloc, h}$ (6PN level) depend only on $\nu$, i.e., they do not involve any $\ln u$ contribution.
[The logarithmic contributions come from the $2\frac{G H_{}}{c^{5}}{\mathcal F}^{\rm split}_{\rm 2PN}(t,t)  \ln \left( \frac{r_{12}^h(t)}{s}\right)$ term in Eq. \eq{Hnonloch} and start contributing to $q_8$ at the 7PN level.]

Though we have already given $q_{83}^{\rm nonloc, h}(\nu)$ in Table VI of \cite{Bini:2020nsb}, let us, for clarity,
list here all the PN coefficients of $q_8$
\bea
q_{81}^{\rm nonloc,h}(\nu)&=&\left(\frac{21668992}{45} \ln(2)+\frac{6591861}{350} \ln(3)-\frac{27734375}{126} \ln(5)-\frac{35772}{175} \right)\nu \nonumber\\
q_{82}^{\rm nonloc,h}(\nu)&=&\left(\frac{703189497728}{33075}\ln(2)+\frac{869626}{525}+\frac{332067403089}{39200}\ln(3)-\frac{468490234375}{42336}\ln(5)-\frac{13841287201}{4320}\ln(7)\right)\nu^2 \nonumber\\
&&+\left(\frac{5788281}{2450}-\frac{16175693888}{1575}\ln(2)-\frac{393786545409}{156800}\ln(3)+\frac{875090984375}{169344}\ln(5)\right. \nonumber\\
&&\left.+\frac{13841287201}{17280}\ln(7)\right)\nu \nonumber\\
q_{83}^{\rm nonloc,h}(\nu)&=&\left(-\frac{154862}{21}+\frac{57604236136064}{99225}\ln(2)+\frac{10467583300341}{39200}\ln(3)-\frac{73366198046875}{381024}\ln(5)\right. \nonumber\\
&&\left.-\frac{7709596970957}{38880}\ln(7)\right) \nu^3  \nonumber\\
&&+\left(-\frac{1746293}{70}-\frac{177055674739808}{297675}\ln(2)-\frac{43719724468071}{156800}\ln(3)+\frac{366449151015625}{1524096}\ln(5)\right. \nonumber\\
&&\left.+\frac{26506549233199}{155520}\ln(7)\right)\nu^2\nonumber\\
&&+\left(-\frac{709195549}{132300}+\frac{5196312336176}{35721}\ln(2)+\frac{17515638027261}{313600}\ln(3)-\frac{63886617280625}{1016064}\ln(5)\right. \nonumber\\
&&\left.-\frac{29247366220639}{933120}\ln(7)\right)\nu 
\eea
For completeness, let us also mention that our self-force computation of the full (local-plus-nonlocal) $q_8$
potential  has given the result
\beq
q_{8,\le \rm 6PN}^{\rm loc+nonloc}=\nu (B_1u +B_2u^2 +B_3u^3) + O(\nu^2)\,,
\eeq
where
\bea
B_1&=&-\frac{27734375}{126}\ln(5)+\frac{6591861}{350}\ln(3)+\frac{21668992}{45}\ln(2)-\frac{35772}{175} ,\nonumber\\
B_2&=&\frac{13841287201}{17280}\ln(7)-\frac{393786545409}{156800}\ln(3)-\frac{16175693888}{1575}\ln(2)+\frac{875090984375}{169344}\ln(5)+\frac{5790381}{2450}, \nonumber\\
B_3&=&-\frac{29247366220639}{933120}\ln(7)-\frac{63886617280625}{1016064}\ln(5)+\frac{5196312336176}{35721}\ln(2)\nonumber\\
&&+\frac{17515638027261}{313600}\ln(3)-\frac{2843819611}{529200} \,.
\eea
The difference 
\beq
\Delta q_{8,\le \rm 6PN}= q_{8,\le 6PN}^{\rm loc+nonloc}-q_{8,\le \rm 6PN}^{\rm nonloc, h}\,,
\eeq
was one of our sources of information for deriving the local part of the Hamiltonian, and is equal to
\beq
\Delta q_{8,\le \rm 6PN}=\nu \left(\frac{6}{7} u^2 -\frac{7447}{560} u^3 \right)+ O(\nu^2)\,.
\eeq


\section{Computing the Delaunay near-zone nonlocal Hamiltonian associated with the $\ln(r^h_{12}/s)$ term along ellipticlike motion} \label{Hlog}

Let us consider the 4+5+6PN nonlocal, h-route (unflexed) Hamiltonian \eqref{Hnonloch}.
We compute here the  Delaunay-average (along an ellipticlike motion) of  the $\ln(r^h_{12}/s)$ contribution
to  $H^{\rm nonloc, h}$, i.e.,
\beq
\langle  H^{\rm nonloc, \ln, h}_{4+5+6 \rm PN}\rangle =\frac{1}{\oint dt}\oint 2\frac{G H}{c^{5}}{\mathcal F}^{\rm split}_{\rm 2PN}(t,t)  \ln \left( \frac{r_{12}^h(t)}{s}\right)dt\,,
\eeq
where
\beq
H=Mc^2(1+\nu \bar E\eta^2)\,,
\eeq
and
\beq
\bar E=-\frac{1}{2a_r^h}-\frac12\left(-\frac74+\frac14\nu\right)\frac{\eta^2}{(a_r^h)^2}+O(\eta^4)\,.
\eeq
It can be written as
\begin{eqnarray}
\langle  H_{\rm nonloc, \ln, h}^{4+5+6 \rm PN}\rangle&=&
\frac{\nu^2}{(a_r^h)^5}\left[{\mathcal A}_{\ln{}}^{\rm 4PN}(e_t^h)+{\mathcal B}_{\ln{}}^{\rm 4PN}(e_t^h)\ln \left(\frac{a_r^h}{s}\right)\right]\nonumber\\
&+&
\frac{\nu^2}{(a_r^h)^6}\left[{\mathcal A}_{\ln{}}^{\rm 5PN}(e_t^h)+{\mathcal B}_{\ln{}}^{\rm 5PN}(e_t^h)\ln \left(\frac{a_r^h}{s}\right)\right]
\nonumber\\
&+&
\frac{\nu^2}{(a_r^h)^7}\left[{\mathcal A}_{\ln{}}^{\rm 6PN}(e_t^h)+{\mathcal B}_{\ln{}}^{\rm 6PN}(e_t^h)\ln \left(\frac{a_r^h}{s}\right)\right]\,.
\end{eqnarray}
Here the non-logarithmic coefficients, ${\mathcal A}_{\ln{}}^{\rm nPN}$, were obtained as expansions in powers of $e_t^h$ up to the order $O((e_t^h)^{10})$ included,
\begin{eqnarray}
{\mathcal A}_{\ln{}}^{\rm 4PN}(e_t^h)&=&
-\frac{176}{5}(e_t^h)^2-\frac{2681}{15}(e_t^h)^4-\frac{90017}{180}(e_t^h)^6-\frac{306433}{288}(e_t^h)^8-\frac{18541327}{9600}(e_t^h)^{10}
\,,\nonumber\\
{\mathcal A}_{\ln{}}^{\rm 5PN}(e_t^h)&=&
\left(\frac{18964}{105}\nu+\frac{2539}{35}\right)(e_t^h)^2+\left(\frac{55521}{35}\nu-\frac{524087}{840}\right)(e_t^h)^4+\left(\frac{456341}{72}\nu-\frac{11468869}{2520}\right)(e_t^h)^6\nonumber\\
&&
+\left(\frac{2526889}{144}\nu-\frac{140341413}{8960}\right)(e_t^h)^8+\left(\frac{251185649}{6400}\nu-\frac{1320019027}{33600}\right)(e_t^h)^{10}
\,,\nonumber\\
{\mathcal A}_{\ln{}}^{\rm 6PN}(e_t^h)&=&
\left(-\frac{9448}{27}\nu^2-\frac{907927}{630}\nu+\frac{2489}{45}\right)(e_t^h)^2+\left(-\frac{44830903}{7560}\nu^2-\frac{3709639}{1680}\nu+\frac{460759}{4536})(e_t^h\right)^4\nonumber\\
&&
+\left(-\frac{1067440939}{30240}\nu^2+\frac{56364713}{2016}\nu-\frac{1114216909}{68040}\right)(e_t^h)^6\nonumber\\
&&
+\left(-\frac{699238489}{5376}\nu^2+\frac{28209572539}{161280}\nu-\frac{76207852937}{725760}\right)(e_t^h)^8\nonumber\\
&&
+\left(-\frac{586193581933}{1612800}\nu^2+\frac{325106833717}{537600}\nu-\frac{76717484827}{201600}\right)(e_t^h)^{10}
\,,
\end{eqnarray}
 whereas the logarithmic coefficients, ${\mathcal B}_{\ln{}}^{\rm nPN}$, are given by the following closed-form expressions
\begin{eqnarray}
{\mathcal B}_{\ln{}}^{\rm 4PN}(e_t^h)&=&
\frac1{(1-e_t^2)^{7/2}}\left[
\frac{64}{5}+\frac{584}{15}(e_t^h)^2+\frac{74}{15}(e_t^h)^4
\right]
\,,\nonumber\\
{\mathcal B}_{\ln{}}^{\rm 5PN}(e_t^h)&=&
\frac1{(1-e_t^2)^{9/2}}\left[
-\frac{11708}{105}-\frac{112}{5}\nu+\left(-\frac{5308}{15}\nu+\frac{1378}{7}\right)(e_t^h)^2+\left(-\frac{1857}{5}\nu+\frac{8941}{10}\right)(e_t^h)^4\right.\nonumber\\
&&\left.
+\left(-\frac{74}{3}\nu+\frac{12539}{140}\right)(e_t^h)^6
\right]
\,,\nonumber\\
{\mathcal B}_{\ln{}}^{\rm 6PN}(e_t^h)&=&
\frac1{(1-e_t^2)^{11/2}}\left[
\frac{32}{5}\nu^2+\frac{179234}{315}\nu+\frac{1445692}{2835}+\left(\frac{13547}{15}\nu^2+\frac{821056}{315}\nu-\frac{10378222}{2835}\right)(e_t^h)^2\right.\nonumber\\
&&
+\left(\frac{220447}{60}\nu^2+\frac{3723539}{1890}-\frac{1062751}{105}\nu\right)(e_t^h)^4+\left(\frac{9393}{5}\nu^2+\frac{15416687}{1260}-\frac{7764587}{840}\nu\right)(e_t^h)^6\nonumber\\
&&
+\left(\frac{4979519}{5040}-\frac{204661}{420}\nu+74\nu^2\right)(e_t^h)^8\nonumber\\
&&\left.
+\left(-96+1060(e_t^h)^2+1863(e_t^h)^4+148(e_t^h)^6\right)\left(1-\frac{2}{5}\nu\right)(1-e_t^2)^{1/2}
\right]
\,.
\end{eqnarray}
\end{widetext}
The latter coefficients are  related via ${\mathcal B}_{\ln{}}^{\rm nPN}(e_t^h)=-2{\mathcal B}^{\rm nPN}(e_t^h)+O((e_t^h)^{11})$, to those entering the full Delaunay-averaged h-route nonlocal Hamiltonian \eqref{Hnonloch} (see Eq. \eqref{H6delaunay}).


\end{document}